\documentclass[11pt,onecolumn]{article}

\usepackage{graphicx}
\usepackage{geometry}
\usepackage{amsmath,amssymb}
\usepackage{graphicx}
\usepackage{cite}
\usepackage{hyperref}
\usepackage{url}

\usepackage{authblk}
\usepackage{adjustbox}
\usepackage{subcaption}
\usepackage{caption}
\usepackage{enumitem}
\usepackage{float}
\usepackage{ragged2e}
\usepackage{microtype}
\usepackage{changepage}
\usepackage{times}

\title{\textbf{Public Perception of AI: Sentiment and Opportunity}}
\author{Jayshree Seth}
\affil{3M}
\date{July 22, 2024}

\begin{document}

\maketitle

\vspace{\baselineskip}
\vspace{\baselineskip}
\begin{adjustwidth}{1.4cm}{1.4cm}
\begin{abstract}
\noindent As Artificial Intelligence (AI) increasingly influences various aspects of society, there is growing public interest in its potential benefits and risks. In this paper we present results of public perception of AI from a survey conducted with 10,000 respondents spanning ten countries in four continents around the world. The results show that currently an equal percentage of respondents who believe AI will change the world as we know it, also believe AI needs to be heavily regulated. However, our findings also indicate that despite the general sentiment among the global public that AI will replace workers, if a company were to use AI to innovate to improve lives, the public would be more likely to think highly of the company, purchase from them and even be interested in a job in that company. Our results further reveal that the global public largely views AI as a tool for problem solving. These nuanced results underscore the importance of AI directed towards challenges that the public would like science and technology-based innovations to address. We draw on a multi-year 3M study of public perception of science to provide further context on what the public perceives as important problems to be solved. 
\end{abstract}
\end{adjustwidth}

\vspace{\baselineskip}
\vspace{\baselineskip}
\section*{Introduction}
In the last two years with the launch of Generative AI/Large Language Models (LLM), the public’s familiarity and interest in AI has increased. With the accelerated development and increasing discourse around AI, discussion of the benefits of AI and the drawbacks of the technology are also active topics in the media. Understanding the public perception of AI is important because it impacts how the technology is devised, developed and deployed for widespread access, adoption and acceptance. Societal perception can also impact policy agenda, research investments and industry practices. As AI becomes more prevalent in our daily lives, understanding what drives public perception and how it could be improved becomes crucial – it impacts the development of responsible and ethical frameworks and is critical for maximizing the benefits of AI while mitigating risks and concerns. In this paper we discuss results of a global survey and provide recommendations for improving public perception of this technology.

\clearpage
\vspace{\baselineskip}
\section*{Background}
3M has been studying the public perception of science since the first survey was fielded in 2017 in 14 countries around the world (See Appendix A1 for the list and relevant details of surveys commissioned by 3M). Over the years, the survey questions have included the topics of AI, robotics and automation. 

\vspace{\baselineskip}
\noindent{\textit{\textbf{3M State of Science Surveys}}}\\
In the 3M State of Science Index (SOSI) launched in 2018, 64\% of the global respondents thought there would be robots in every workplace during their lifetime \cite{1}. In 2019 3M SOSI, healthcare, climate change and sustainability related challenges were the top picks from a pre-populated list of issues that science should most help to solve, while only 8\% globally wanted science to help with workplace automation/AI/robotics \cite{2}. 

In 2020, pre-pandemic 3M SOSI, 77\% globally thought they knew some or a lot about AI, and 60\% globally considered AI as an improvement (not a threat) to society, with 51\% considering any impact of integration of AI/machine learning on their job as mostly positive \cite{3}. However, 64\% globally wanted the government to be more involved in ethics around automation/AI /robotics and 33\% viewed AI/automation replacing jobs as a threat to STEM careers. In 2021 3M SOSI, featuring at number eight out of ten pre-populated options, “data analytics and digital health records to track and improve patient health outcomes” was chosen by 32\% in their top four choices when asked about healthcare advancements science should prioritize beyond coronavirus/COVID-19. The most common answer chosen was “vaccines for future pandemics” with 52\% putting it in their top four choices \cite{4}. 

In 2022 3M SOSI, 54\% agreed that they trust how private companies are using AI and 65\% globally agreed that AI is an exciting technology that impacts their daily lives \cite{5}. However, 47\% also agreed that they worried that advancements in AI within next five years will cause them to lose their jobs. In 2023 3M SOSI, the percentage that agreed AI is an exciting technology was at 75\% globally, while 68\% of Americans, and 80\% globally, agreed that AI can help build a more sustainable future \cite{6}.

\vspace{\baselineskip}
\noindent{\textit{\textbf{Surveys focused on public perception of AI}}}\\
The last few years have seen an increase in survey-based studies to specifically understand public perception of AI. Survey responses generally indicate contradictory emotions, in that the public sees the technology having a significant impact due to its benefits as well as drawbacks, with typically the more favorable impressions being in emerging and/or Asian markets and more negative impressions in Western countries \cite{7}. 

A 2016 paper examined trends in the perception of AI over 30 years and concluded that the public had been more optimistic than pessimistic, however, the fear of loss of control of AI, had increased over the years while hopes for AI increased over time \cite{8}. A 2021 global survey of 10,000 citizens, spanning eight countries across six continents, reported a mix of positive and negative feelings about AI among respondents \cite{9}.

In another recent survey, 66\% of respondents agreed that products and services using AI will profoundly change their daily lives in the next 3-5 years and 49\% of respondents say that products and services using AI have profoundly changed their daily life in the past 3-5 years \cite{10}. An Ipsos survey reported 46\% of workers in the US, UK, and Australia have used AI tools while at work \cite{11}. 
Although it appears unclear how the impact on jobs will evolve and eventually whether AI will replace or create jobs, job loss continues to be an area of concern. A recent survey showed 49\% of respondents are concerned about the impact of AI on jobs \cite{12}. In the World Economic Forum Future of Jobs Report 2023, 75\% of companies surveyed said they expect to adopt AI and 50\% of the firms expect jobs to be created as a result while 25\% expect job declines \cite{13}.

\clearpage
A recent consensus paper in the journal Science describes extreme risks to humans from advanced AI systems. These include large-scale social harms and injustice, malicious and criminal uses, and an irreversible loss of human control over autonomous AI systems that can facilitate warfare, conduct surveillance, customize mass manipulation and more \cite{14}. With much discussion of these topics in media, not surprisingly interest in AI safety policies, regulation, and support for more intentional approaches in order to safeguard human values and existence is growing  \cite{15}. 

\vspace{\baselineskip}
\noindent{\textit{\textbf{AI as a tool to help humans solve human problems}}}\\
Public perception of AI is currently at a crossroads and there have been some efforts to understand what drives these sentiments \cite{16}. A gap between people’s competency and their perceptions and a lack of accurate understanding of AI, can contribute to the public overestimating the abilities of this emerging technology and underestimating its limitations \cite{17,18}. However, there is also the acceptance and awareness of AI’s transformative power and that if managed carefully, the technology could serve human interests by helping to cure diseases, elevate living standards, and protect the environment \cite{19}. In a recent 21 country survey, 39\% responded positively regarding the value AI can bring particularly to improve their daily lives \cite{20}.

US President’s Council of Advisors on Science and Technology (PCAST) recently published a report that emphasizes that AI should not replace but instead be employed to empower human scientists \cite{21}. The report explores how AI can “supercharge” research and play a crucial role in tackling major societal challenges. Augmenting human creativity, critical thinking and ingenuity with AI to do data intensive analysis is an effective use of AI as a tool for problem-solving \cite{22}.  Especially given that general-purpose AI can be ‘brittle’ in that it can excel in some domains but prone to failure in others because of lack of human capability for contextual understanding and abstract reasoning \cite{23}.

AI directed towards social good has become one of the goals of AI strategies and regulation \cite{24}. Studies have shown that approval of AI can be dependent on the context of application. There is higher acceptance in sectors that people consider critical areas - where the use of advanced technologies, like AI, could ensure greater progress \cite{17}. Experts believe that AI has the potential to make a positive difference for UN Sustainable Development Goals (SDGs) \cite{25}. A recent survey conducted in Taiwan linked AI to UN SDGs and the SDGs for quality education, industry innovation, and good health had the highest support rate amongst the public\cite{26}.

A survey from Australia showed consistently high levels of agreement for using AI for ‘social good’ i.e. to address social, humanitarian and environmental challenges \cite{27}. When respondents were asked to rate their support to the use of AI in 5 different areas of possible implementation, chosen randomly from an overall list of 12 areas, the top 3 highest proportions of ‘strong support’ for AI use were in the areas of health (44.1\%) and medicine (43\%) followed by ‘environmental challenges’ (41.6\%). The researchers note that despite the contentious nature of issues such as environment, the support did not vary along demographic lines of age, status or political preference. 

With AI rapidly advancing and poised to shape virtually every aspect of human lives, it is crucial to consider the public opinion and implications as it relates to widespread acceptance and adoption. Given that, expert recommendations now often reflect not just technical, but social as well as sociotechnical impact and the importance of dialogue, collaboration and action among various stakeholders, to strategically guide the AI landscape \cite{15}. 

The combination of utopian and dystopian narratives that concurrently abound can create expectations and confusion about the technology, its usage and its impact. Continual public participation and education as well as monitoring and understanding of public perception is key for steering the discourse, development and deployment of AI towards societal benefits in a responsible, accessible and ethical manner. Furthermore, incentivizing activities that use ‘AI as a tool’ that can benefit society, should in turn help to improve public perception and acceptance. 

\vspace{\baselineskip}
\section*{Survey Details}
The 2024 3M State of Science Insights survey gauges the public sentiment towards science as well as the public perception of AI \cite{29}. On the topic of AI, the survey included key questions around the transformative nature of the technology and its use as a tool for problem-solving. The survey also uncovers the nuanced understanding of impact on perception if a company were to use AI to innovate to improve lives through products and services.	

From a methodological standpoint, the work is in the genre of public opinion polling: it is an original, third-party research, conducted by Morning Consult. The survey comprising 30 questions, shown in Appendix A2, was fielded from December 13, 2023, to January 10, 2024, and surveyed 10 countries across Asia Pacific, North America, South America, and Europe among 1,000 general population adults (18+) in each of the following countries: Brazil, Canada, China, France, Germany, Japan, Mexico, South Korea, U.K. and the U.S. At the 95\% confidence level, the margin of error is +/- 1 percentage points at the global, 10-country level and +/- 3 percentage points for each individual country. For charting and reporting purposes, the percentages were rounded off to the nearest whole number.   

\section*{Results}
For the purpose of this paper, we will only focus on a subset of the questions and datasets. See Appendix A3 and A4 for more 2024 data and AI related data from 3M SOSI 2018-2023, respectively. (Unpublished data from 2024 survey not shown here is available upon request). 

\vspace{\baselineskip}
\noindent{\textit{\textbf{Integration of AI at work}}}\\
Globally, 50\% of survey respondents are integrating AI “somewhat” or “to a great extent” as shown in Figure 1A, in response to a question around integration of AI at work. China, Brazil and Mexico lead in their responses to AI usage while Japan reported the highest percentage of those responding with “not at all.” US, UK, Canada and Germany are well above global average for those agreeing with ``not at all'' for AI usage while France is closer to the global average of 25\% for those who responded, “not at all.” South Korea has the largest percentage responding with “very little” integration of AI in their work. 

\begin{figure}[H]
\centering
\fbox{\includegraphics[width=0.85\linewidth]{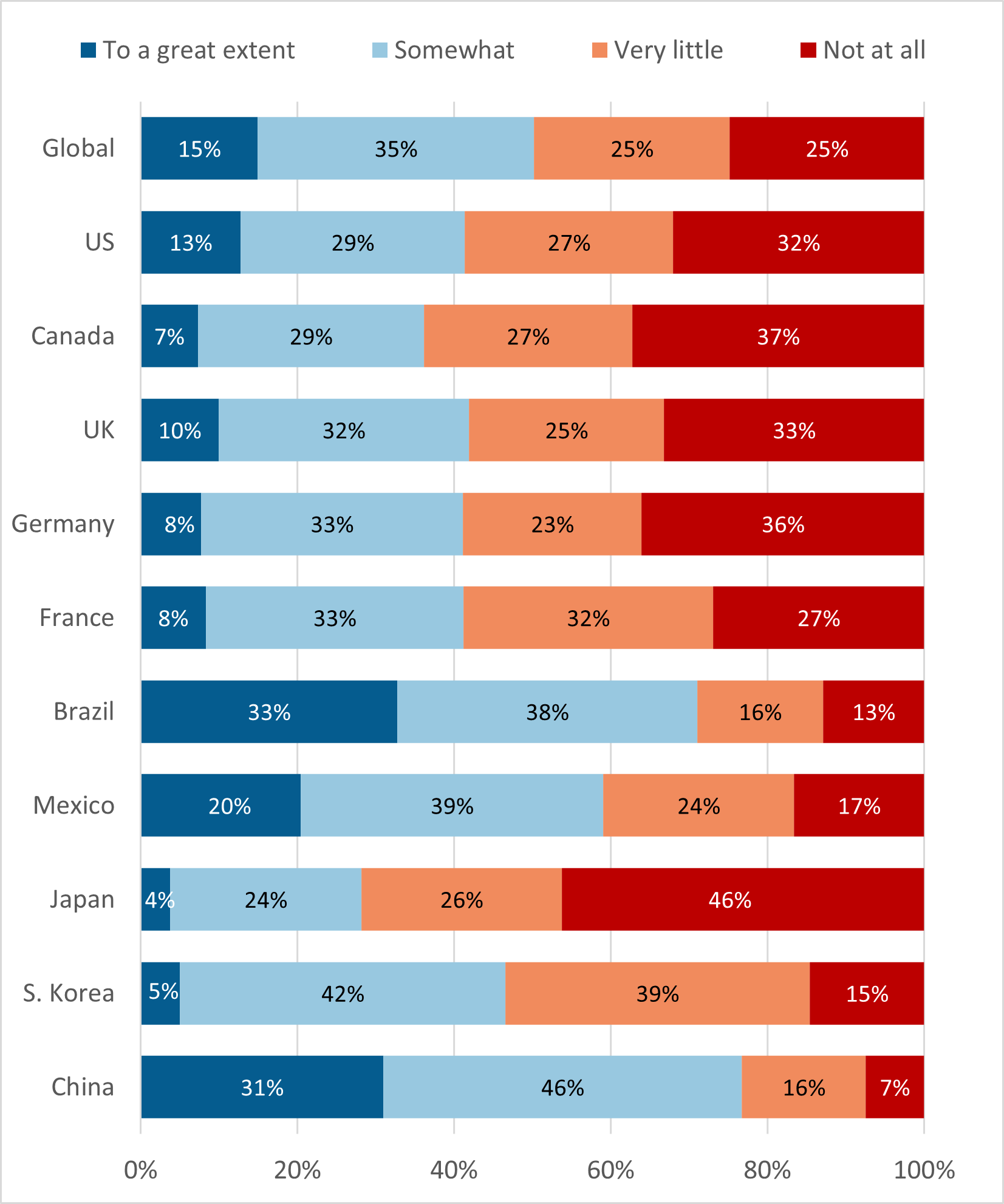}}
\caption*{\textbf{Figure 1A: General population responses for “To what extent are you integrating AI into your work environment and process (i.e. using AI to write emails or basic code, generating content using AI, etc.)?”} (Question 23 from the 2024 3M State of Science Insights)}
\end{figure}
Figure 1B shows the gender demographic data for the question around integration of AI into work. The overall leaders in AI usage, China, Brazil and Mexico, have fairly comparable male and female responses across the four categories. US, UK, Germany and France have the largest differences between genders, with only 4\% of female respondents in the US using AI “to a great extent” and 42\% “not at all,” while 20\% US male respondents are using it “to a great extent” and only 24\% responding “not at all.” Except China, a higher percentage of female respondents report using AI “not at all” in comparison to 
\clearpage
\noindent male. 
 Japan leads in the percentage of females (51\%) responding “not at all” for AI usage and has the lowest percentage (1\%) of female respondents who use it “to a great extent” followed by South Korea (3\%) and US (4\%). In contrast to US, Canada notably has similar percentages for the genders in those who use AI “to a great extent” and those who use it “not at all.”

\begin{figure}[H]
\centering
\fbox{\includegraphics[width=0.85\linewidth]{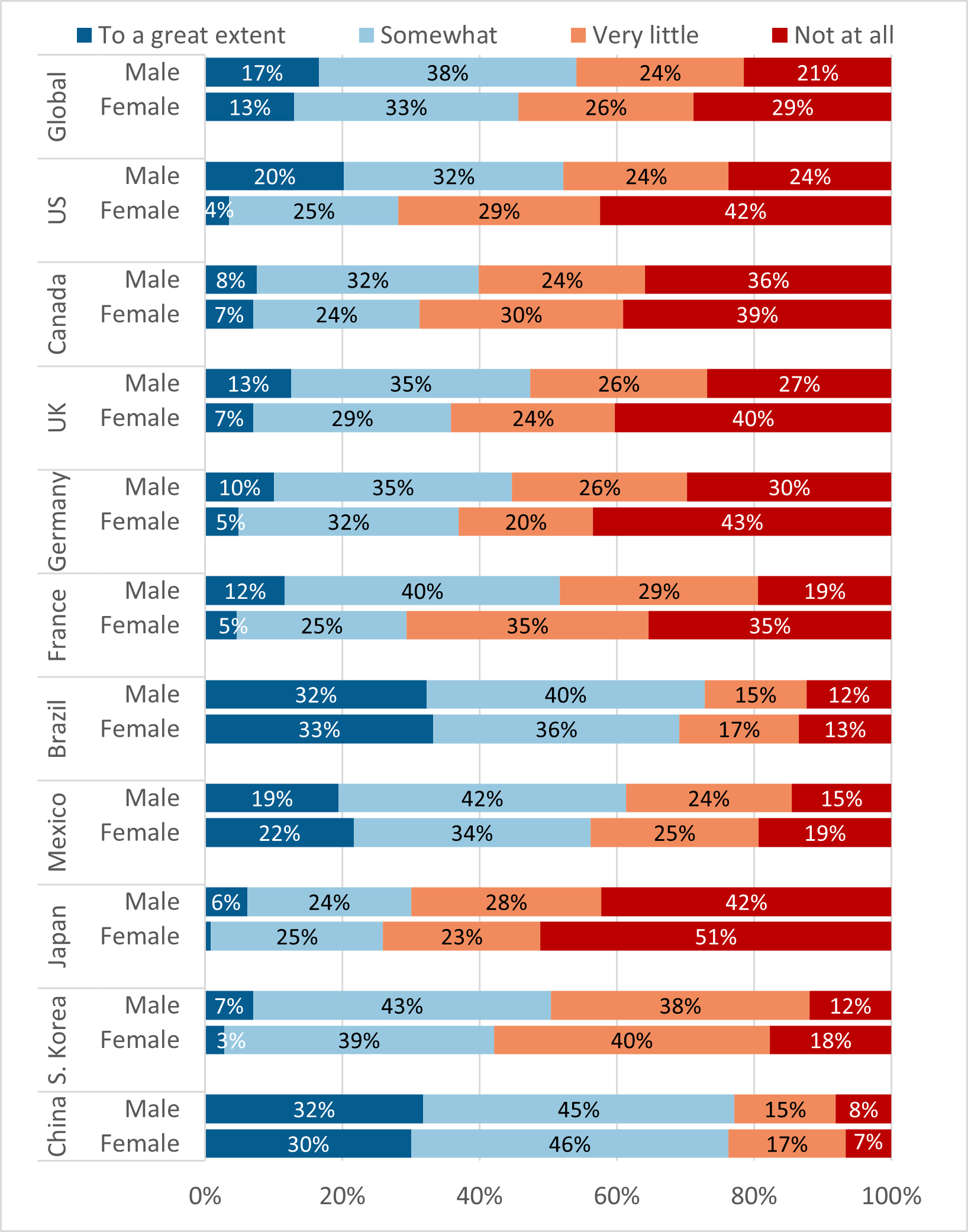}}
\caption*{\textbf{Figure 1B: Responses grouped by gender for “To what extent are you integrating AI into your work environment and process (i.e. using AI to write emails or basic code, generating content using AI, etc.)?”} (Question 23 from the 2024 3M State of Science Insights)}
\end{figure}

Figure 1C shows the responses to the same question depicted by age groups of 18-34, 35-54, and 55+ for each country, except for US, which is arranged into age groups of 18-34, 35-44, 45-64, and 65+. Here Brazil and Mexico show the younger demographic reporting AI usage “to a great extent” while in China the older demographic, the age group of 55+, leads in AI usage. US, Canada, UK, and Germany show the same pattern where the usage of AI “to a great extent” is highest for the youngest demographic and no usage of AI is highest for the oldest demographic. In South Korea and Japan, the younger demographic stands out for those who use AI “somewhat” or “to a great extent” when compared with the two older age groups that are comparable to each other. 

\begin{figure}[H]
\centering
\fbox{\includegraphics[width=0.85\linewidth]{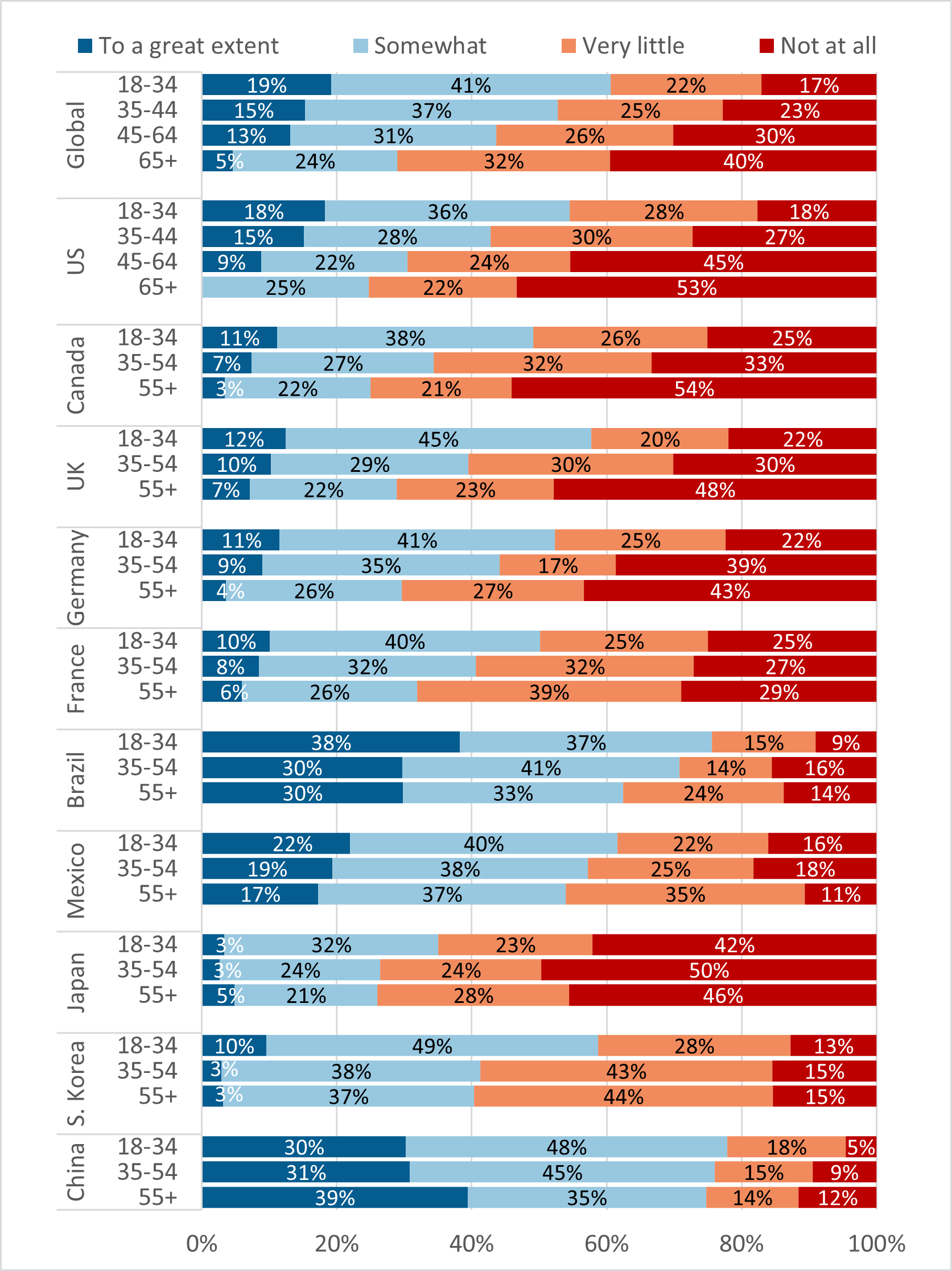}}
\caption*{\textbf{Figure 1C: Responses grouped by age for “To what extent are you integrating AI into your work environment and process (i.e. using AI to write emails or basic code, generating content using AI, etc.)?”} (Question 23 from the 2024 3M State of Science Insights)}
\end{figure}
\clearpage
Figure 1D shows the responses to the same question, this time grouped by employment status. Since this question pertains to using AI in a work environment, it was only asked to respondents who were currently employed. Among employed respondents, the same trend seen before for AI usage follows, in that China, Brazil and Mexico lead in integration of AI at work and Japan reported the highest percentage (46\%) of those responding with a “not at all.” 

\begin{figure}[H]
\centering
\fbox{\includegraphics[width=0.85\linewidth]{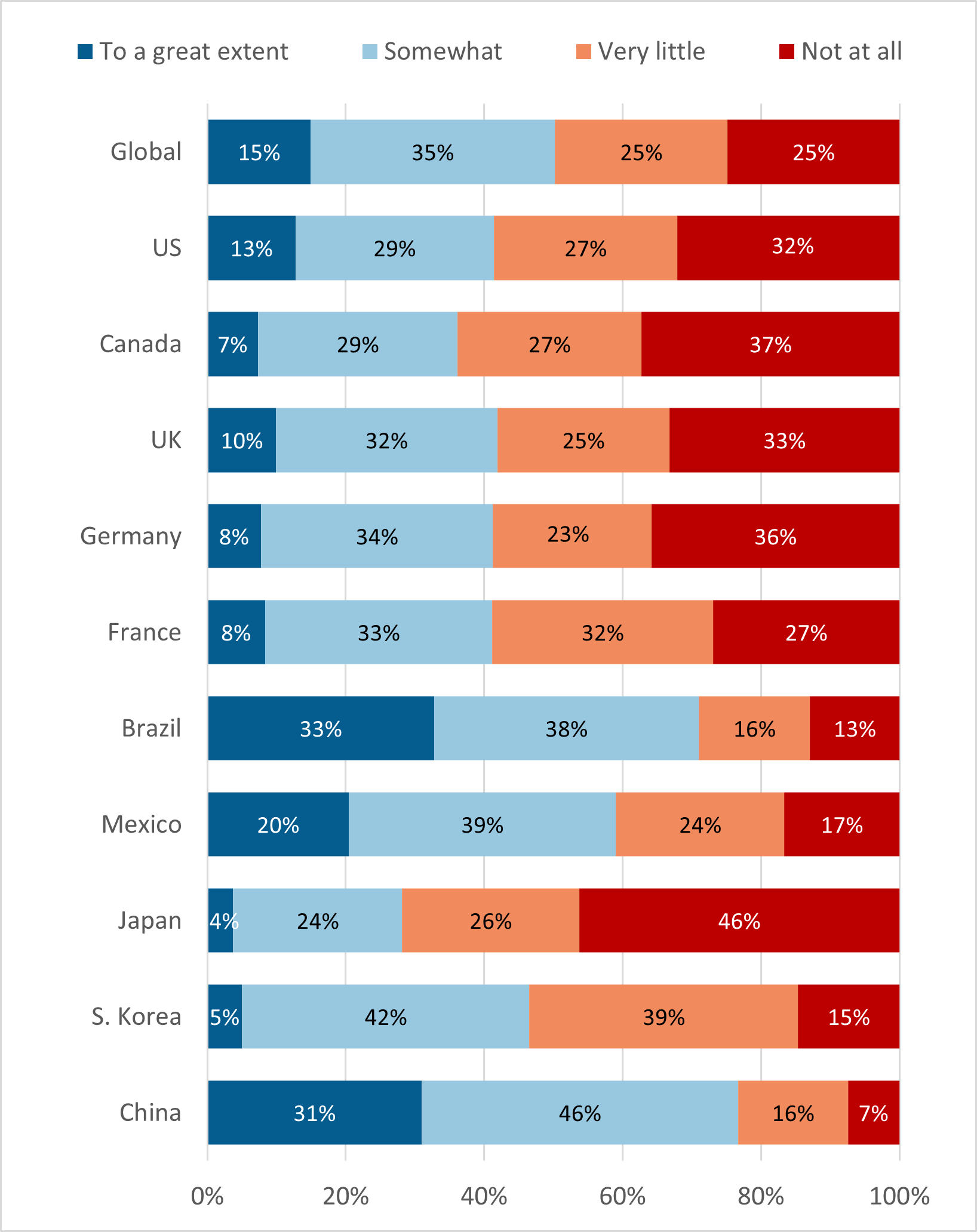}}
\caption*{\textbf{Figure 1D: Responses for employed subset for “To what extent are you integrating AI into your work environment and process (i.e. using AI to write emails or basic code, generating content using AI, etc.)?”} (Question 23 from the 2024 3M State of Science Insights)}
\end{figure}

\noindent{\textit{\textbf{Overall sentiments around AI}}}\\
Figure 2 below shows the responses to a series of questions (Q21\_1 to Q21\_7) that deal with awareness, usage, expectations and perception of AI.  The percentages reported are the sum of percentages that responded, “strongly agree” and “somewhat agree” when provided with these options alongside three other options: “somewhat disagree,” “strongly disagree” and “don’t know/no opinion.”

The data shows that Mexico, South Korea, China and Brazil are the most worried about their jobs being replaced by AI. Japan at 47\% is closer to the global average of 48\% i.e. almost one in two are worried about their job being replaced. The rest of the countries are below the global average in being worried about job loss to AI. 

In terms of the impact of AI in changing “the world as we know it,” again China, Brazil, South Korea and Mexico have the highest agreement, while Canada, Germany, UK and US are not too far behind ranging from 77-72\%. Japan is below 70\% in believing that AI will “change the world as we know it” with France at the lowest – 66\% agreement.

Regarding perception of “AI as a tool for problem solving,” China, Brazil, South Korea and Mexico are well above the global average of 68\% with China leading at an 89\% agreement with this statement. France has the lowest agreement level (51\%) which still means one out of two agree that AI is a tool to help solve problems. 

In terms of seeing themselves using AI in their daily life China leads the pack with 89\% agreement, followed by a distant second Brazil at 72\% and then Japan, Mexico and South Korea. Germany at 58\% is close to the global average of 59\% while UK, US and France trail here with lowest agreement for France at 41\%.

As reported in many other surveys there is agreement across the countries that AI needs to be heavily regulated given the global average at 77\%. China and South Korea lead with 90\% and 84\% respectively and the lowest agreement at 71\% for Germany. This is the highest percentage for low agreement across this series of questions, the second one being “AI will change the world as we know it” (66\% lowest – France) and third is the perception that “AI is a tool for problem solving” (51\% - France). All other questions have one or multiple countries with 50\% or less than 50\% agreement. 

China leads by far, at 88\%, in the agreement regarding an increased presence of AI in their daily lives followed by Brazil at a distant second with 70\% agreement. US, UK, Canada, Germany and France range from 50-53\% and well below the global average at 61\%.

Finally in terms of understanding where AI is being used in everyday life, China leads with 79\% agreement followed by South Korea (68\%) and Brazil (64\%) well above the global average of 55\%. Mexico and Germany are above the global average whereas Canada, US, UK and France are well below with France only at 41\%. 

\begin{figure}[H]
\centering
\fbox{\includegraphics[width=0.91\linewidth]{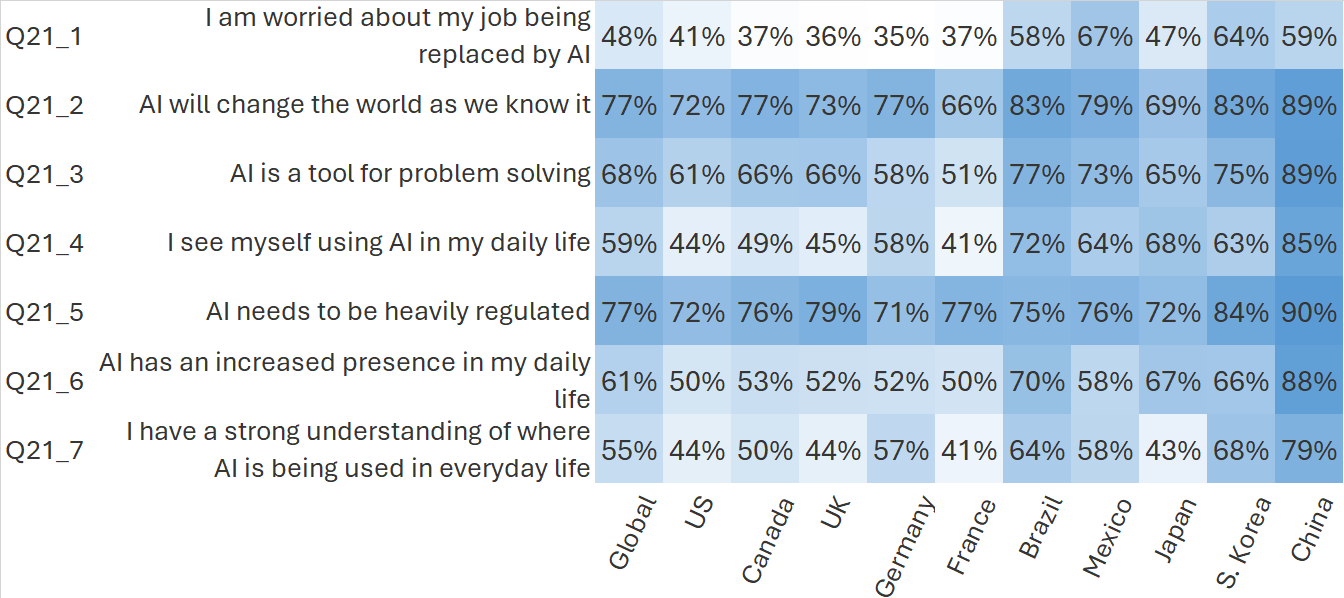}}
\caption*{\textbf{Figure 2: Percentage of respondents who “strongly agreed” or “somewhat agreed” to the question “thinking about AI, how much do you agree or disagree with each of the following.”} (Question 21 from the 2024 3M State of Science Insights)}
\end{figure}

Figure 3A shows the general population responses to “AI will change the world as we know it.” Globally 36\% of respondents “strongly agree” to this statement. France and Japan are the farthest below this average with 23\% and 21\% strong agreement, respectively. France and Japan also have the highest amount of disagreement with the statement with 20\% of respondents who “somewhat disagree” or “strongly disagree.” On the other hand, the countries with the most respondents that “strongly agree” with this statement are Brazil (52\%), Mexico (45\%), and Canada (41\%). US (16\%) leads, followed by France (14\%) and UK (13\%) for the percentage that “Don’t know/No opinion.”

\begin{figure}[H]
\centering
\fbox{\includegraphics[width=0.85\linewidth]{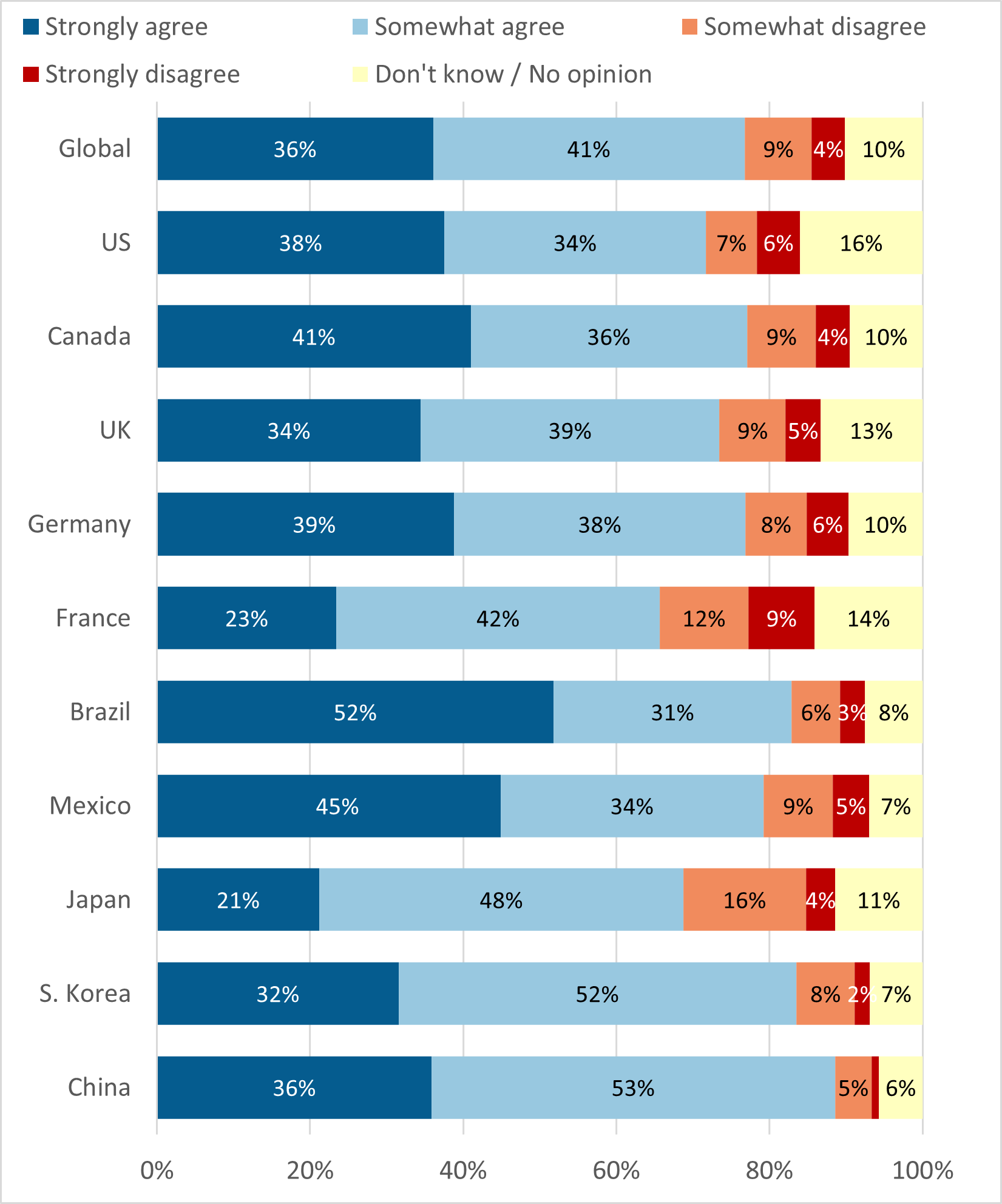}}
\caption*{\textbf{Figure 3A: General population response for “Thinking about AI, how much do you agree or disagree with each of the following? \--- AI will change the world as we know it.”} (Question 21\_2 from the 2024 3M State of Science Insights)}
\end{figure}

\clearpage
Figure 3B shows the data on responses to the same question but grouped by male and female demographic. Globally, the percentage of male respondents who strongly agreed was 38\% and for female respondents it was 34\%. In 9 out of the 10 countries surveyed, male respondents showed higher or equal percentages of strong agreement. Germany was the exception where 41\% of female respondents “strongly agree” while only 36\% of male respondents “strongly agree,” however when looking at total agreement (“somewhat agree” and “strongly agree”), the percentage for male respondents is greater than female respondents in every country. The difference between male and female who “strongly agree” was the highest in South Korea where 38\% of male respondents “strongly agree” but only 25\% of female respondents “strongly agree” that “AI will change the world as we know it.”

\begin{figure}[H]
\centering
\fbox{\includegraphics[width=0.78\linewidth]{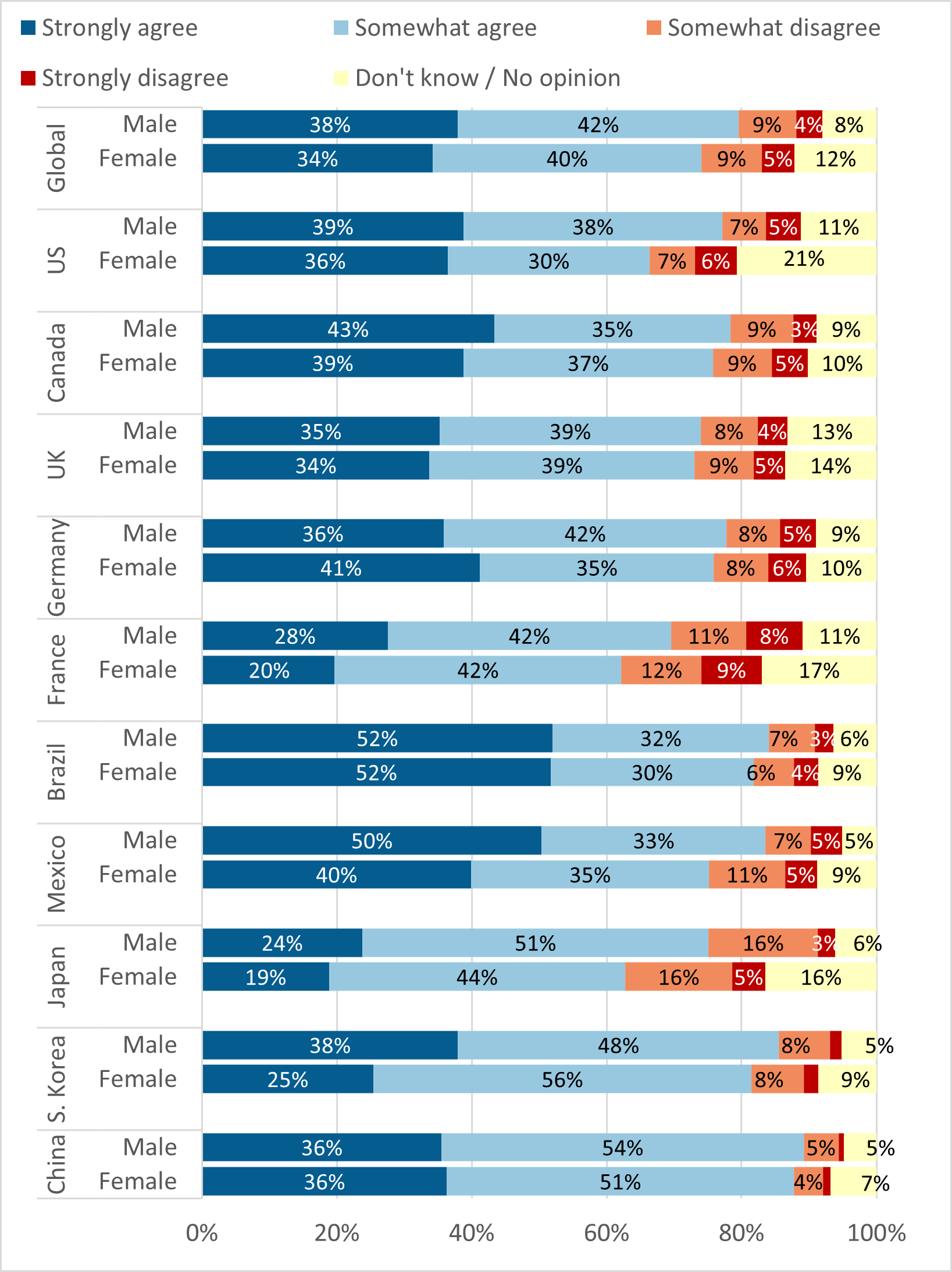}}
\caption*{\textbf{Figure 3B: Responses grouped by gender for “Thinking about AI, how much do you agree or disagree with each of the following? \--- AI will change the world as we know it.”} (Question 21\_2 from the 2024 3M State of Science Insights)}
\end{figure}

\clearpage
In Figure 3C the responses to the same question are grouped by age. On average, 39\% of 18-34-year-olds “strongly agree” and the age groups 35-44, 45-64, and 65+ all have 35\% of respondents “strongly agree.” US, UK, France, Brazil and Japan all follow a similar trend where the high and low ages agree the most and the middle age groups agree less that “AI will change the world as we know it.” Brazil has the biggest differences in agreement between age groups with 59\% of 55+ year olds who “strongly agree” but 47\% of 35-54-year-olds who “strongly agree.” Japan has the biggest differences in overall agreement between age groups where the total disagreement (“somewhat disagree” and “strongly disagree”) for ages 18-34 is 25\%, but it is only 17\% for ages 55+. 

\begin{figure}[H]
\centering
\fbox{\includegraphics[width=0.83\linewidth]{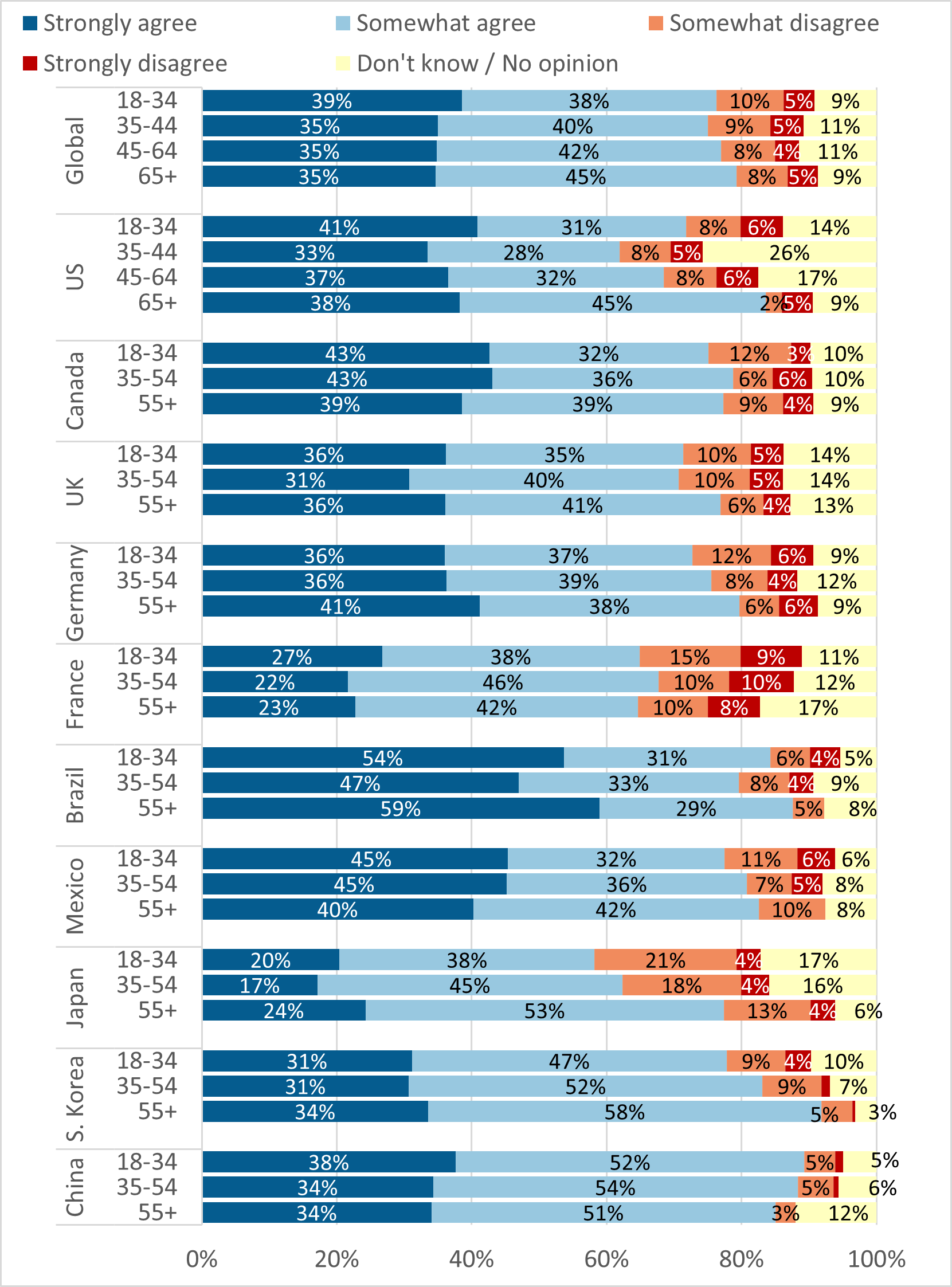}}
\caption*{\textbf{Figure 3C: Responses grouped by age for “Thinking about AI, how much do you agree or disagree with each of the following? \--- AI will change the world as we know it.” } (Question 21\_2 from the 2024 3M State of Science Insights)}
\end{figure}

Figure 3D shows the responses to the same question grouped by employment status. Globally, 37\% of employed respondents “strongly agree” and 34\% of not employed respondents “strongly agree.” In 7 of the 10 countries the percentage for employed and not employed respondents who “strongly agree” are within 3\%. However, for Mexico, Japan and Brazil the percentages of employed respondents who “strongly agree” were much higher than the percentages of not employed respondents who “strongly agree.” The differences between employed and not employed respondents who “strongly agree” in Mexico, Japan and Brazil were 13\%, 10\% and 17\% respectively. None of the countries had notably high percentages of “strongly disagree,” France was the highest with 9\% for employed and 8\% for not employed. 

\begin{figure}[H]
\centering
\fbox{\includegraphics[width=0.85\linewidth]{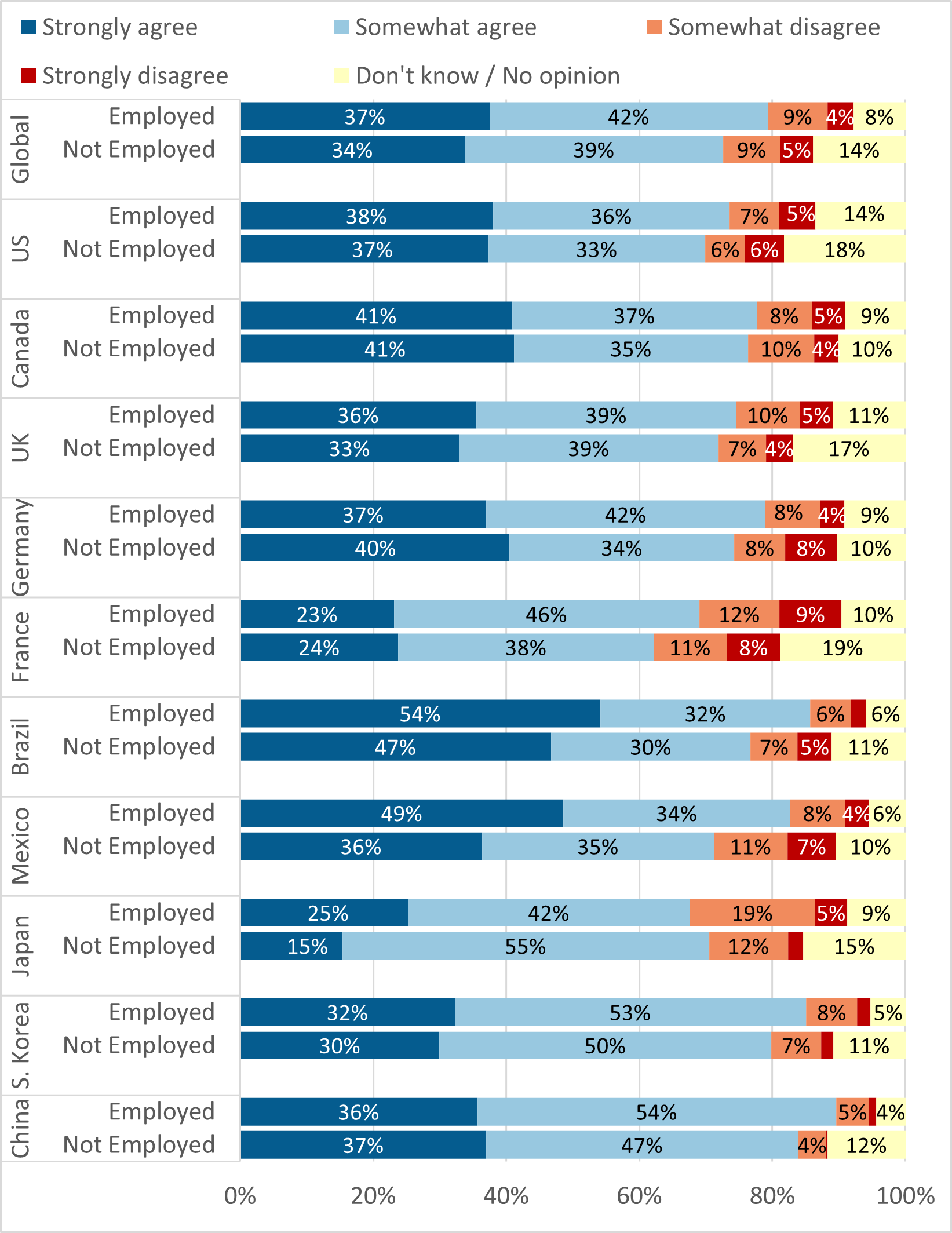}}
\caption*{\textbf{Figure 3D: Responses grouped by employment for “Thinking about AI, how much do you agree or disagree with each of the following? \--- AI will change the world as we know it.”} (Question 21\_2 from the 2024 3M State of Science Insights)}
\end{figure}

\clearpage
\noindent{\textit{\textbf{AI is a tool}}}\\
Figure 4A shows the responses to “AI is a tool for problem solving” from the general population. Globally, 24\% of respondents “strongly agree.” Brazil, Mexico, South Korea and China are again above this average at 36\%, 32\%, 75\%, and 38\% respectively. China also has by far the smallest percent of respondents that “somewhat disagree” (5\%) or “strongly disagree” (1\%). On the other end of the spectrum, France has the lowest total agreement with only 51\% of respondents who “somewhat agree” or “strongly agree” that “AI is a tool for problem solving.” 
\vspace{\baselineskip}

\begin{figure}[H]
\centering
\fbox{\includegraphics[width=0.85\linewidth]{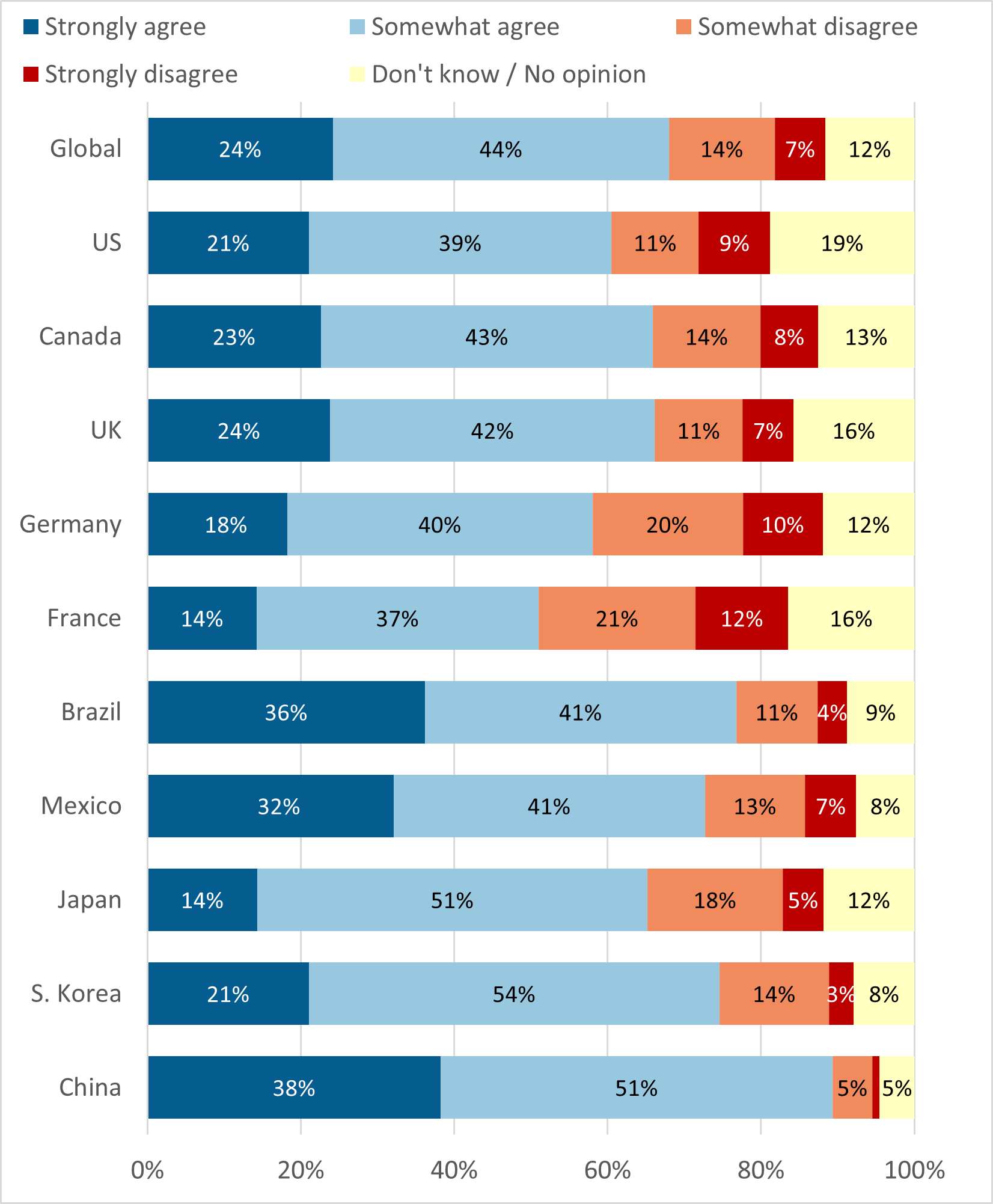}}
\caption*{\textbf{Figure 4A: General population responses for “Thinking about AI, how much do you agree or disagree with each of the following? \--- AI is a tool for problem solving.”} (Question 21\_3 from the 2024 3M State of Science Insights)}
\end{figure}

\clearpage
Figure 4B shows the data for the responses to the same question as above but grouped for male and female demographic. Of all the male respondents 29\% “strongly agree” that “AI is a tool for problem solving” while only 20\% of the female respondents feel the same.” For every country surveyed, the percentage that “strongly agree” is higher for male respondents than female respondents. The US has the greatest difference between agreement of male and female responses with 28\% of male answering “strongly agree” but only 15\% of females with the same response. 

\begin{figure}[H]
\centering
\fbox{\includegraphics[width=0.85\linewidth]{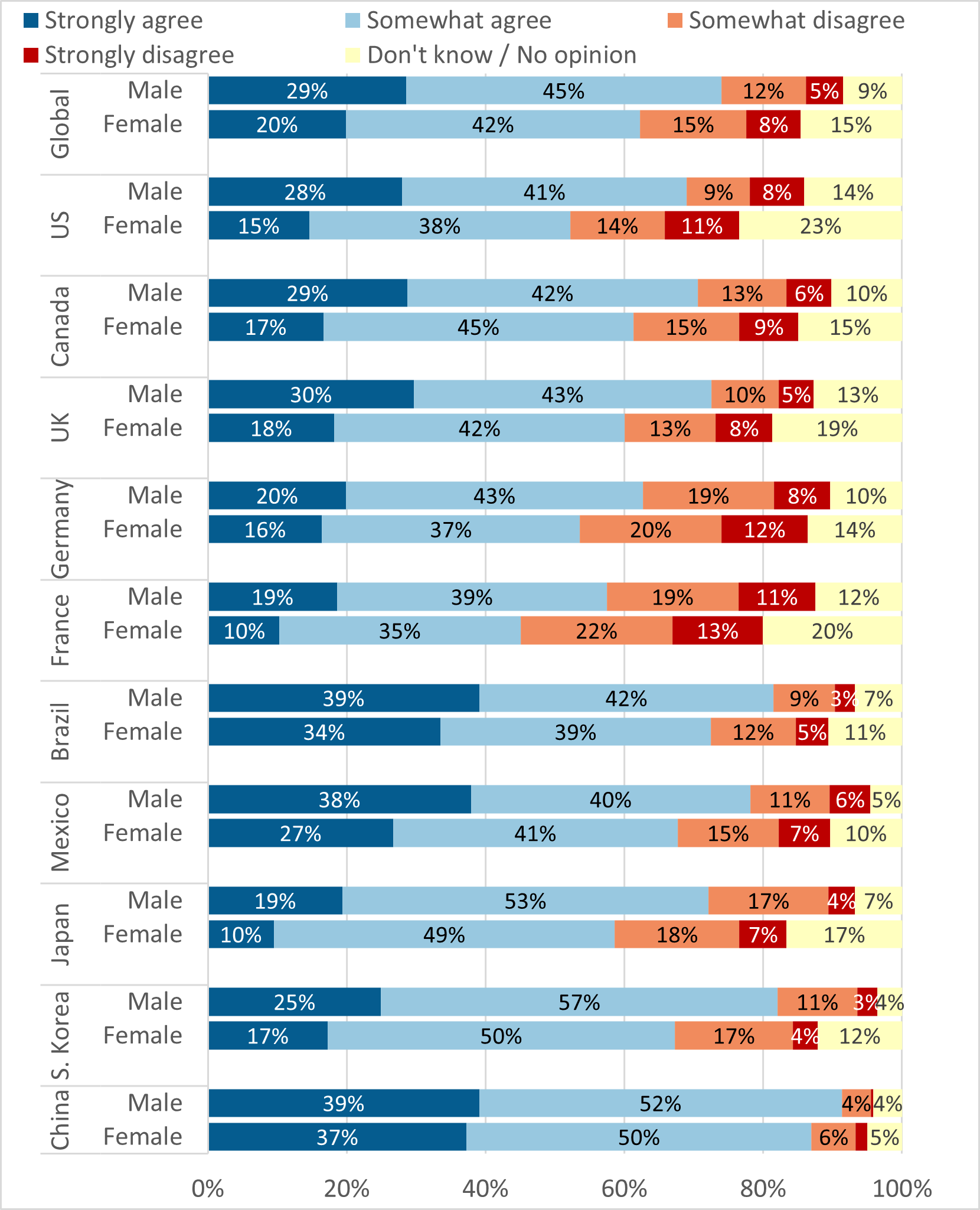}}
\caption*{\textbf{Figure 4B: Responses grouped by gender for “Thinking about AI, how much do you agree or disagree with each of the following? \--- AI is a tool for problem solving.”} (Question 21\_3 from the 2024 3M State of Science Insights)}
\end{figure}

\clearpage
In Figure 4C the responses to the same question are grouped by age brackets. Globally the trend is that the younger age groups “strongly agree” more than the older age groups. 30\% of 18-34-year-olds “strongly agree” that “AI is a tool for problem solving, which only 16\% of 65+ “strongly agree.” For this question there is not much imbalance between age demographics. The biggest difference in age group responses was in China where 41\% of 18-34-year-olds “strongly agree,” while only 18\% of the 55+ year olds felt the same. In every country except Japan, there is also a clear trend that as age increases, the percentage who “strongly agree” decreases. 

\begin{figure}[H]
\centering
\fbox{\includegraphics[width=0.85\linewidth]{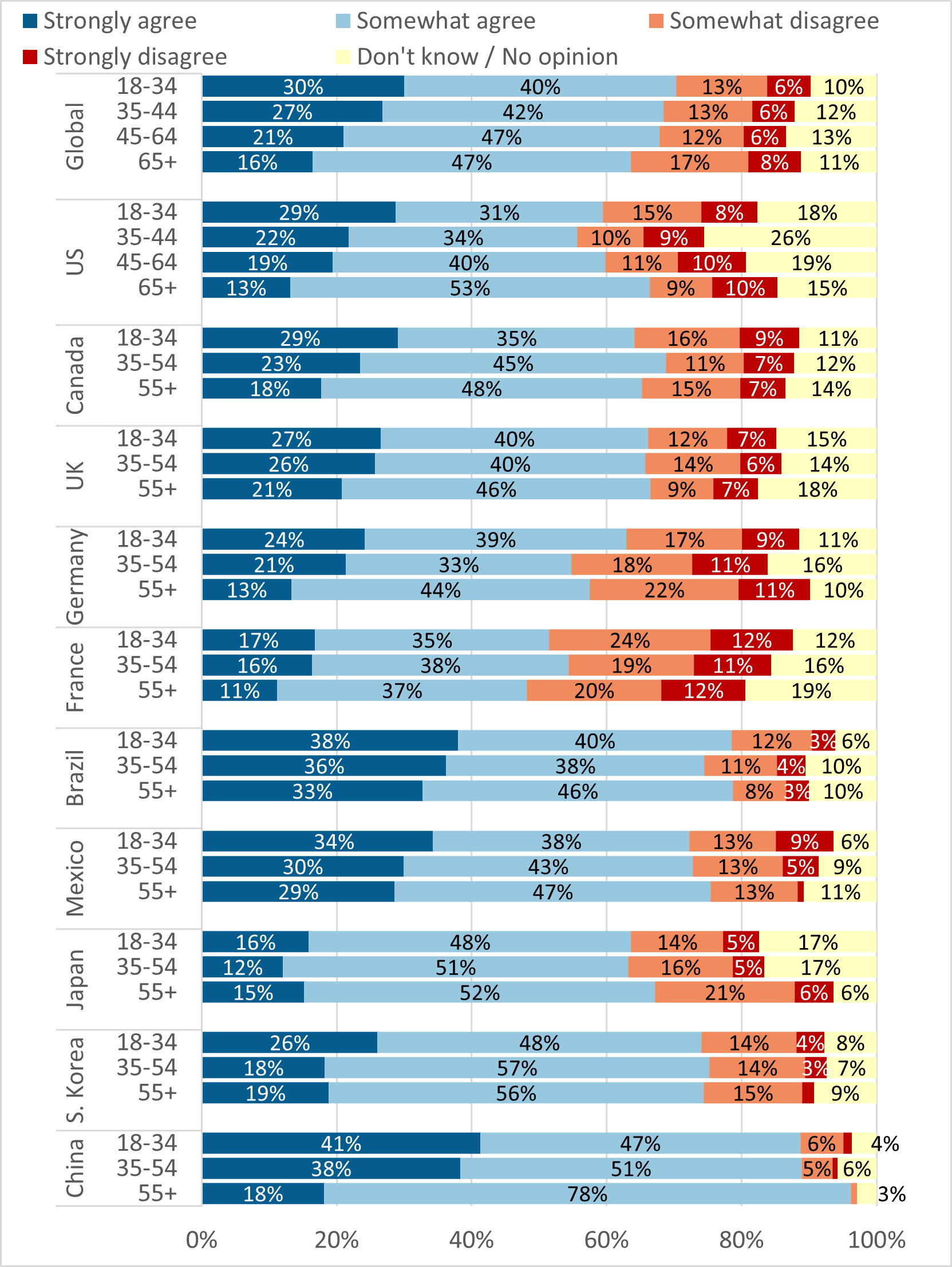}}
\caption*{\textbf{Figure 4C: Responses grouped by age for “Thinking about AI, how much do you agree or disagree with each of the following? \--- AI is a tool for problem solving.”} (Question 21\_3 from the 2024 3M State of Science Insights)}
\end{figure}

\clearpage
Figure 4D addresses the same question again but this time the responses are grouped by employment status. Globally and in each country individually, a higher percentage of employed respondents(27\%) “strongly agree” that “AI is a tool for problem solving” than respondents that are not employed(20\%). Once again Brazil, Mexico and South Korea stand out in this data set. All three have a larger difference between the two groups of respondents with Brazil having the biggest difference of 13\% between employed respondents and not employed respondents who answered, “somewhat agree” or ``strongly agree.''

\begin{figure}[H]
\centering
\fbox{\includegraphics[width=0.85\linewidth]{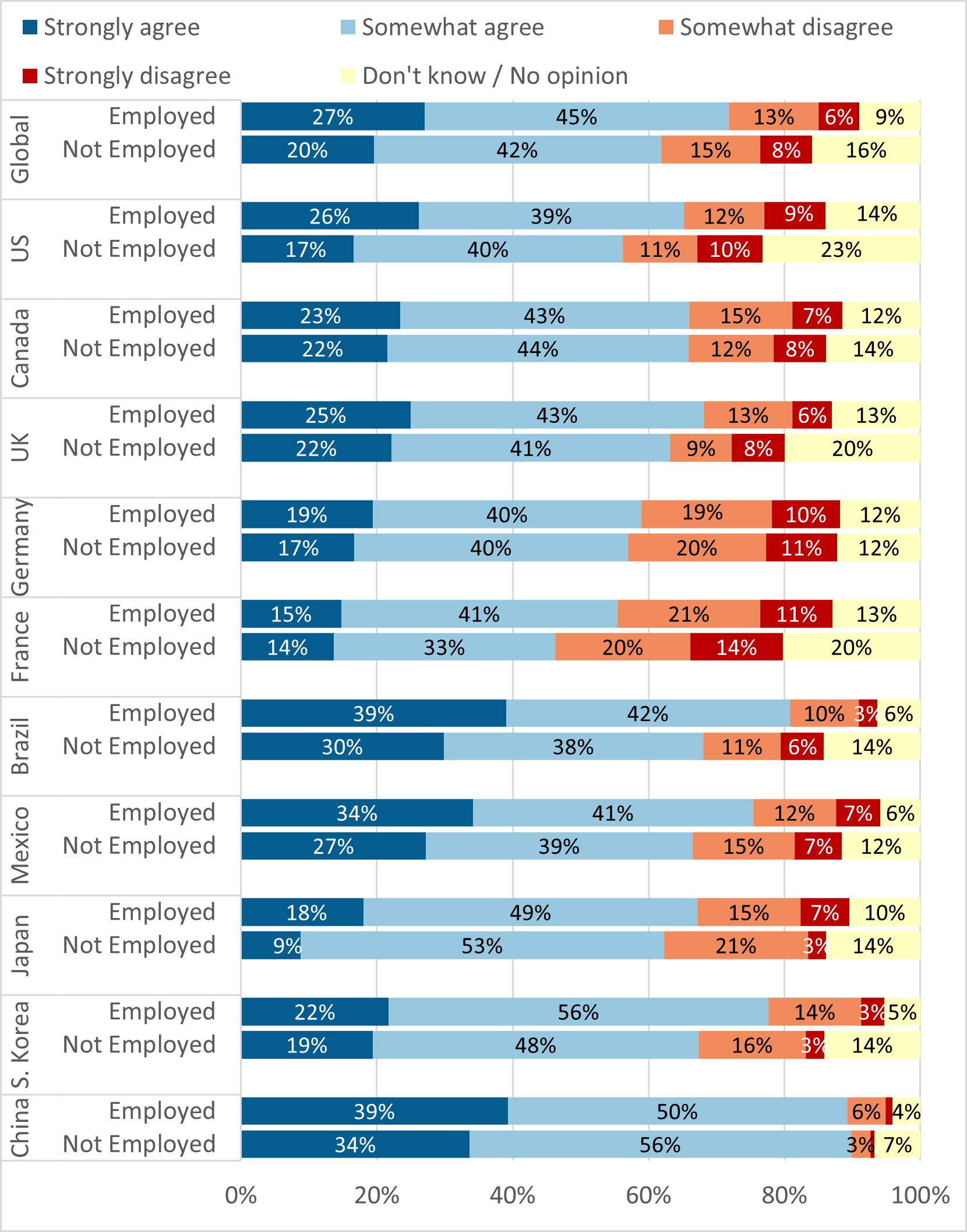}}
\caption*{\textbf{Figure 4D: Responses grouped by gender for “Thinking about AI, how much do you agree or disagree with each of the following? \--- AI is a tool for problem solving.”} (Question 21\_3 from the 2024 3M State of Science Insights)}
\end{figure}

\clearpage
\vspace{\baselineskip}
\noindent{\textit{\textbf{Perception of companies that use AI to improve daily lives}}}\\
Figure 5 below shows the responses to a series of questions (Q22\_1 to Q22\_4) that probe the public perception of companies that are using AI to innovate to improve daily lives.  The percentages reported are the sum of percentages that responded, “much more likely” and “somewhat more likely” when provided with these options alongside three other options: “no impact,” “somewhat less likely” and “much less likely.”

The data below shows that China, Brazil, Mexico and South Korea lead in a positive response to such companies (Q22\_1) while the rest of the countries are all well below the global average of 45\% with US, Canada, UK, Germany, France and Japan all at 29-33\% in thinking highly of a company using AI to innovate to improve lives.

As for intent to purchase from the company, it mirrors closely the responses for Q22\_1 in that China, Brazil, Mexico and South Korea show more intent compared to all the rest of the countries are between 25\%-31\% with global average at 41\% showing intent to purchase from the company.

Regarding interest in the company (Q22\_3) the global average is higher compared to previous questions while the trends remain the same. China, Brazil, Mexico and South Korea lead the pack while US, Canada and France are at 38\% followed by Japan at 36\% and Germany at 31\%. While for interest in pursuing a job in the company (Q22\_4) the global average is 41\% more likely with again China, Brazil, Mexico and South Korea leading the positive intent and the rest of the countries between 25-30\% more likely to pursue a job with a company using AI to innovate to improve daily lives. 

\begin{figure}[H]
\centering
\fbox{\includegraphics[width=0.95\linewidth]{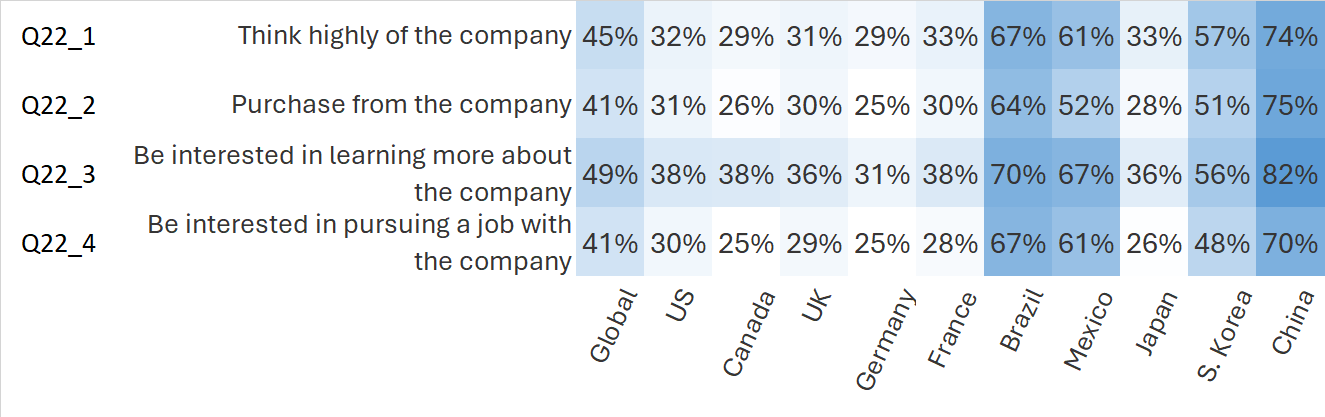}}
\caption*{\textbf{Figure 5: The percentage of respondents who answered “much more likely” or “somewhat more likely” to the question “If a company were to use AI to innovate the products and services you use on a daily basis to improve your life, would you be more or less likely to—.“} (Question 22 from the 2024 3M State of Science Insights)}
\end{figure}
Figure 6A is the breakdown of the general population response for the most positive sentiment in Figure 5, which was Q22\_3, where almost one in two survey respondents would be ``more interested'' in learning about a company if they used AI to innovate to improve daily lives. Figures 6B-6D are the demographic data grouped by gender, age and employment status respectively. 

\clearpage
Figure 6A shows that China and Brazil are comparable for those who would be much more likely to want to learn more about the company. Also, France and Japan have an equal percentage who will be “somewhat more likely” to be interested in the company. Japan also leads in the percentage, at 56\%, that responded with “no impact” on their interest level in a company using AI to innovate.

\begin{figure}[H]
\centering
\fbox{\includegraphics[width=0.85\linewidth]{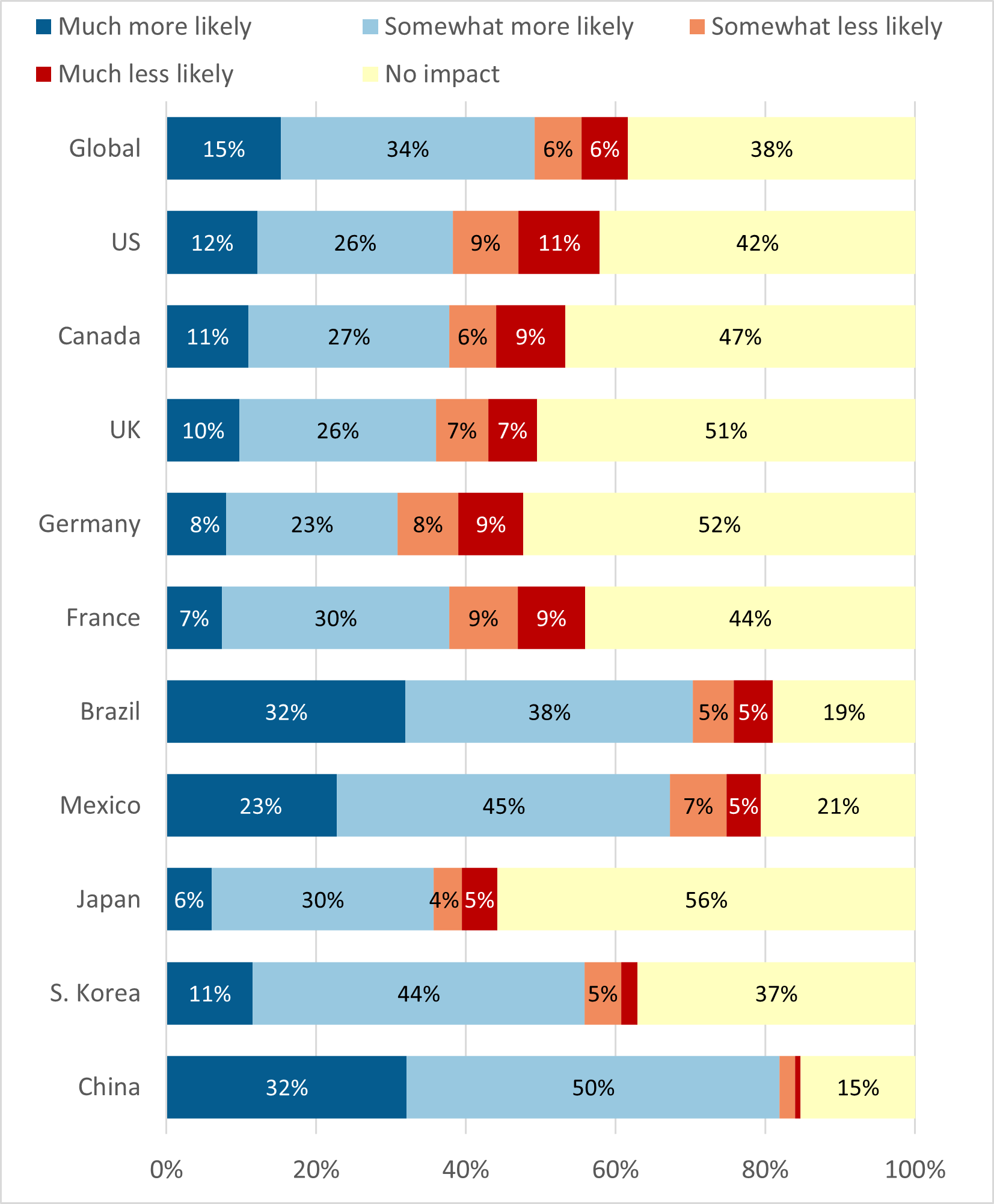}}
\caption*{\textbf{Figure 6A: General population responses for “If a company were to use AI to innovate the products and services you use on a daily basis to improve your life, would you be more or less likely to \-- Be interested in learning more about the company.”} (Question 22\_3 from the 2024 3M State of Science Insights)}
\end{figure}
\clearpage
Figure 6B shows that the male and female responses are fairly close for “much more likely,” globally 17\% of male respondents and 14\% of female respondents answered “much more likely.” This trend follows for many countries except US, Canada, UK and South Korea --– with the disparity being the largest in US where 17\% males would be “much more likely” while only 8\% for females. In addition, US also led in the percentage of females, 13\%, who said they would be “much less likely” to be interested in learning more about the company compared to males at 8\%.

\begin{figure}[H]
\centering
\fbox{\includegraphics[width=0.85\linewidth]{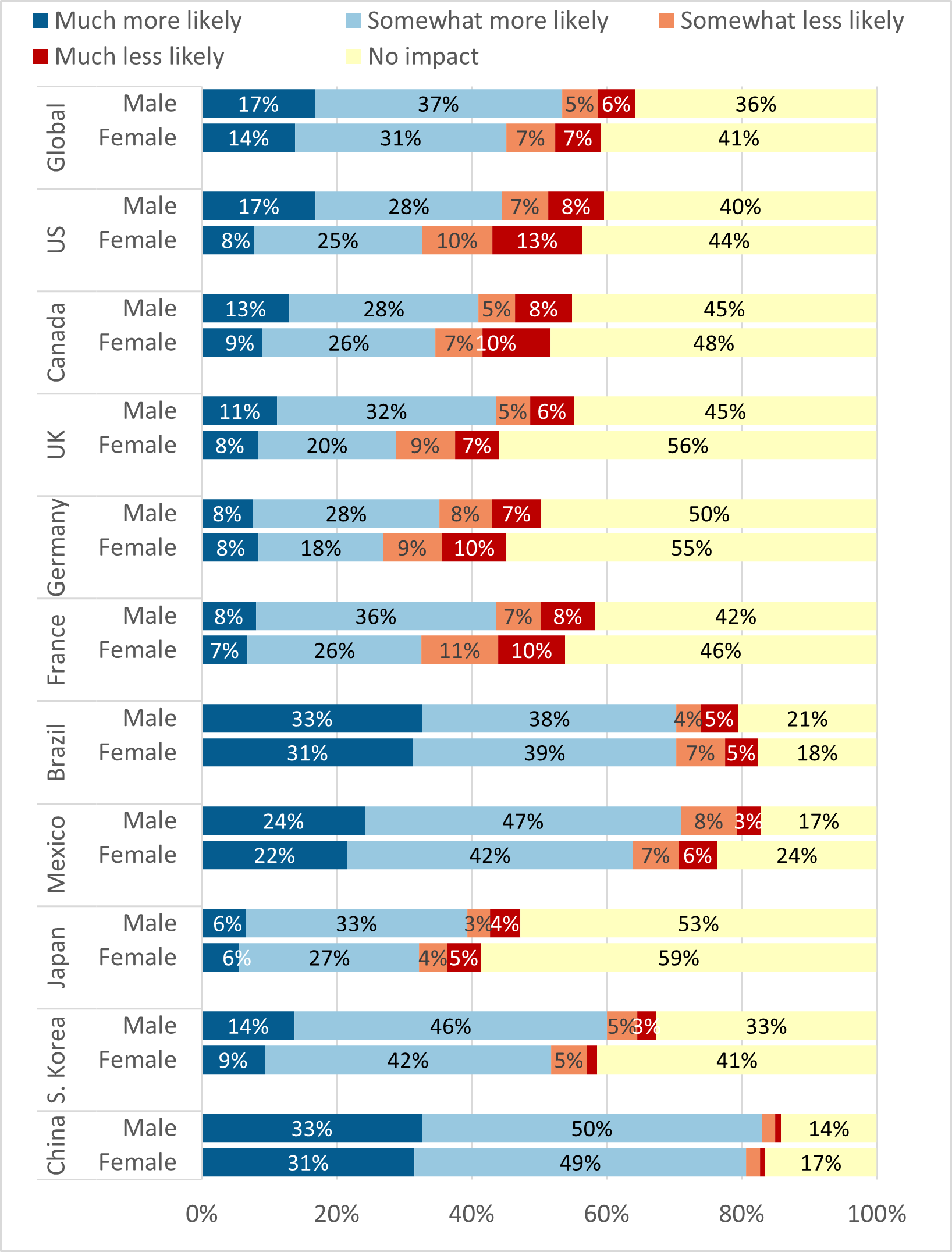}}
\caption*{\textbf{Figure 6B: Responses grouped by gender for “If a company were to use AI to innovate the products and services you use on a daily basis to improve your life, would you be more or less likely to \-- Be interested in learning more about the company.”} (Question 22\_3 from the 2024 3M State of Science Insights)}
\end{figure}
\clearpage
Figure 6C shows the data grouped by age. Globally, 19\% of 18-34, 17\% of 35-44, 14\% of 45-64, and 8\% of 65+ year-olds are ``much more likely'' to ``be interested in learning more about the company.'' Again, for US the categories are different compared to the rest of the countries. The data shows that the two younger demographic groups are “much more likely” to have a greater interest in a company using AI to innovate, compared to the two older groups.

The trend of more interest from younger age group is generally consistent across all countries except Brazil and Mexico where the 55+ age group showed the strongest response to “much more likely,” while in Japan and Korea the “much more likely” and “somewhat more likely” put together had a higher percentage for the 55+ demographic. Interestingly, for several countries the less likely was also higher for the younger demographic. Japan, Germany and UK had amongst the largest percentage for “no impact” and in all three cases the percentage was higher for the older age group. 

\begin{figure}[H]
\centering
\fbox{\includegraphics[width=0.75\linewidth]{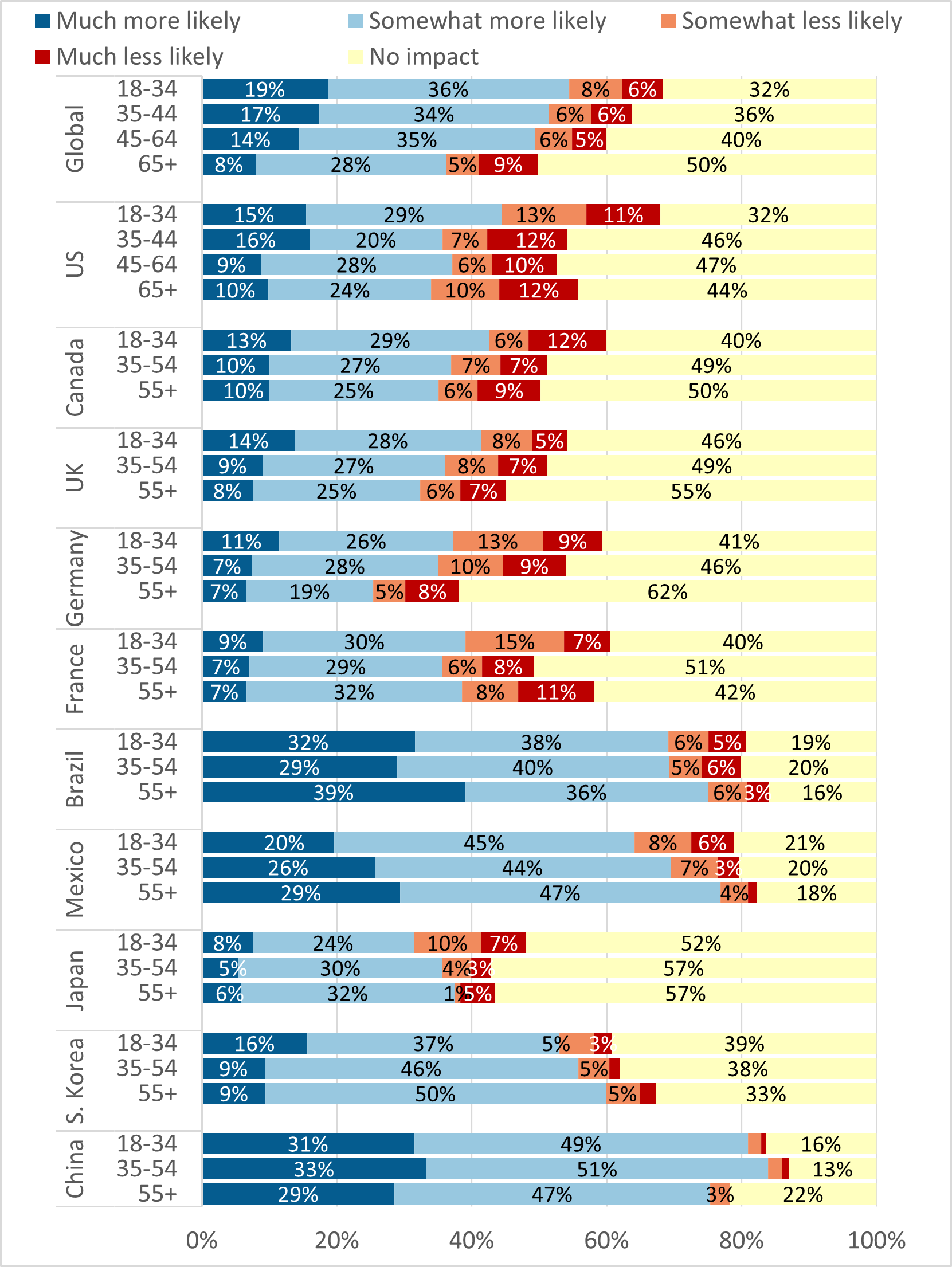}}
\caption*{\textbf{Figure 6C: Responses grouped by age for “If a company were to use AI to innovate the products and services you use on a daily basis to improve your life, would you be more or less likely to \-- Be interested in learning more about the company.”} (Question 22\_3 from the 2024 3M State of Science Insights)}
\end{figure}
\clearpage
Figure 6D shows the responses grouped by employment status for interest in company using AI to innovate to improve everyday lives. Across the board, for all countries, the employed(18\% globally) are “much more likely” to be more interested in the company than the unemployed (11\% employed). US, France, Brazil and Mexico have the highest difference between employed and not employed for “somewhat less likely” and “much less likely.”

\begin{figure}[H]
\centering
\fbox{\includegraphics[width=0.85\linewidth]{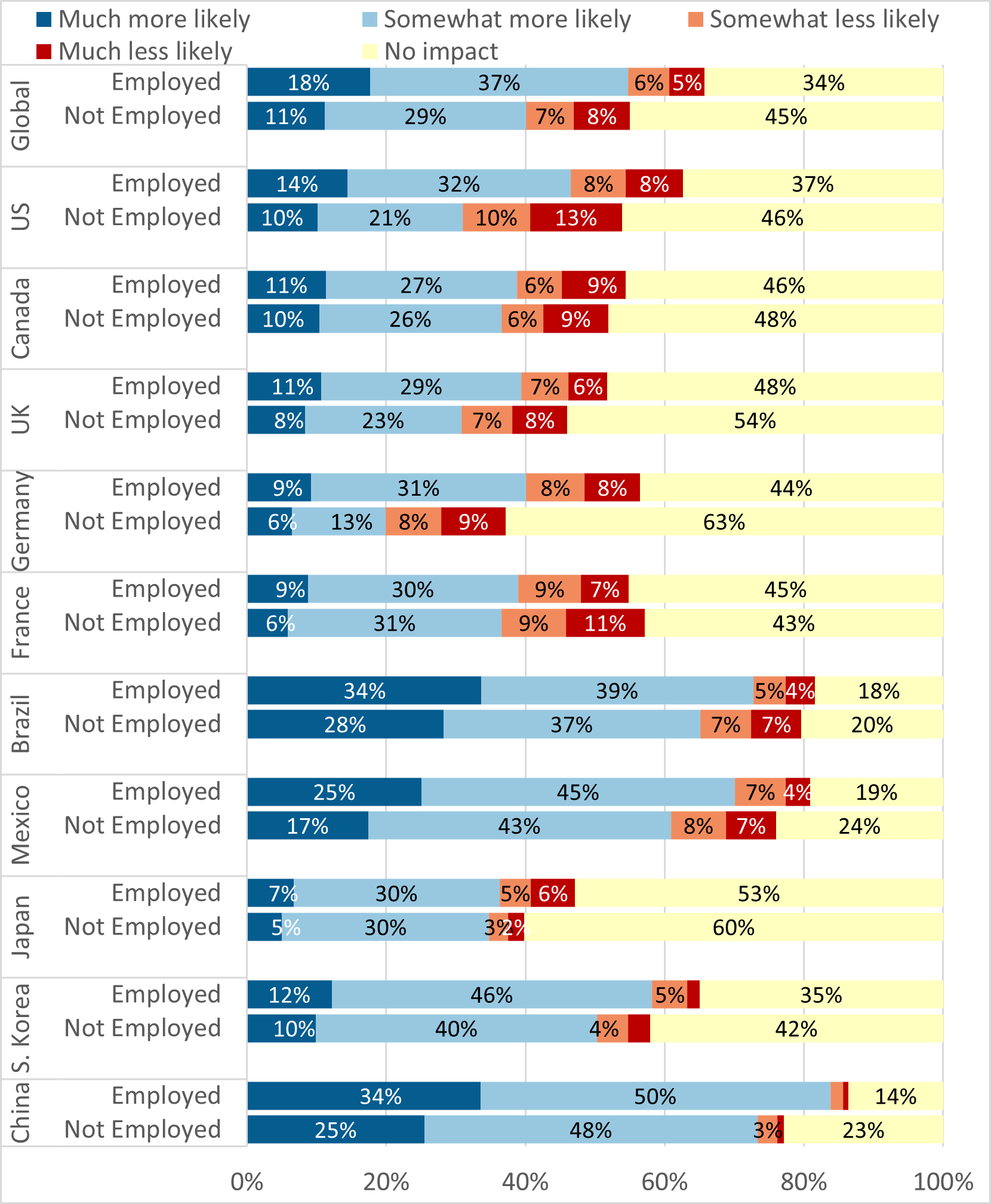}}
\caption*{\textbf{Figure 6D: Responses grouped by employment for “If a company were to use AI to innovate the products and services you use on a daily basis to improve your life, would you be more or less likely to \-- Be interested in learning more about the company.” } (Question 22\_3 from the 2024 3M State of Science Insights)}
\end{figure}

\clearpage
Figure 7 depicts data for the two job-related questions in Figure 2 and Figure 5. The countries with a higher worry about losing jobs to AI also have a higher likelihood for pursuing job in companies that use AI to innovate for everyday lives. 

\begin{figure}[H]
\centering
\fbox{\includegraphics[width=0.95\linewidth]{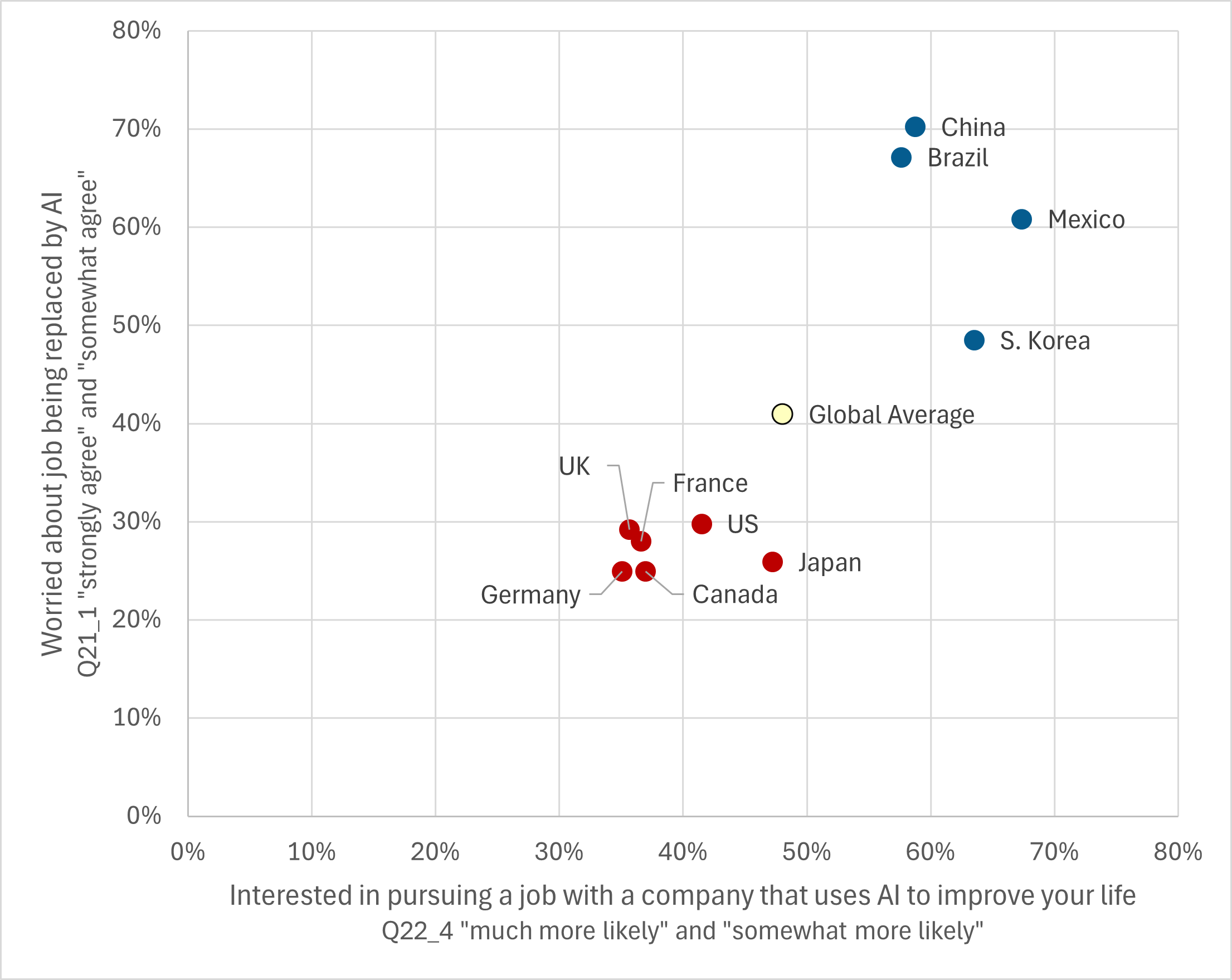}}
\caption*{\textbf{Figure 7: Data from two questions: 21\_1 (Figure 2) “Thinking about AI, how much do you agree or disagree with each of the following? --- I am worried about my job being replaced by AI” and 22\_4 (Figure 5) “If a company were to use AI to innovate the products and services you use on a daily basis to improve your life, would you be more or less likely to \--- Be interested in pursuing a job with the company.” }}
\end{figure}
Figure 8 shows the response to Q10\_8 asked in the same survey and specifically relates to ‘green jobs,’ defined as jobs that contribute to preserving or restoring the environment. Brazil (61\%) and Mexico (57\%) lead China (50\%) in terms of strong agreement while Japan is the lowest at 17\% who “strongly agree.” Japan, along with Germany also has the largest percentage (17\%) of those who disagree. Germany, US and Japan also lead for “Don’t know/No opinion” category with 22\%, 20\% and 18\% respectively.

\begin{figure}[H]
\centering
\fbox{\includegraphics[width=0.85\linewidth]{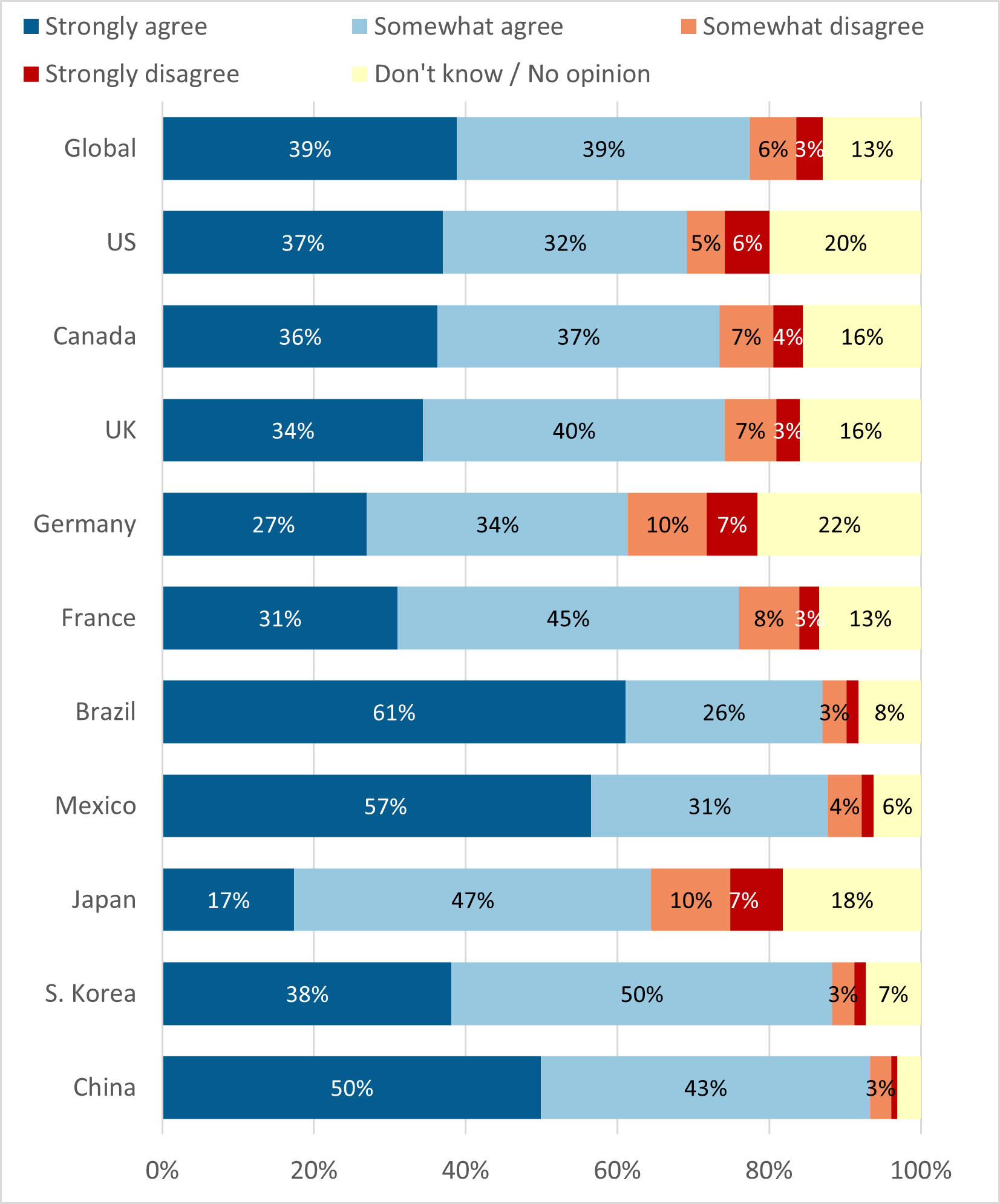}}
\caption*{\textbf{Figure 8: General population responses for “How much do you agree or disagree with each of the following statements?\-- ‘Green jobs’ have a positive impact on society.” } (Question 10\_8 from the 2024 3M State of Science Insights)}
\end{figure}

\section*{Discussion}
Understanding public perception of AI is crucial not just for shaping product offerings and the potential of the technology, but also public policy. Several recent public opinion polls and surveys conducted by academics and other stakeholders in the AI ecosystem have shown the current, somewhat broad, and relatively shallow understanding of AI by the public. This is perhaps due to the ambiguous nature of the technology itself, the level of common knowledge regarding its usage and the media hype specifically around Generative AI. But given the potentially long-lasting impacts of AI – understanding public perception and its chronological evolution is key. 

\clearpage
The 2024 3M State of Science Insights survey results show the paradoxical nature of the public perception of AI as an opportunity, and a potential threat that needs to be heavily regulated. While concerns about job displacement and the need for regulation of AI have been consistent themes over time, and are understandable, it is perhaps important to approach this technology with a sense of possibility. It is especially important to demonstrate the ability to solve problems with AI given that the public agrees “it will change the world as we know it.”

There are national differences across the data set, as well as demographic differences, and both warrant more research to examine the social and cultural factors that can impact impressions, sentimentality and expectations of AI \cite{30}. Consistent with other datasets, developing countries such as China, South Korea, Brazil and Mexico show higher awareness and general positive sentiments compared to developed countries like US, UK, Canada, France, Germany and Japan.  Previous survey data from 2021 has shown that in developing countries “exciting” was the dominant sentiment surrounding AI \cite{9}. China stands out with the most positive perception \cite{10}.

It is important to note that the 2024 3M State of Science Insights survey is amongst the first survey showing that a large percentage of the global public currently views AI as a tool for problem solving and there is a positive response to companies that use AI to innovate to improve daily lives. We suggest that with the agreement associated with the use of AI as a tool, and the positive perception of ‘green jobs’ and their impact on society, there is an opportunity to shape the societal perception of AI by gearing it towards the problems that need to be solved and the jobs that can help with that.  

When a modality can solve a problem its positive perception is likely to increase. This offers an opportunity for businesses to focus on the benefits AI can bring to society – by using AI to help solve the problems that matter the most to the public. According to the Edelman Trust Barometer report, 59\% of people do trust businesses with the introduction of innovations into society \cite{31}.

Our own research over the years, polling the public perception of science, shows that demonstration of problem-solving ability can serve to improve perception. Trust in science from 2020 3M SOSI pre-pandemic was at 85\% and by 2023 rose to 88\%, while trust in scientists was 80\% pre-pandemic and went up to 86\% by 2023 \cite{3,6}. During the pandemic, science was in the public discourse and scientists were center stage as the pandemic unfolded. Moreover, science and scientists played a crucial role in the development of vaccines –-- science was having a `problem-solving moment' and our data showed a more positive public perception.

\vspace{\baselineskip}
\noindent{\textit{\textbf{Problems people want science and technology to solve}}}\\
Climate change and environmental sustainability related topics feature heavily amongst the problems to solve that have been top of mind for the global public. (See Appendix A5 for list of problems and the public response data from 3M commissioned 2018-2024 surveys). In 2024, in addition to 75\% of the global respondents agreeing that ‘green jobs,’ defined as jobs that contribute to preserving or restoring the environment, are crucial to addressing climate change --- 89\% globally agree science and technological innovation should advance the planet and 81\% think climate change is one of the biggest threats to humankind \cite{29}. 

Experts agree that AI has the potential to make a positive difference for UN Sustainable Development Goals \cite{32}. AI modalities can help address climate change and its implications, however, among the problems that will need to be addressed is energy usage. In a 2024 survey, some respondents cited environmental concerns about the increased energy use needed to train Gen AI models as the reason for not embracing Gen AI \cite{33}. Judicious and focused use of AI to help with climate tech, green jobs and science-based scalable innovations could help to solve challenges such as energy usage, and that in turn can facilitate a broader use of AI to solve for other climate related issues, thereby creating a virtuous cycle. 

A recent study has shown ‘personal relevance’ as an important motivational factor that shapes information processing and attitude formation, indicating the extent to which an individual commits to cognitive involvement in a topic such as AI \cite{34}. Moreover, the 2022 3M SOSI results show that many who do choose STEM (Science, Technology, Engineering and Math) fields do so with pro-social and communal goals such as “solving world’s biggest challenges” or “making a difference in the world” \cite{35}. This sense of purpose in solving problems and human agency in tackling challenges can be a win-win. Through augmentation of human creativity and ingenuity with the power of AI, businesses can accelerate innovation and drive progress in areas such as climate change, healthcare and social inequality. By focusing on the potential of AI to solve the problems that matter most, we can ensure that AI is used to benefit society as a whole, while inspiring the next generation of jobs and workers. 

Public support can be important for emerging technologies, but the interpretation of public opinion can admittedly be complex \cite{36}. Since technology can shape public opinion while public opinion can shape technology, it is important to continually measure and monitor engagement and perceptions to understand the nuances of evolving public perspective, the sentiments shaping it and determine collective course of action. Harnessing the positive and mitigating the negative impacts of AI can only be achieved through robust support of stakeholders and effective polices. Deeper understanding of the public perception of AI through more research is warranted for alignment of future development, innovation and governance with societal needs. We recommend additional surveys and research into further understanding the problems public seeks solutions to and the connections between AI and SDGs. 

\section*{Conclusion}
This study of the public perception of AI sheds light on the global landscape of current sentiments surrounding AI. The public largely views AI as a tool for problem solving and has a positive perception of companies that use AI for innovations to improve daily lives. Results from our current survey, and the research into public perception of science, emphasize the ways in which AI may have a favorable impact on public opinion by using AI for addressing public concerns such as climate change. 

Our results align with calls to develop technology that supports public values and serve to build a foundation for fostering beneficial AI in accordance with societal expectations. Accounting for public opinion and emphasizing social context in the discussion around the future of AI is important --- it can help to ensure science and technology can serve society and potential social ills such as "science without humanity" can be avoided.  

With AI permeating every aspect of our lives, reshaping how we think, what we know, and our overall perceptions of the world around us, it is clear that the public perception of this technology is complex, multifaceted and continually evolving. Given that, continual public engagement and monitoring of public perception is key, along with incorporation of the knowledge gained into shaping the technology landscape.

\section*{Acknowledgements}
I would like to thank Katherine Schneider for her tremendous help with the manuscript and Aadarsh Padiyath for his guidance on how to structure it. Thanks also to Robert Brittain, Lauren Cox, and Stephen Priest from 3M, Devon Bottomley and Justine McGarrigle from our extended SOSI team and Gina Gapp, Stephanie Potocny and their team at Morning Consult. 

\clearpage
\RaggedRight
\bibliographystyle{unsrt}
\bibliography{Bibliography.bib}

\begin{center}
\clearpage
\section*{Appendix A1}
Summary of year, survey partner, countries surveyed, and dates fielded 
\par for all 3M State of Science surveys (2018-2024) 
\end{center}

\begin{center}
\label{tab:Appendix A1}
\begin{tabular}{|p{2cm}|p{2cm}|p{7.5cm}|p{2.5cm}|}
\hline

\textbf{Year} &  \textbf{Partner} & \textbf{Countries Surveyed} & \textbf{Dates Fielded} \\ \hline

2018 \cite{1} & Ipsos & \underline{14 countries:} Brazil, Canada, China, France, Germany, India, Japan, Mexico, Poland, Saudi Arabia, Singapore, South Africa, UK, US & Jun. 14 –- Aug. 26, 2017\\ \hline

2019 \cite{2} & Ipsos & \underline{14 countries:} Brazil, Canada, China, Germany, India, Japan, Mexico, Poland, Singapore, South Africa, South Korea, Spain, UK, US & Jul. 13 -- Sep. 10, 2018\\ \hline

2020a \cite{3} (Pre-pandemic) & Ipsos & \underline{14 countries:} US, Canada, UK, Germany, Poland, Spain, Brazil, Mexico, Japan, Singapore, South Korea, China, India, South Africa & Aug. 19 –- Oct. 22, 2019\\ \hline

2020b \cite{3} (pandemic pulse) & Ipsos & \underline{11 countries:} US, Canada, UK, Germany, Poland, Spain, Brazil, Japan, Singapore, South Korea, China & Jul. 22 -– Aug. 16, 2020\\ \hline

2021 \cite{4} & Ipsos & \underline{17 countries:} US, Canada, UK, Germany, Poland, Brazil, Mexico, Japan, Singapore, South Korea, China, India, France, UAE, Italy, Colombia, Australia & Feb. 2 –- Mar. 23, 2021 \\ \hline

2022 \cite{5} & Ipsos & \underline{17 countries:} US, Canada, UK, Germany, Poland, Brazil, Mexico, Japan, Singapore, South Korea, China, India, France, UAE, Italy, Colombia, Australia & Sep. 27 –- Dec. 17, 2021\\ \hline

2023 \cite{6} & Ipsos & \underline{17 countries:} US, Canada, UK, Germany, Taiwan, Brazil, Mexico, Japan, South Korea, Spain, China, India, France, Hong Kong, Italy, Thailand, Australia & Sep. 21 –- Dec. 7, 2022\\ \hline

2024 \cite{29} & Morning Consult & \underline{10 countries:} Brazil, Canada, China, France, Germany, Japan, Mexico, South Korea, UK, US & Dec. 13, 2023 –- Jan. 10, 2024\\ \hline
\end{tabular}
\end{center}

\clearpage
\begin{center}
\section*{Appendix A2}
Survey questions from 3M 2024 State of Science Insights
\end{center}

\begin{enumerate}

  \item Thinking about the present\-day, how important do you feel science is\---
    \begin{enumerate}[label=\theenumi\_\arabic*]
      \item To you in your everyday life
      \item To society in general
      \item[] \textit{[Answer choices: very important, somewhat important, not important at all, no opinion/don’t care]}
    \end{enumerate}

  \item How much do you agree or disagree with each of the following statements? \---
    \begin{enumerate}[label=\theenumi\_\arabic*]
      \item If science didn't exist, my everyday life wouldn't be all that different
      \item I trust scientists	
      \item I trust science
      \item In the future, we will be more dependent on scientific knowledge than ever before
      \item I am more skeptical of science than I was five years ago
      \item[] \textit{[Answer choices: completely agree, somewhat agree, somewhat disagree, completely disagree]}
    \end{enumerate}

  \item How much do you agree or disagree with each of the following statements? \---
    \begin{enumerate}[label=\theenumi\_\arabic*]
      \item Science and technological innovation should advance people. End poverty and zero hunger, health and well-being, gender and racial equality, quality education.
      \item Science and technological innovation should advance the planet. Clean water and sanitation, clean air, wildlife conservation, environmental conservation,\\
      sustainable consumption and production, natural resource management, and \\climate change solutions.
      \item Science and technological innovation should advance prosperity. Economic growth, job creation, clean and affordable energy. Ensure that all human beings can enjoy prosperous and fulfilling lives and that economic, social, and technological progress occurs in harmony with nature.
      \item[] \textit{[Answer choices: very well, somewhat well, not very well, not well at all]}
    \end{enumerate}

  \item To what extent have you incorporated these science and technology based innovations into your life? \---
    \begin{enumerate}[label=\theenumi\_\arabic*]
      \item Artificial Intelligence (AI) (ChatGPT, Bard, etc.)
      \item Augmented Reality/Virtual Reality (AR/VR)
      \item Alternative energy sources (Solar, Wind, etc.)
      \item Clean transportation (Electric vehicles, increased usage of public transit or biking)
      \item[] \textit{[Answer choices: to a great extent, somewhat, very little, not at all, not aware of this technology]}
    \end{enumerate}
 
  \item To you personally, how important is addressing climate change?
    \begin{enumerate}[label=\theenumi\_\arabic*]
      \item[] \textit{[Answer choices: very important, somewhat important, not important at all, no opinion/don’t care]}
    \end{enumerate}

  \item How much do you agree or disagree with each of the following statements? \---
    \begin{enumerate}[label=\theenumi\_\arabic*]
      \item Climate change is one of the biggest threats to humankind
      \item I demand transparency from brands that claim sustainable commitments and expect to know how they are impacting the world
      \item I will actively seek out businesses, establishments or environments (hotels, cafes, office buildings, etc.) that have a mission of sustainability
      \item I will actively seek out products made of sustainable materials or from brands with a commitment to the environment
      \item Businesses need to be more committed to using sustainable materials and technologies in their manufacturing process
      \item Single-use plastics are a major threat to the environment
      \item Fossil fuels are a major threat to the environment
      \item Science can play an essential role in combatting climate change.
      \item Science can help to protect people from the impacts of climate change
      \item Science can help to protect the planet from the impacts of climate change
      \item[] \textit{[Answer choices: strongly agree, somewhat agree, somewhat disagree, strongly disagree, don’t know/no opinion]}
    \end{enumerate}    

  \item Which, if any, of the following consequences of climate change are you most concerned about? Select top three.
    \begin{enumerate}[label=\theenumi\_\arabic*]
      \item[] \textit{[Answer choices: Poor air quality, Lack of clean water, Increase in global hunger, Negative \\impact on the economy (i.e., increased poverty, job loss, etc.), Displacement from my home, Extreme weather events (i.e., severe hurricanes, fires, floods, etc.), Long-term shift in temperatures and weather patterns, Loss of biodiversity (i.e., the variety of plants and\\ animals in a particular habitat), Health issues stemming from changes to the climate, Lessening quality of life for future generations, Other, please specify, I am not concerned about any consequences of climate change]}
      \item[\theenumi b.] When do you think your local community will experience the consequences of climate change?
      \item[] \textit{[Answer choices: I believe climate change is already impacting my community, Within the next five years, In the next 6-10 years, In the next 11-25 years, More than 25 years from now, I believe climate change will have no impact on my community]}
    \end{enumerate}  
   
  \item How familiar are you with the term 'green jobs'?
    \begin{enumerate}[label=\theenumi\_\arabic*]
      \item[] \textit{[Answer choices: very familiar, somewhat familiar, I’ve heard of it but am not familiar, I’ve never heard of the term ‘green jobs’]}
      \item[\theenumi b.] How familiar are you with the term 'green skills'?
      \item [] \textit{[Answer choices: very familiar, somewhat familiar, I’ve heard of it but am not familiar, I’ve never heard of the term ‘green jobs’]}
    \end{enumerate}  

  \item A 'green job' is a job that contributes to preserving or restoring the environment, be that in traditional sectors such as manufacturing and construction, or in new, emerging green sectors such as renewable energy and energy efficiency. If a company were to note that they were dedicated to developing 'green skills' and creating 'green jobs,' how would that affect your perception of the company? \---
    \begin{enumerate}[label=\theenumi\_\arabic*]
      \item Think highly of the company
      \item Purchase from the company
      \item Be interested in learning more about the company
      \item Be interested in pursuing a job with the company
      \item[] \textit{[Answer choices: much more likely, somewhat more likely, no impact, somewhat less likely,\\ much less likely]}
    \end{enumerate}

  \item How much do you agree or disagree with each of the following statements? \---
    \begin{enumerate}[label=\theenumi\_\arabic*]
      \item ‘Green jobs' provide exciting new opportunities for work
      \item I expect the 'green jobs' market to grow in the next five years
      \item Looking for 'green jobs' is not a generational trend, people in all career stages should be looking for 'green jobs'
      \item I would like to learn 'green skills' that would help me obtain a 'green job'
      \item 'Green jobs' are crucial to addressing climate change
      \item The workforce needs more 'green job' workers
      \item I respect people who work in 'green jobs'
      \item 'Green jobs' have a positive impact on society
      \item I feel empowered to make choices in my current role that advance sustainability and/or climate action
      \item It is crucial to build momentum among employees and employers for 'green skill' development
      \item Companies are well equipped to promote and transition to 'green jobs'
      \item[10b\_1] There are opportunities in my country to pursue educational paths with a focus on environmental preservation
      \item[10b\_2] My country's government has an interest in promoting 'green jobs'
      \item[] \textit{[Answer choices: strongly agree, somewhat agree, somewhat disagree, strongly disagree, don’t know/no opinion]}
    \end{enumerate}  

  \item How familiar are you with the term 'carbon footprint'?
    \begin{enumerate}[label=\theenumi\_\arabic*]
      \item[] \textit{[Answer choices: very familiar, somewhat familiar, I’ve heard of it, but am not familiar, I’ve never heard of carbon footprint]}
    \end{enumerate}

  \item A carbon footprint is the amount of greenhouse gas (GHG) emissions released into the atmosphere by an individual, organization, process, product, or event from within a specified boundary. Many environmental initiatives aim to neutralize, erase, or reduce carbon footprints for the benefit of the climate. Greenhouse gases, such as carbon dioxide, methane, nitrous oxide, and certain synthetic chemicals, trap some of the Earth's outgoing energy, thus retaining heat in the atmosphere. If a company were to note that it had a small, neutral or negative carbon footprint, would you be more or less likely to \---
    \begin{enumerate}[label=\theenumi\_\arabic*]
      \item Think highly of the company
      \item Purchase from the company
      \item Be interested in learning more about the company
      \item Be interested in pursuing a job with the company
      \item[] \textit{[Answer choices: much more likely, somewhat more likely, no impact, somewhat less likely,\\ much less likely]}
    \end{enumerate}  

\item[12b.] Setting aside cost for a moment, how interested would you be in participating in each of these actions to address climate change? \---
    \begin{enumerate}[label=\theenumi\_\arabic*]
      \item[12b\_1] Paying a monthly utility fee to neutralize or 'offset' your personal carbon footprint
      \item[12b\_2] Driving an electric vehicle (EV)
      \item[12b\_3] Joining a community solar garden (i.e. sourcing your electricity from solar panels located somewhere other than your property)
      \item[12b\_4] Installing solar panels on your home
      \item[] \textit{[Answer choices: very interested, somewhat interested, neither interested nor \\uninterested, somewhat uninterested, very uninterested, don’t know/no opinion]}
    \end{enumerate}  

  \item How familiar are you with each of the following as a means to address climate change? \---
    \begin{enumerate}[label=\theenumi\_\arabic*]
      \item Carbon capture and sequestration (a process where carbon dioxide, a greenhouse gas, is filtered and stored securely away from air and water to prevent it from affecting the atmosphere)
      \item Renewable energy and fuels (wind power, solar power, green hydrogen)
      \item Clean transportation (bikes, public transport, and electric vehicles, including cars, buses, and trains)
      \item Sustainable buildings and infrastructure (energy efficiency, green and resilient materials)
      \item Regenerative agriculture (resource management, crop selection and rotation, land management)
      \item Waste reduction (renewable materials, reduced plastic use, compostable materials)
      \item[] \textit{[Answer choices: very familiar, somewhat familiar, I’ve heard of it, but am not familiar, I’ve\\ never heard of it]}
    \end{enumerate}

  \item And how interested would you be in learning more about each of the following as a means to addressing climate change? \---
    \begin{enumerate}[label=\theenumi\_\arabic*]
      \item Carbon capture and sequestration (a process where carbon dioxide, a greenhouse gas, is filtered and stored securely away from air and water to prevent it from affecting the atmosphere)
      \item Renewable energy and fuels (wind power, solar power, green hydrogen)
      \item Clean transportation (bikes, public transport, and electric vehicles, including cars, buses, and trains)
      \item Sustainable buildings and infrastructure (energy efficiency, green and resilient materials)
      \item Regenerative agriculture (resource management, crop selection and rotation, land management)
      \item Waste reduction (renewable materials, reduced plastic use, compostable materials)
      \item[] \textit{[Answer choices: extremely interested, very interested, somewhat interested, not very\\ interested, not at all interested]}
    \end{enumerate}

  \item Carbon capture is a process where carbon dioxide, a greenhouse gas, is filtered and stored securely away from air and water to prevent it from affecting the atmosphere. This process results in carbon dioxide being captured either from the air or during an industrial process. After reading this definition, how supportive are you of carbon capture as a method for addressing climate change?
    \begin{enumerate}[label=\theenumi\_\arabic*]
      \item[] \textit{[Answer choices: very supportive, somewhat supportive, not very supportive, not at all supportive, no opinion]}
    \end{enumerate}

  \item How familiar are you with what a data center is?
    \begin{enumerate}[label=\theenumi\_\arabic*]
      \item[] \textit{[Answer choices: very familiar, somewhat familiar, somewhat unfamiliar, not at all familiar]}
    \end{enumerate}

  \item Data centers, commonly used to provide the power necessary to power modern technology, contain many servers which produce heat, meaning that the facilities must be heavily cooled, and that process of cooling can use large quantities of water and energy. After learning more about the environmental impact of data centers, how willing would you be to do the following things? 
    \begin{enumerate}[label=\theenumi\_\arabic*]
      \item Change your email address to a provider who uses energy and water efficient data centers
      \item Transfer your documents to a provider who uses energy and water efficient data centers
      \item Leave a popular social media platform that does NOT use energy and water efficient data centers
      \item Stop using a streaming platform that does NOT use energy and water efficient data centers
      \item[] \textit{[Answer choices: very willing, somewhat willing, somewhat unwilling, not at all willing, don’t know/no opinion]}
    \end{enumerate}

  \item Thinking about the next vehicle you plan to purchase or lease, setting aside cost, would \\you prefer to have?
    \begin{enumerate}[label=\theenumi\_\arabic*]
      \item[] \textit{[Answer choices: Gas or diesel engine, Hybrid, Electric, I have no intention of purchasing or leasing a vehicle]}
    \end{enumerate}

  \item How much do you agree or disagree with the following statement: My country has the infrastructure required to support widespread usage of electric vehicles.
    \begin{enumerate}[label=\theenumi\_\arabic*]
      \item[] \textit{[Answer choices: strongly agree, somewhat agree, somewhat disagree, strongly disagree, don’t know/no opinion]}
    \end{enumerate}

  \item Thinking about the adoption of electric vehicles, how concerned are you about each of the following:\---
    \begin{enumerate}[label=\theenumi\_\arabic*]
      \item Battery range (distance vehicle can travel without needing to recharge)
      \item Ease of finding a charger while traveling
      \item Compatibility of chargers with the electric vehicle
      \item Cost of vehicles
      \item Type of electric vehicles available
      \item Battery fires
      \item Safety concerns (other than battery fires)
      \item Environmental impact of lithium batteries
      \item[] \textit{[Answer choices: very concerned, somewhat concerned, not very concerned, not at all concerned]}
    \end{enumerate}

  \item Thinking about AI, how much do you agree or disagree with each of the following? \---
    \begin{enumerate}[label=\theenumi\_\arabic*]
      \item I am worried about my job being replaced by AI
      \item AI will change the world as we know it
      \item AI is a tool for problem solving
      \item I see myself using AI in my daily life
      \item AI needs to be heavily regulated
      \item AI has an increased presence in my daily life
      \item I have a strong understanding of where AI is being used in everyday life
      \item[] \textit{[Answer choices: strongly agree, somewhat agree, somewhat disagree, strongly disagree, don’t know/no opinion]}
    \end{enumerate}

  \item If a company were to use AI to innovate the products and services you use on a daily basis to improve your life, would you be more or less likely to \---
    \begin{enumerate}[label=\theenumi\_\arabic*]
      \item Think highly of the company
      \item Purchase from the company
      \item Be interested in learning more about the company
      \item Be interested in pursuing a job with the company
      \item[] \textit{[Answer choices: much more likely, somewhat more likely, no impact, somewhat less\\ likely, much less likely]}
    \end{enumerate}

  \item To what extent are you integrating AI into your work environment and processes (i.e. using AI to\\ write emails or basic code, generating content using AI, etc.) ?
    \begin{enumerate}[label=\theenumi\_\arabic*]
      \item[] \textit{[Answer choices: to a great extent, somewhat, very little, not at all]}
    \end{enumerate}

  \item How much do you agree or disagree with each of the following? \---
    \begin{enumerate}[label=\theenumi\_\arabic*]
      \item Jobs in STEM (science, technology, engineering, or math) are the future of work
      \item I am worried about jobs being replaced by automation/machines/robots
      \item It is important to have a diverse workforce
      \item New technologies will lead to more job opportunities
      \item New technologies will lead to fewer job opportunities
      \item My workplace is diverse
      \item It is important to have a college degree
      \item I would be excited to work with emerging technologies like AI, robots and automated machinery
      \item It will be more important to have a college degree in the future
      \item Companies play a significant role in helping to develop STEM talent
      \item[] \textit{[Answer choices: strongly agree, somewhat agree, somewhat disagree, strongly disagree, don’t know/no opinion]}
    \end{enumerate}

  \item How likely would you be to advise a younger person to pursue each of the following jobs? \---
    \begin{enumerate}[label=\theenumi\_\arabic*]
      \item A job in STEM (science, technology, engineering, math)
      \item A job in the service industry
      \item A 'green job'; any job that contributes to preserving or restoring the environment
      \item A job in the humanities/arts
      \item A job in a skilled trade
      \item[] \textit{[Answer choices: very likely, somewhat likely, not very likely, not at all likely]}
    \end{enumerate}

  \item You mentioned you would be likely to advise a younger person to pursue a job in STEM. Which,\\ if any, of the following, are reasons you say that?
    \begin{enumerate}[label=\theenumi\_\arabic*]
      \item[] \textit{[Answer choices: Higher pay, Job security, Personal fulfillment, The multitude of career opportunities, Opportunities for problem-solving and innovation, Respect \\for careers, Good career trajectories/growth opportunities, STEM is a growing field, The world needs more STEM workers, STEM skills will be increasingly important to all jobs. Other]}
    \end{enumerate}

  \item Which, if any, of the following are reasons you might NOT encourage a younger person to pursue a job in STEM?
    \begin{enumerate}[label=\theenumi\_\arabic*]
      \item[] \textit{[Answer choices: Time required to learn STEM skills, Cost of STEM degrees and/or certifications, Lack of gender diversity in STEM fields, Lack of racial diversity in STEM fields, It seems too mentally challenging, I'm not sure what steps to recommend taking to secure a STEM job, Location of STEM jobs, I believe humanities/social sciences roles are more important, Other, None of the above]}
    \end{enumerate}

  \item You mentioned you would be likely to advise a younger person to pursue a green job, a job that contributes to preserving or restoring the environment, be they in traditional sectors such as manufacturing and construction, or in new, emerging green sectors such as renewable energy and energy efficiency. Which, if any, of the following, are reasons you say that?
    \begin{enumerate}[label=\theenumi\_\arabic*]
      \item[] \textit{[Answer choices: Higher pay, Job security, Personal fulfillment, Multitude of career opportunities, Opportunities for problem solving and innovation, Respect for careers, Good career trajectories/growth opportunities, Sustainable/'green' jobs are a growing field, The world needs more green job workers, Other]}
    \end{enumerate}

  \item Which, if any, of the following are reasons you might NOT encourage a younger person to pursue a green job?
    \begin{enumerate}[label=\theenumi\_\arabic*]
      \item[] \textit{[Answer choices: Time required to learn 'green' skills, It seems too challenging, I'm not sure what steps to recommend taking to secure a 'green' job, Location of green jobs, The field is too niche, It's a trend, there's no long-lasting careers options, Other, None of the above]}
    \end{enumerate}

  \item In addition to schools and teachers, who do you think is responsible for stimulating interest of young people in STEM careers?
    \begin{enumerate}[label=\theenumi\_\arabic*]
      \item[] \textit{[Answer choices: Government, Parents, Friends and peers, Other family members, Companies, Community groups, Other, None of the above]}
    \end{enumerate}
\end{enumerate}

\clearpage
\begin{center}

\section*{Appendix A3}
Additional data for AI related questions from 2024 3M State of Science Insights \cite{29} 
\end{center}

\begin{figure}[H]
\fbox{\includegraphics[width=0.47\linewidth]{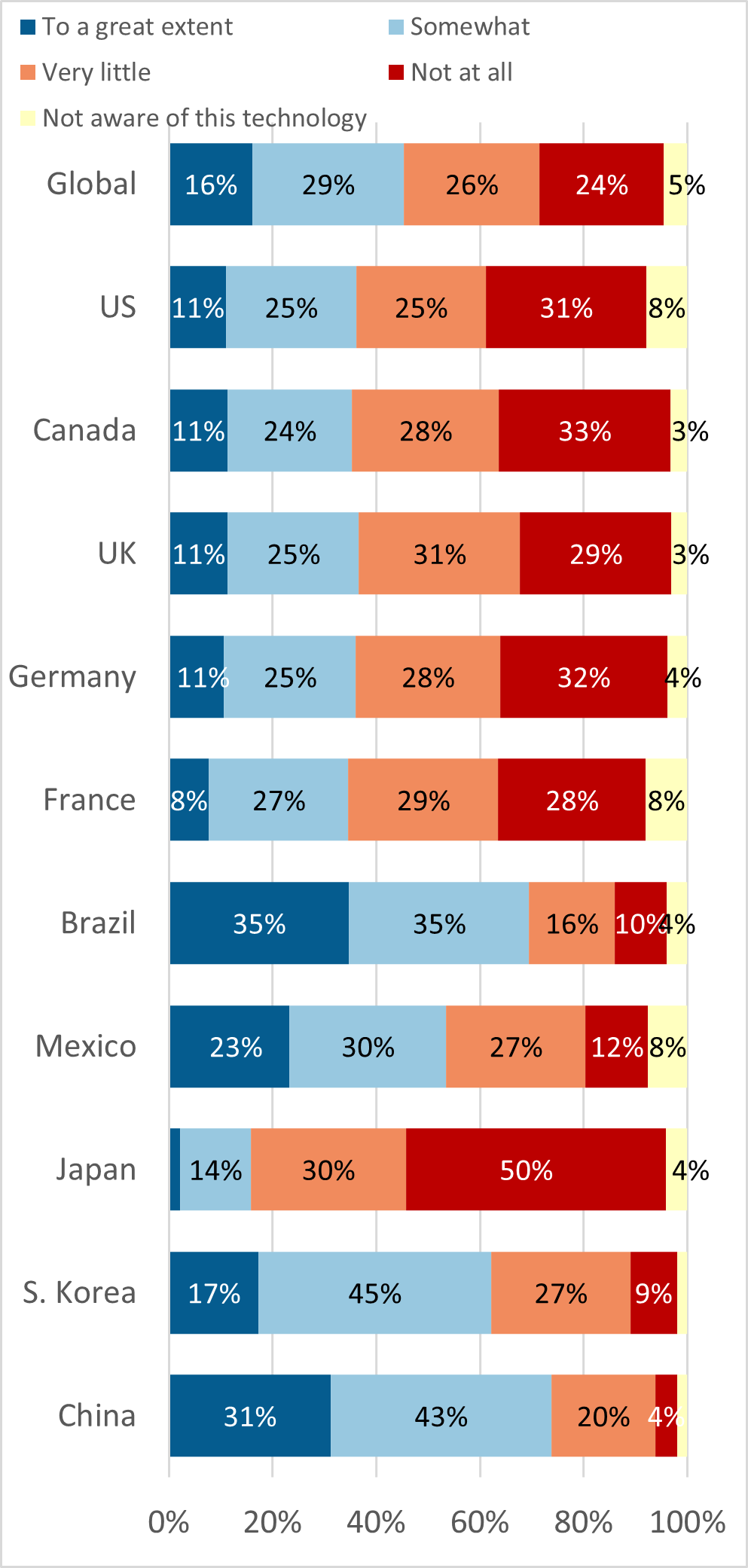}}
\fbox{\includegraphics[width=0.46699\linewidth]{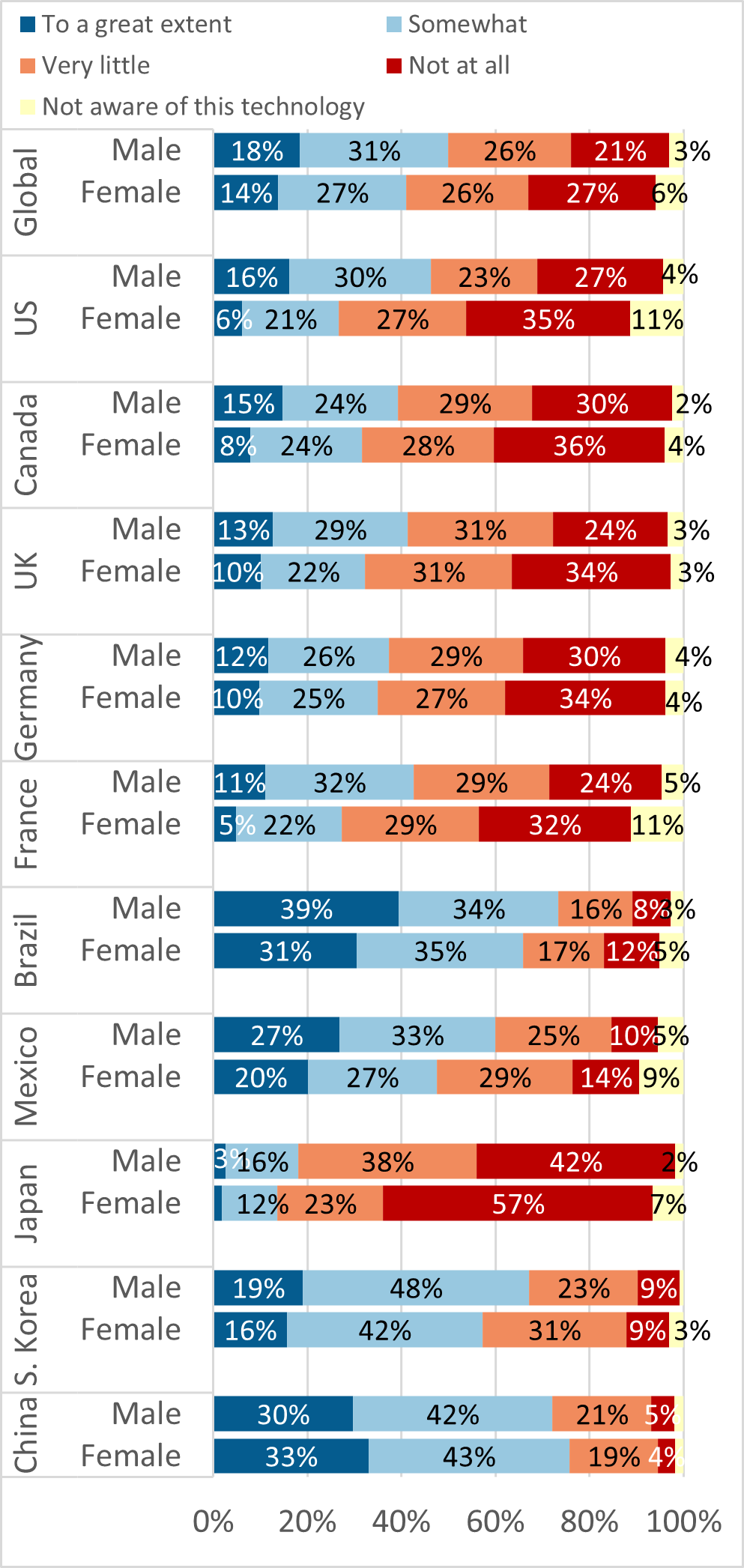}}
\caption*{\textbf{Figure A1A and A1B: Responses for “To what extent have you incorporated these science and technology based innovations into your life? Artificial intelligence (AI) (ChatGPT, Bard, etc.)”} Figure A1A(left): Results from general population. Figure A1B(right): Results grouped by gender. (Question 4\_1 from the 2024 3M State of Science Insights)}
\end{figure}
\begin{figure}[H]
\fbox{\includegraphics[width=0.46\linewidth]{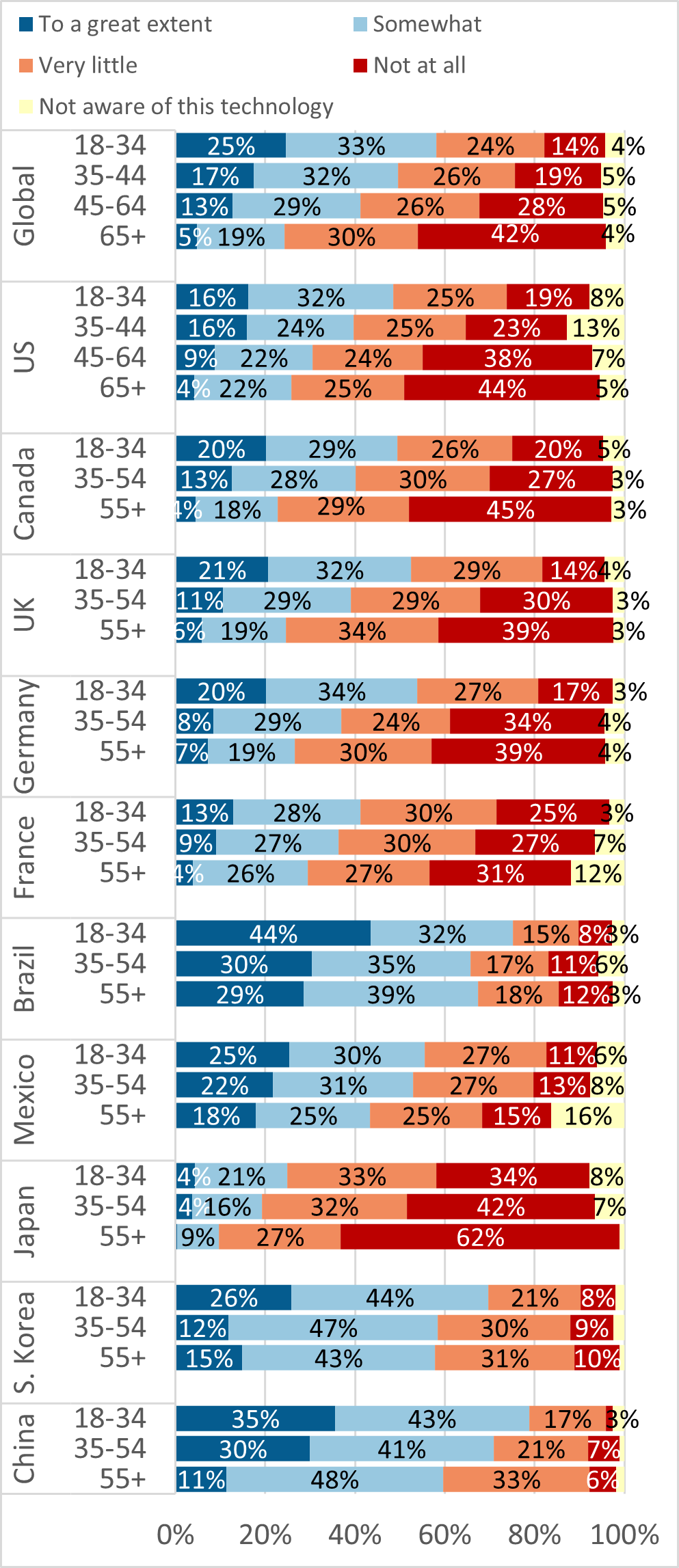}}
\fbox{\includegraphics[width=0.47554\linewidth]{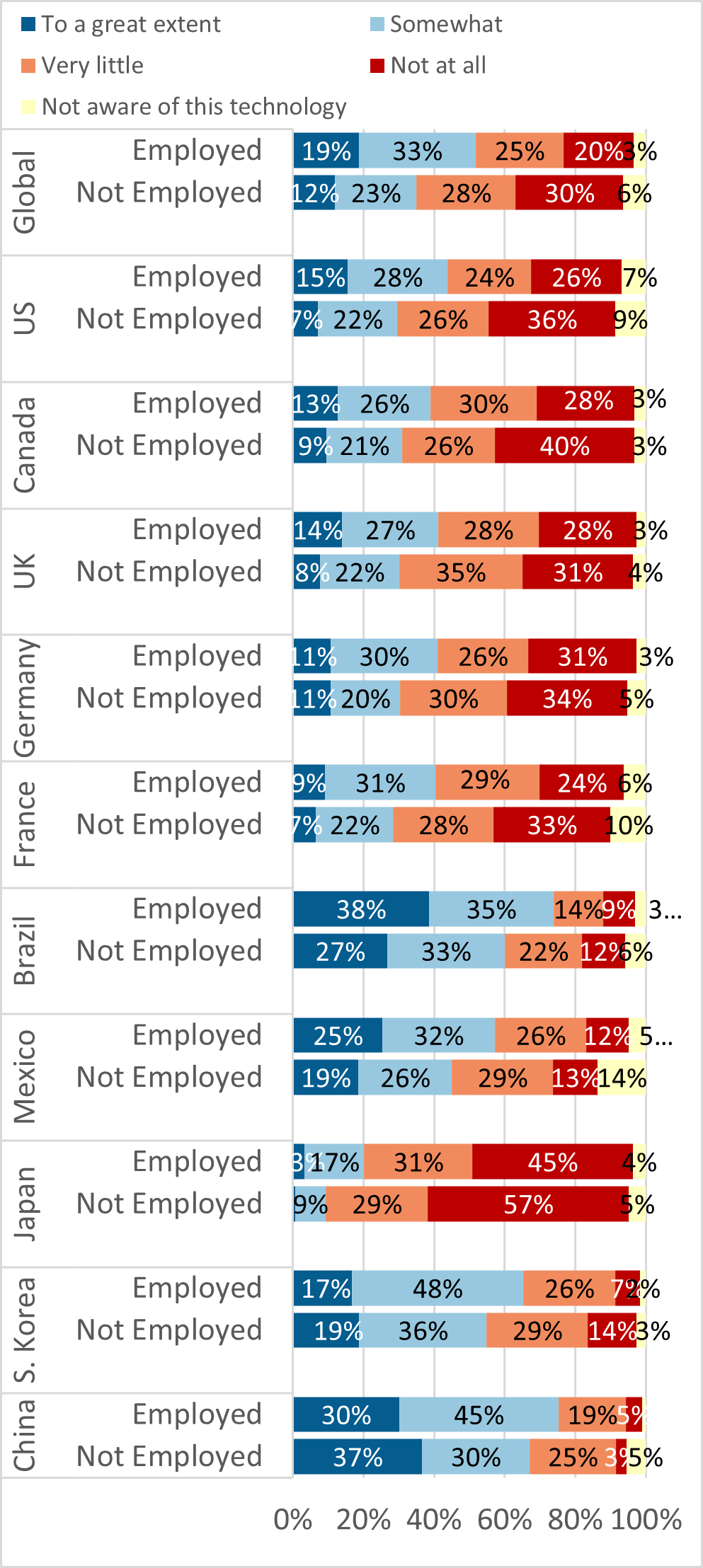}}
\caption*{\textbf{Figure A1C and A1D: Responses for “To what extent have you incorporated these science and technology based innovations into your life? Artificial intelligence (AI) (ChatGPT, Bard, etc.)” }Figure A1C(left): Results grouped by age. Figure A1D(right): Results grouped by employment. (Question 4\_1 from the 2024 3M State of Science Insights)}
\end{figure}

\begin{figure}[H]
\fbox{\includegraphics[width=0.466\linewidth]{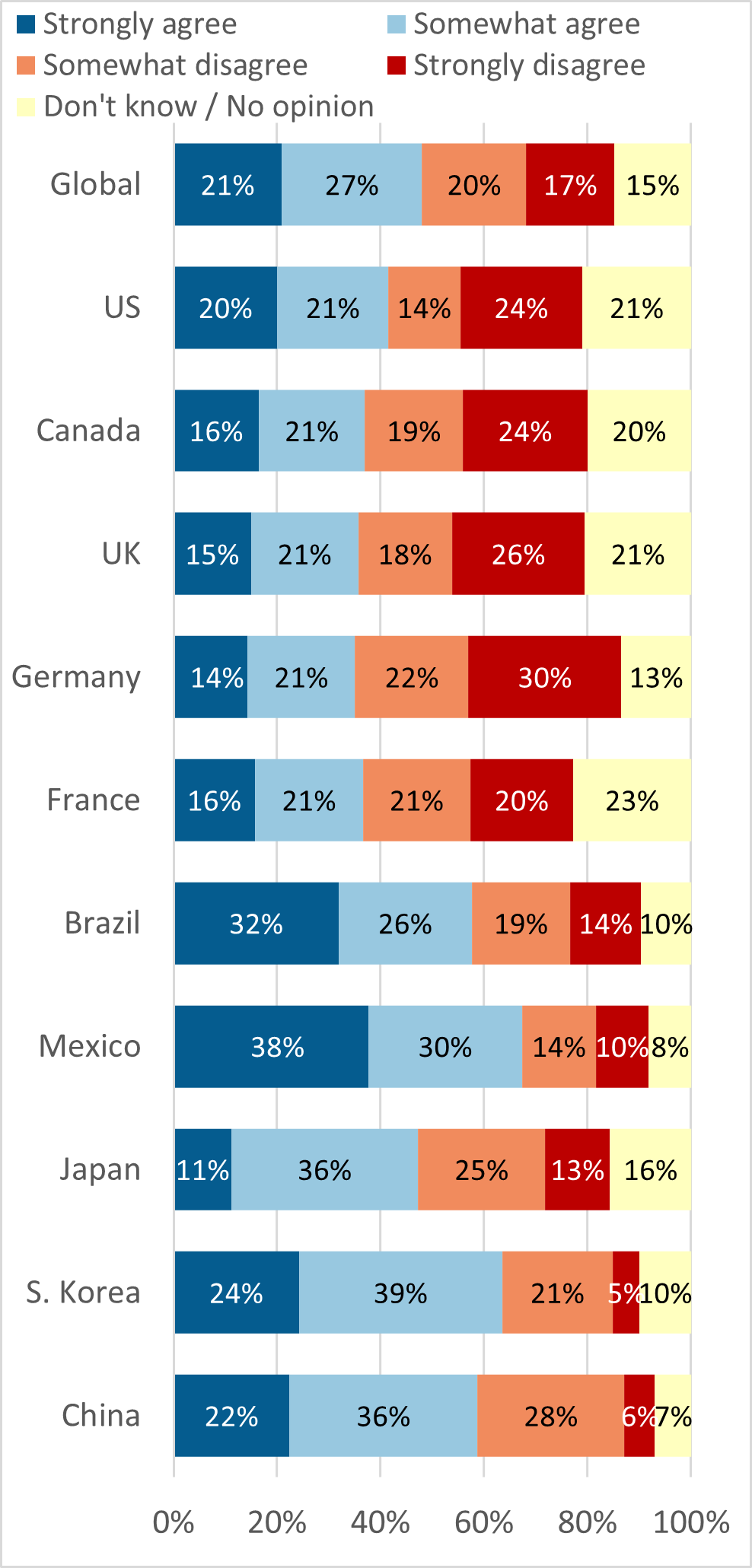}}
\fbox{\includegraphics[width=0.466\linewidth]{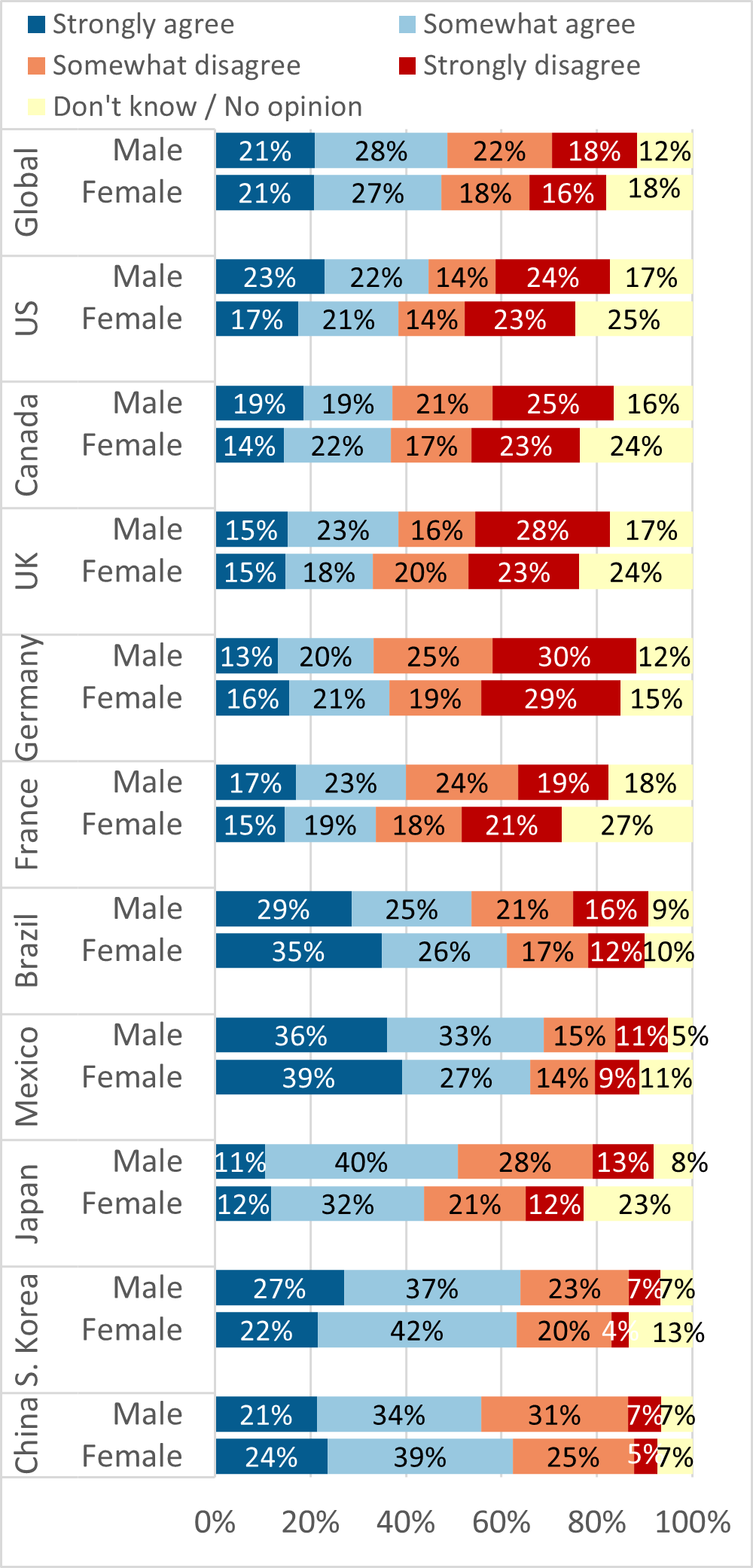}}
\caption*{\textbf{Figure A2A and A2B: Responses for “Thinking about AI, how much do you agree or disagree with the following?—--I am worried about my job being replaced by AI” }Figure A2A(left): Results from general population. Figure A2B(right): Results grouped by gender. (Question 21\_1 from the 2024 3M State of Science Insights) }
\end{figure}
\begin{figure}[H]
\fbox{\includegraphics[width=0.47\linewidth]{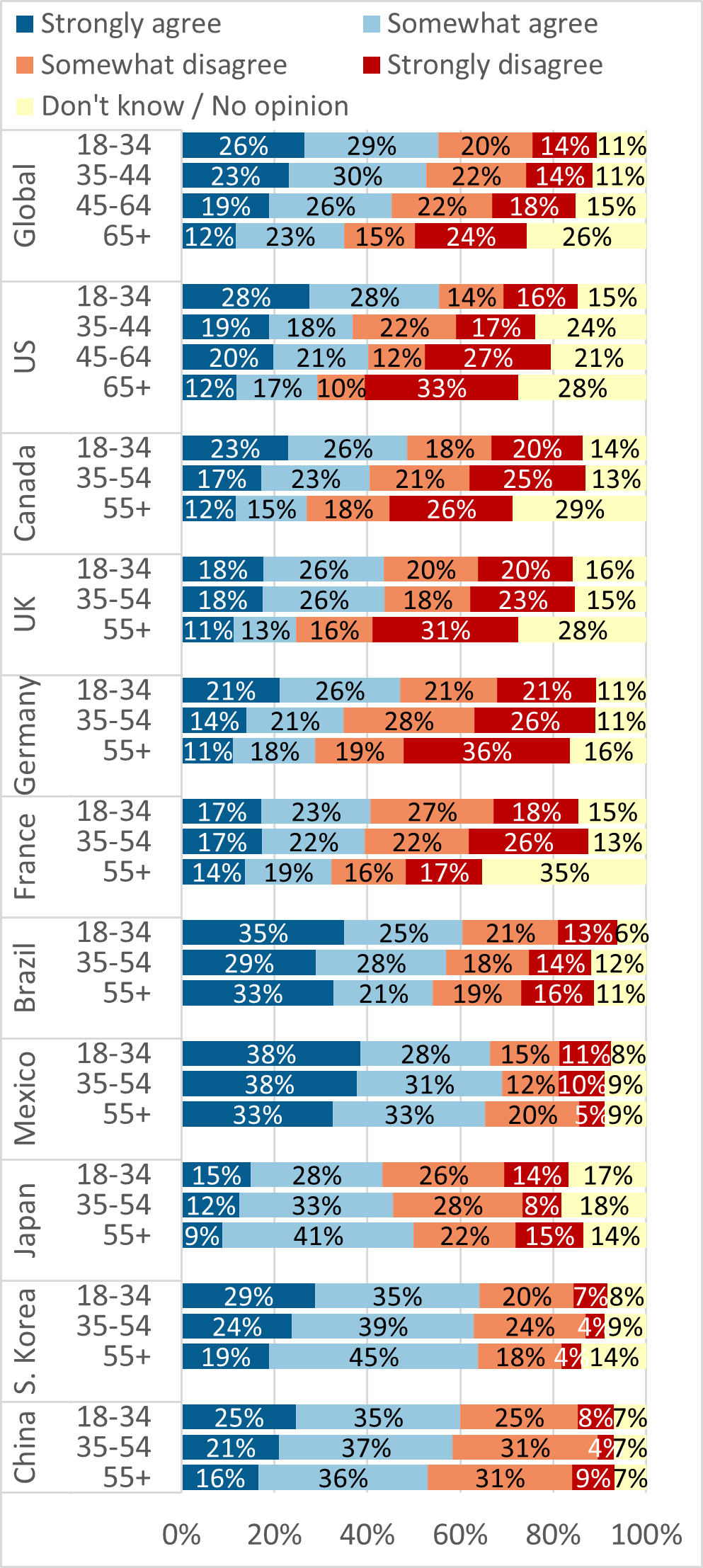}}
\fbox{\includegraphics[width=0.47005\linewidth]{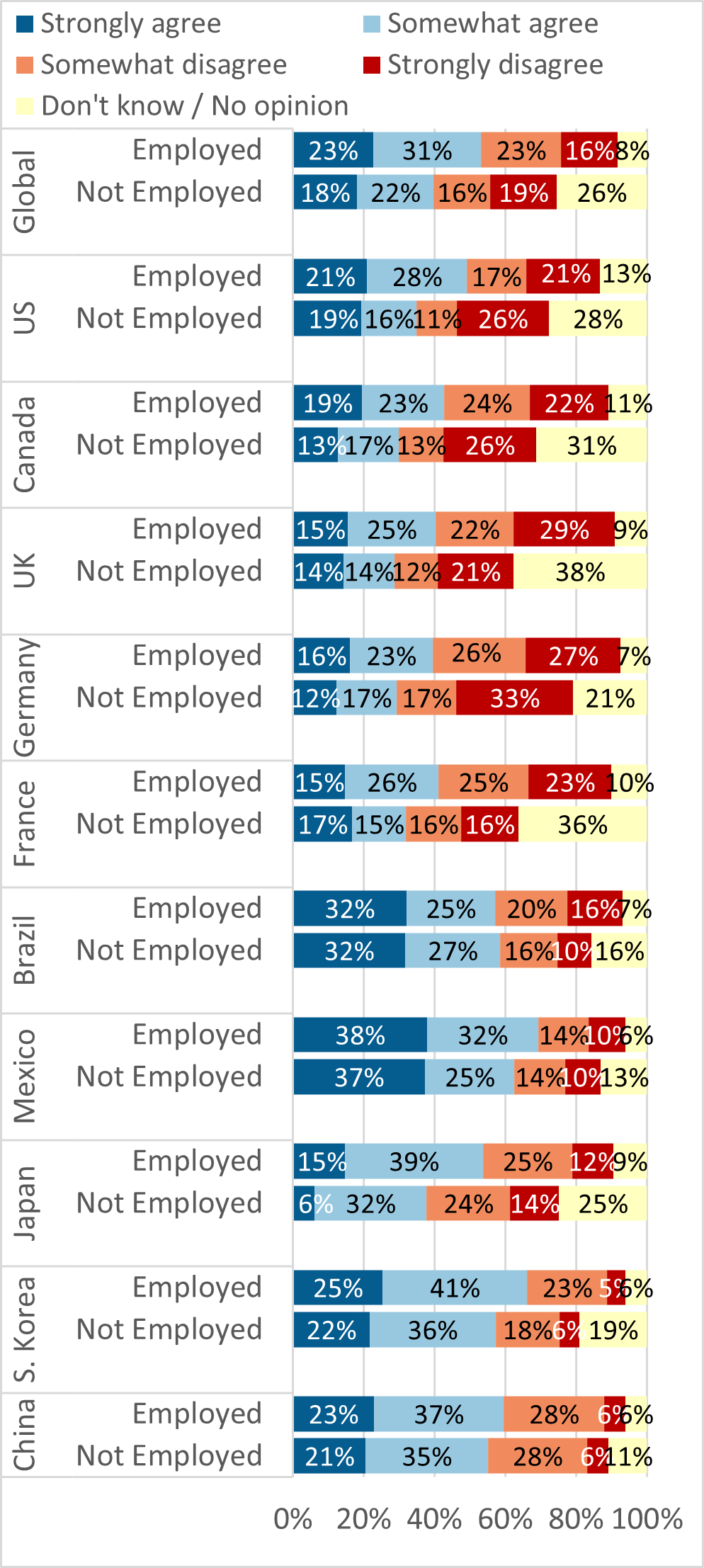}}
\caption*{\textbf{Figure A2C and A2D: Responses for “Thinking about AI, how much do you agree or disagree with the following?--—I am worried about my job being replaced by AI” }Figure A2C(left): Results grouped by age. Figure A2D(right): Results grouped by employment. (Question 21\_1 from the 2024 3M State of Science Insights) }
\end{figure}

\begin{figure}[H]
\fbox{\includegraphics[width=0.47\linewidth]{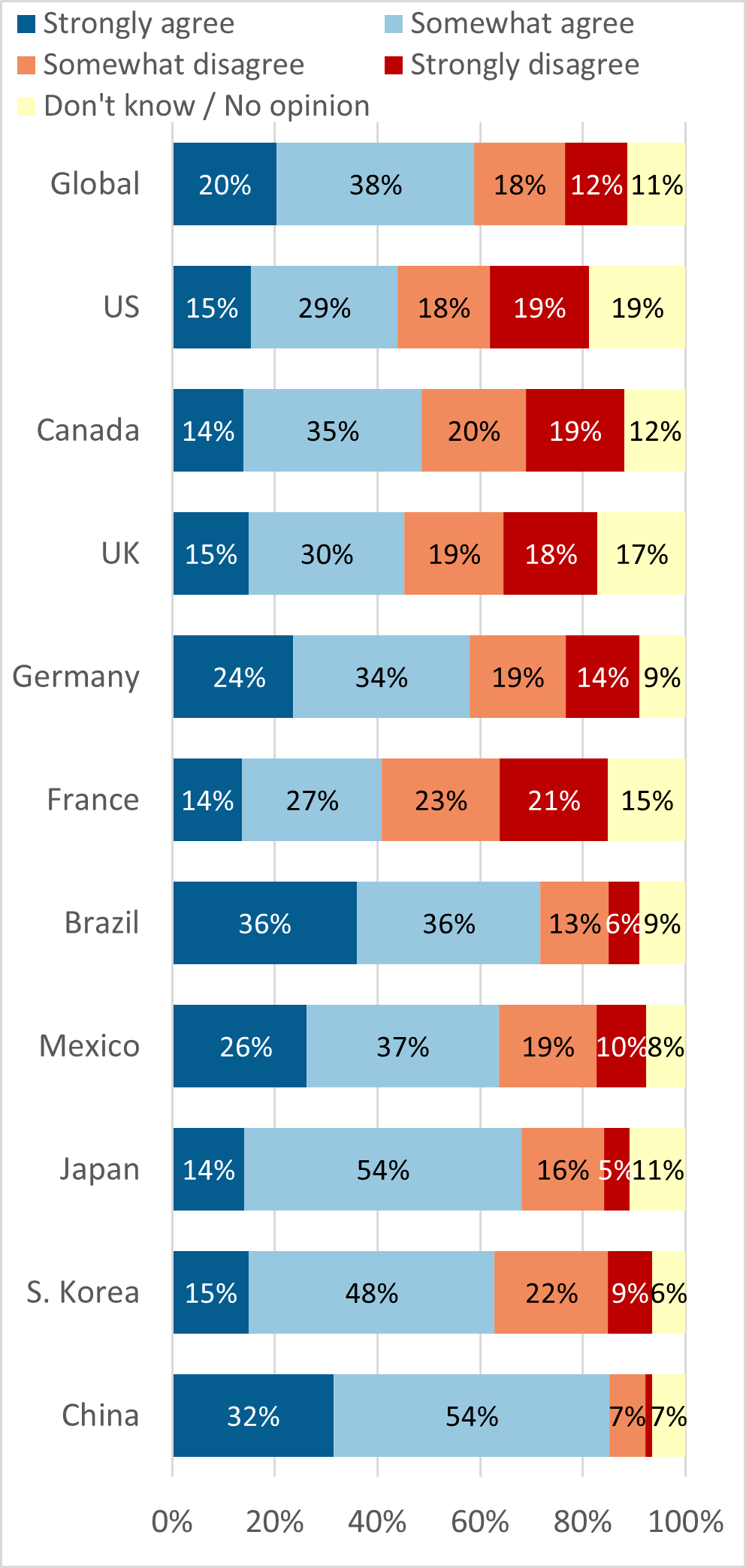}}
\fbox{\includegraphics[width=0.473\linewidth]{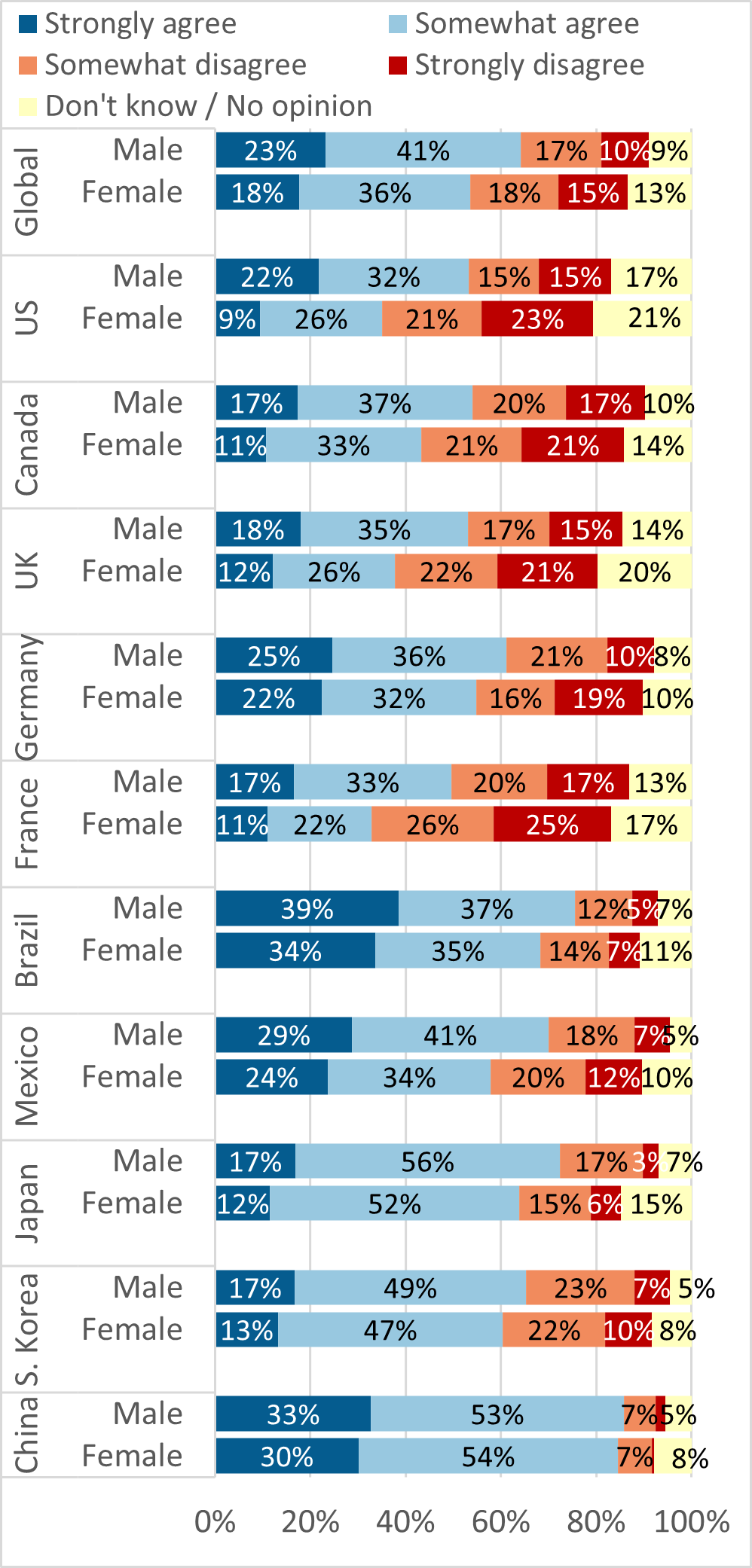}}
\caption*{\textbf{Figure A3A and A3B: Responses for “Thinking about AI, how much do you agree or disagree with the following?--—I see myself using AI in my daily life” }Figure A3A(left): Results from general population. Figure A3B(right): Results grouped by gender. (Question 21\_4 from the 2024 3M State of Science Insights) }
\end{figure}
\begin{figure}[H]
\fbox{\includegraphics[width=0.47\linewidth]{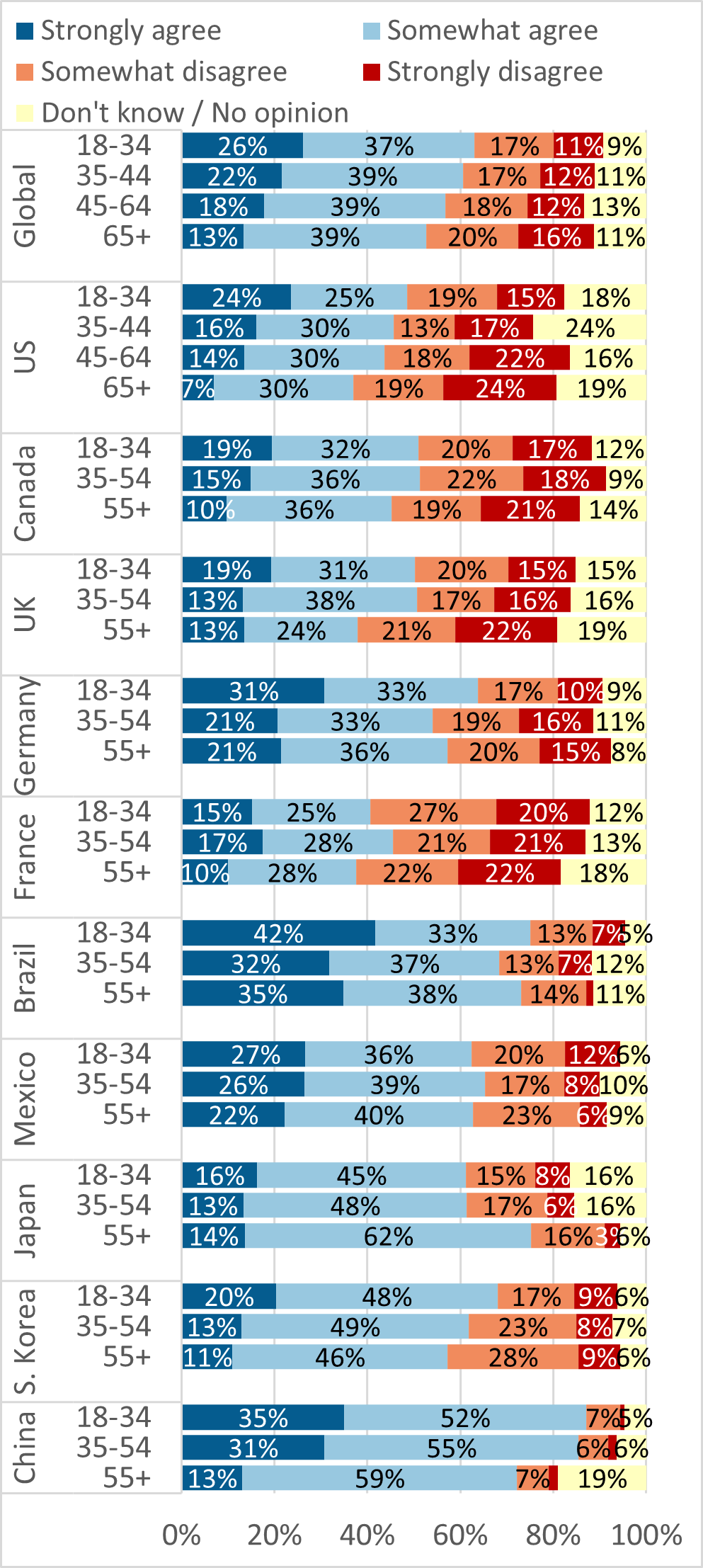}}
\fbox{\includegraphics[width=0.469\linewidth]{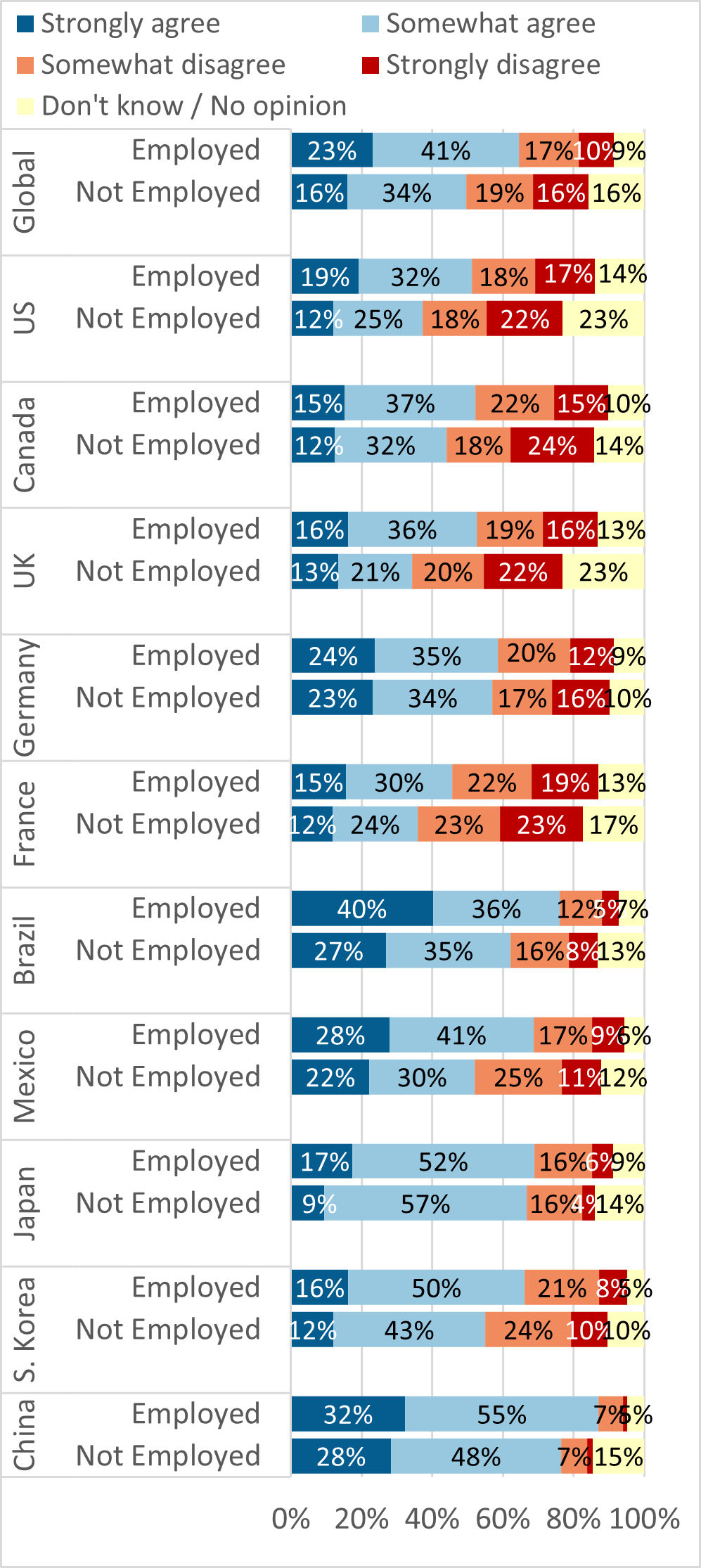}}
\caption*{\textbf{Figure A3C and A3D: Responses for “Thinking about AI, how much do you agree or disagree with the following?--—I see myself using AI in my daily life.” }Figure A3C(left): Results grouped by age. Figure A3D(right): Results grouped by employment. (Question 21\_4 from the 2024 3M State of Science Insights) }
\end{figure}

\begin{figure}[H]
\fbox{\includegraphics[width=0.469\linewidth]{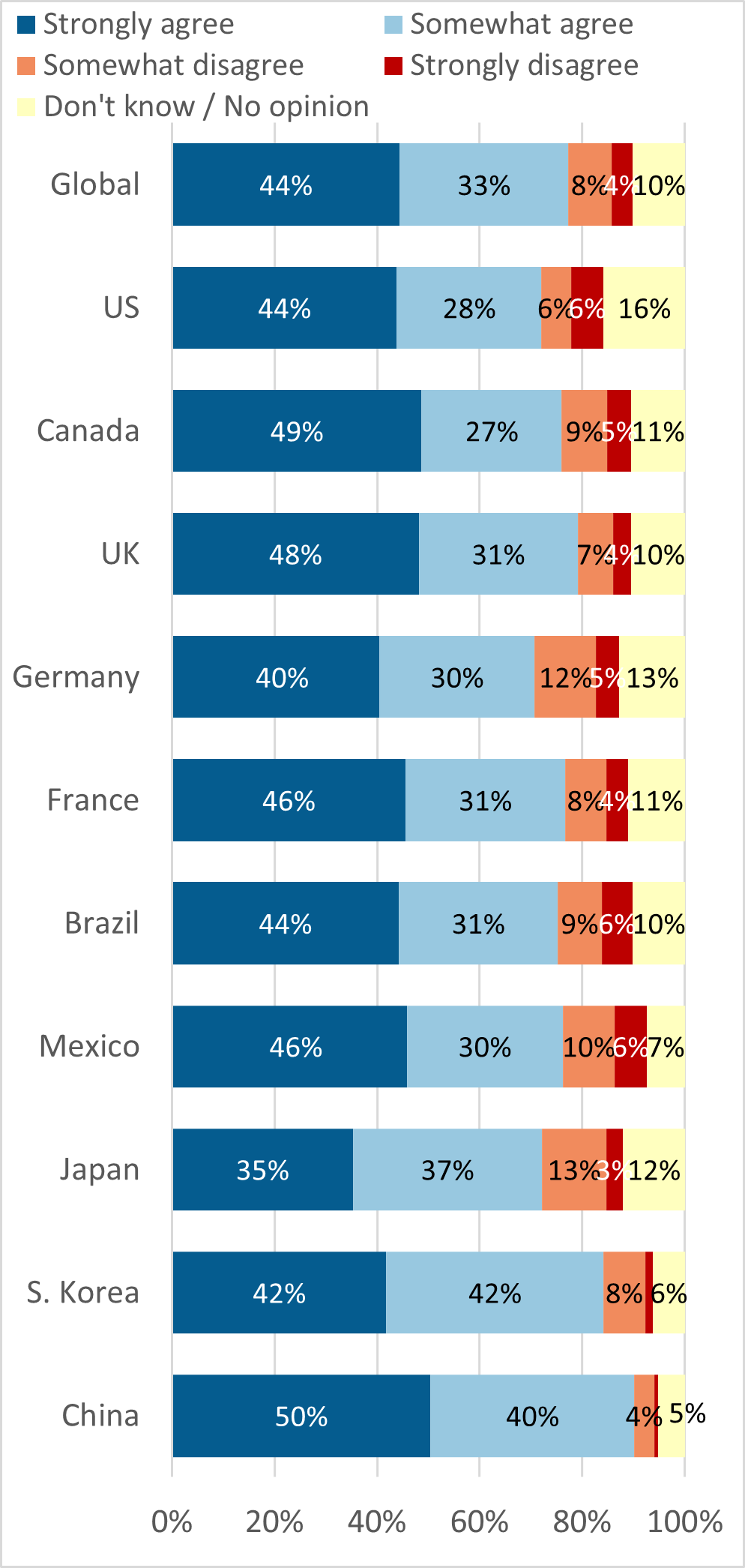}}
\fbox{\includegraphics[width=0.47\linewidth]{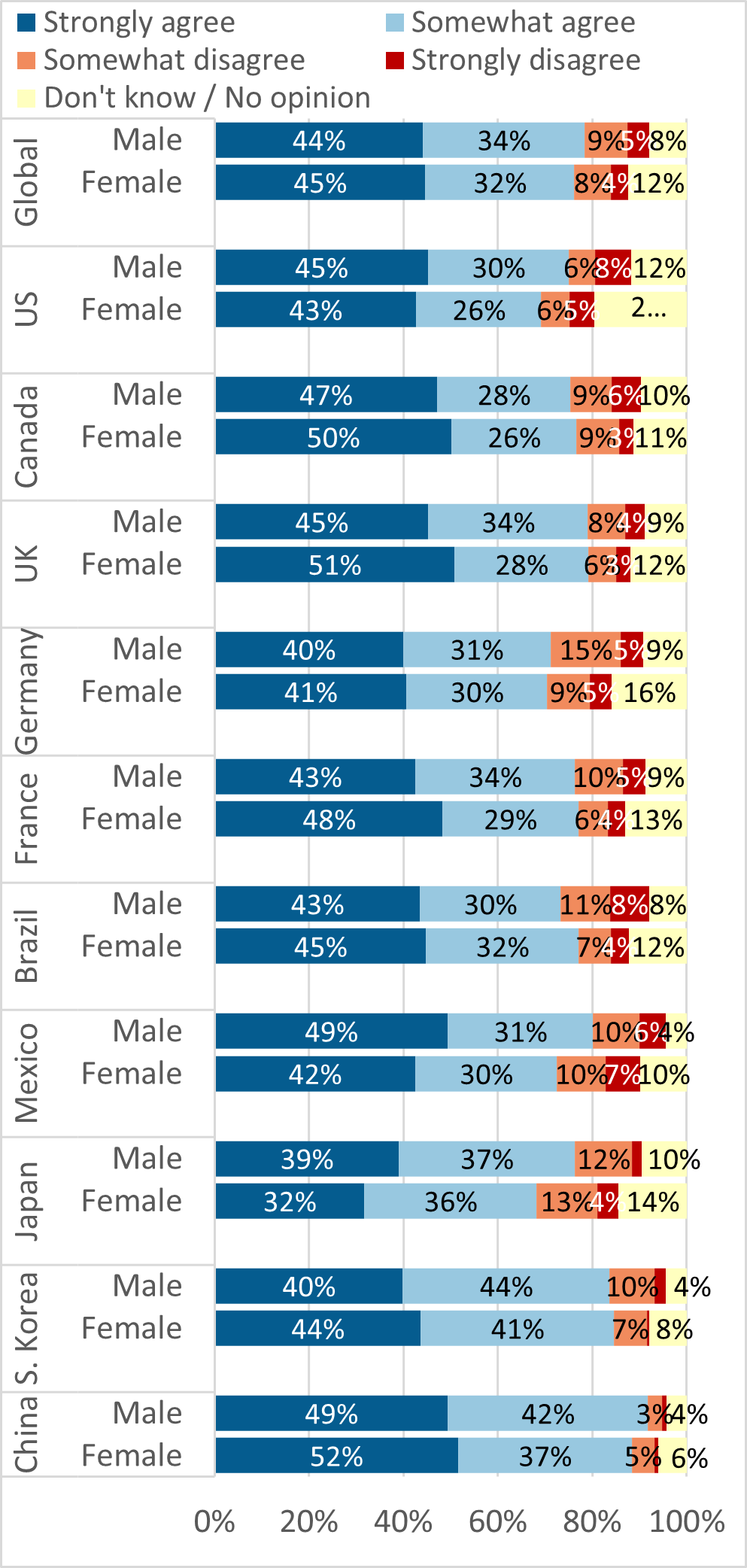}}
\caption*{\textbf{Figure A4A and A4B: Responses for “Thinking about AI, how much do you agree or disagree with the following?--— AI needs to be heavily regulated” }Figure A4A(left): Results from general population. Figure A4B(right): Results grouped by gender. (Question 21\_5 from the 2024 3M State of Science Insights) }
\end{figure}
\begin{figure}[H]
\fbox{\includegraphics[width=0.47\linewidth]{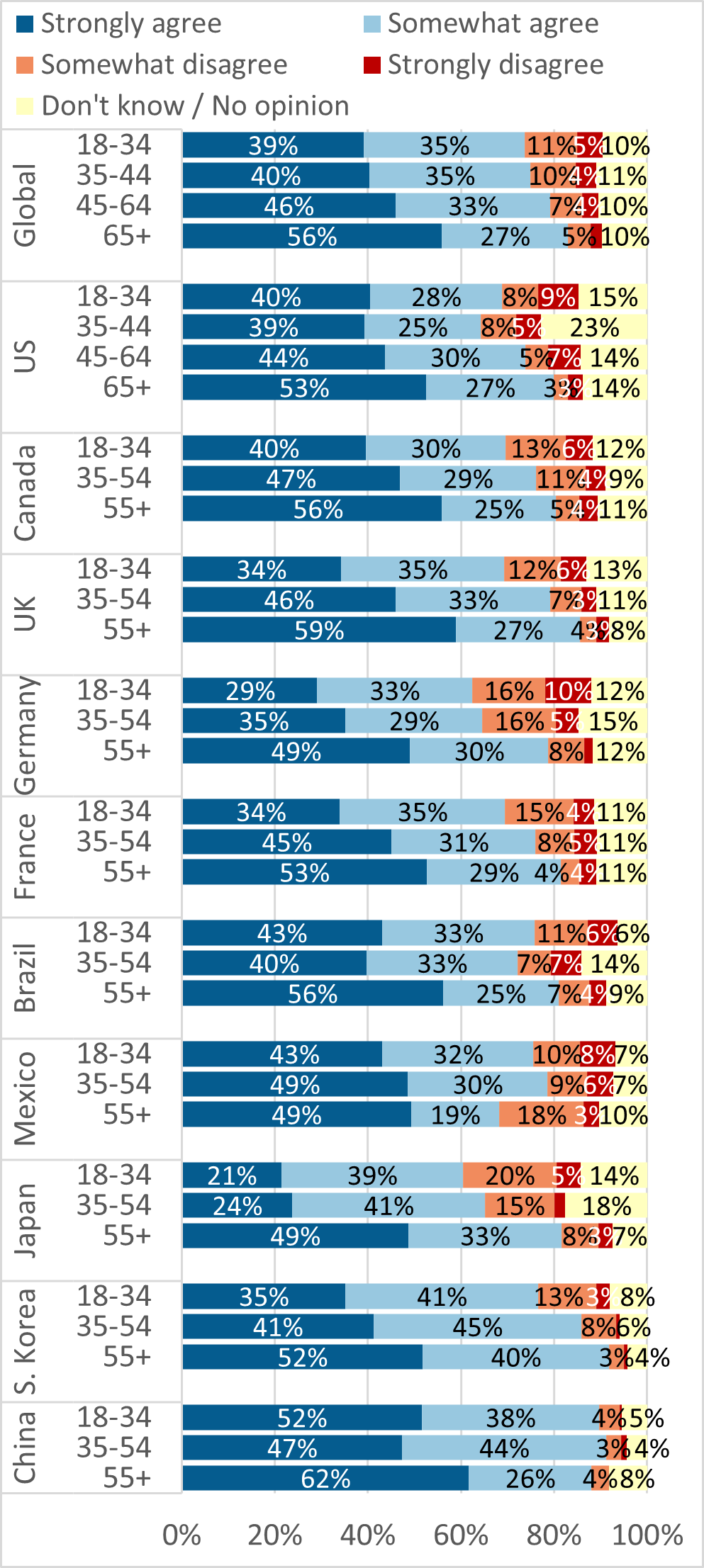}}
\fbox{\includegraphics[width=0.47\linewidth]{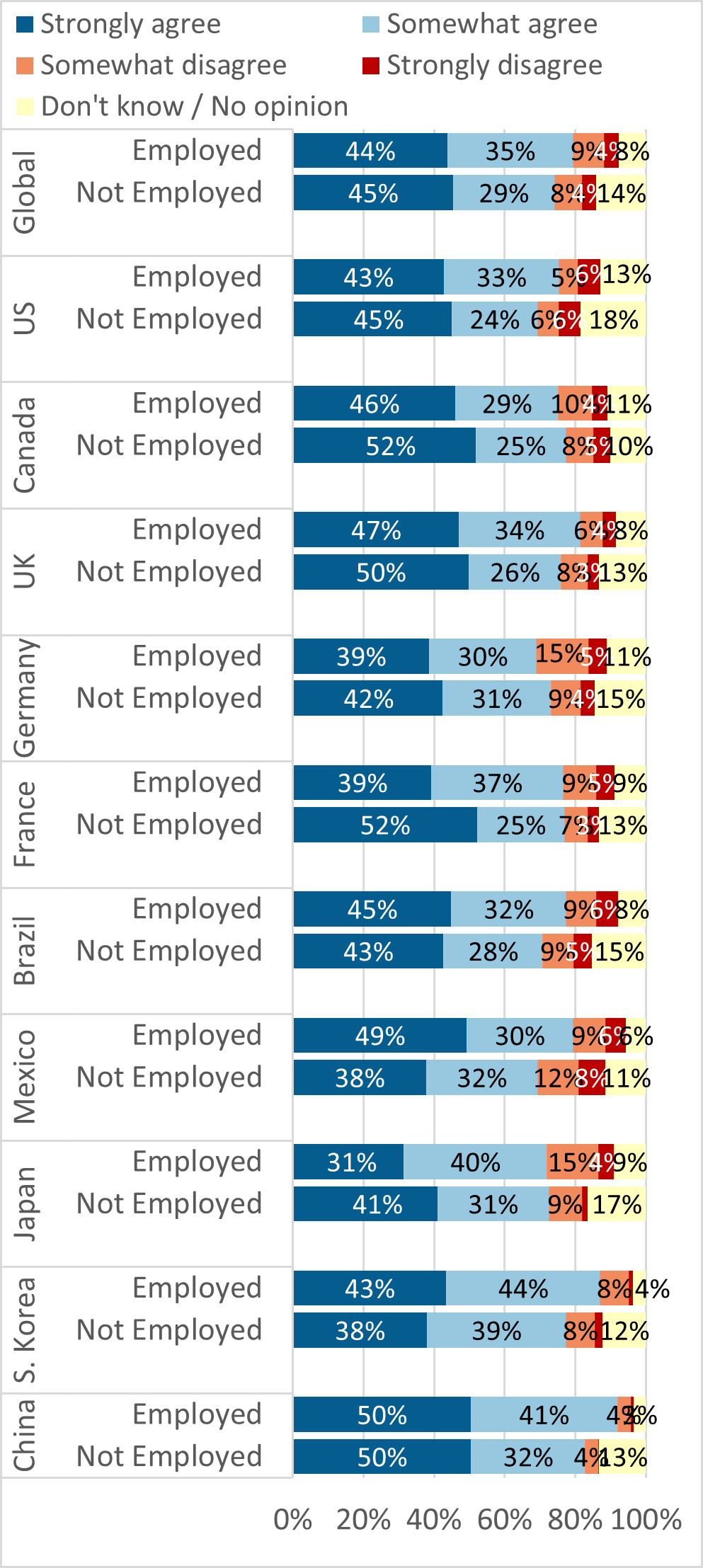}}
\caption*{\textbf{Figure A4C and A4D: Responses for “Thinking about AI, how much do you agree or disagree with the following?--— AI needs to be heavily regulated.” }Figure A4C(left): Results grouped by age. Figure A4D(right): Results grouped by employment. (Question 21\_5 from the 2024 3M State of Science Insights)  }
\end{figure}

\begin{figure}[H]
\fbox{\includegraphics[width=0.47\linewidth]{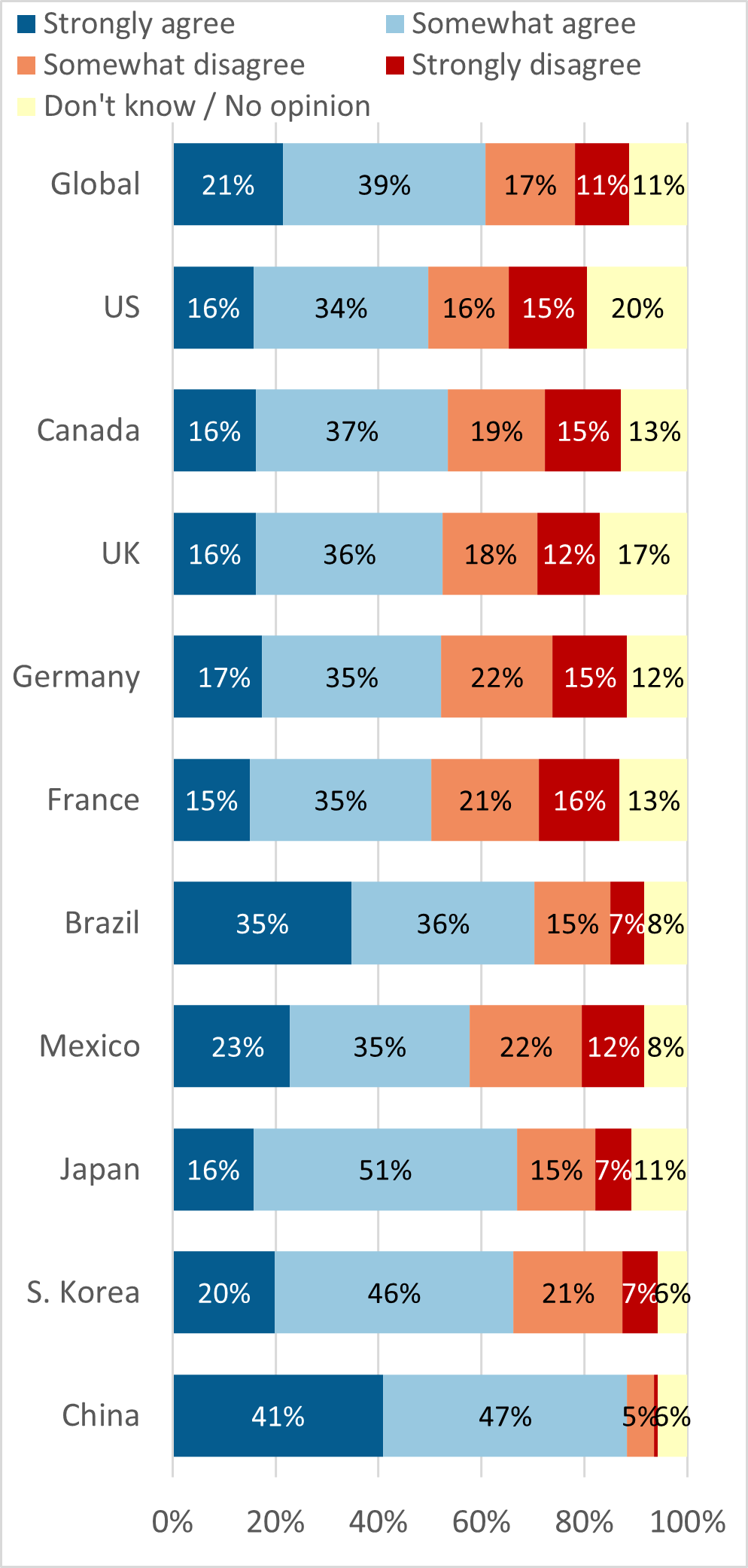}}
\fbox{\includegraphics[width=0.47\linewidth]{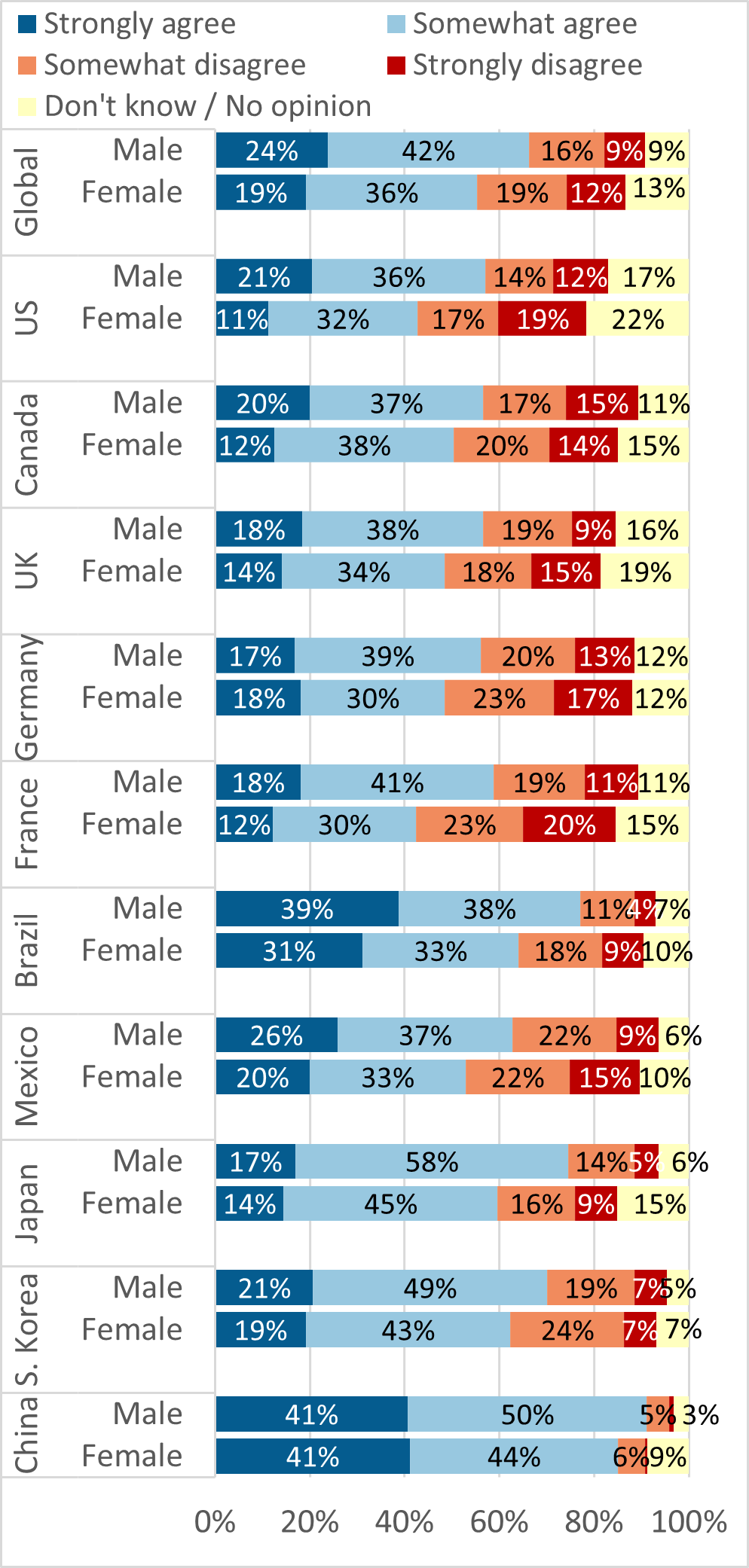}}
\caption*{\textbf{Figure A5A and A5B: Responses for “Thinking about AI, how much do you agree or disagree with the following?--— AI has an increased presence in my daily life” }Figure A5A(left): Results from general population. Figure A5B(right): Results grouped by gender. (Question 21\_6 from the 2024 3M State of Science Insights)  }
\end{figure}
\begin{figure}[H]
\fbox{\includegraphics[width=0.47\linewidth]{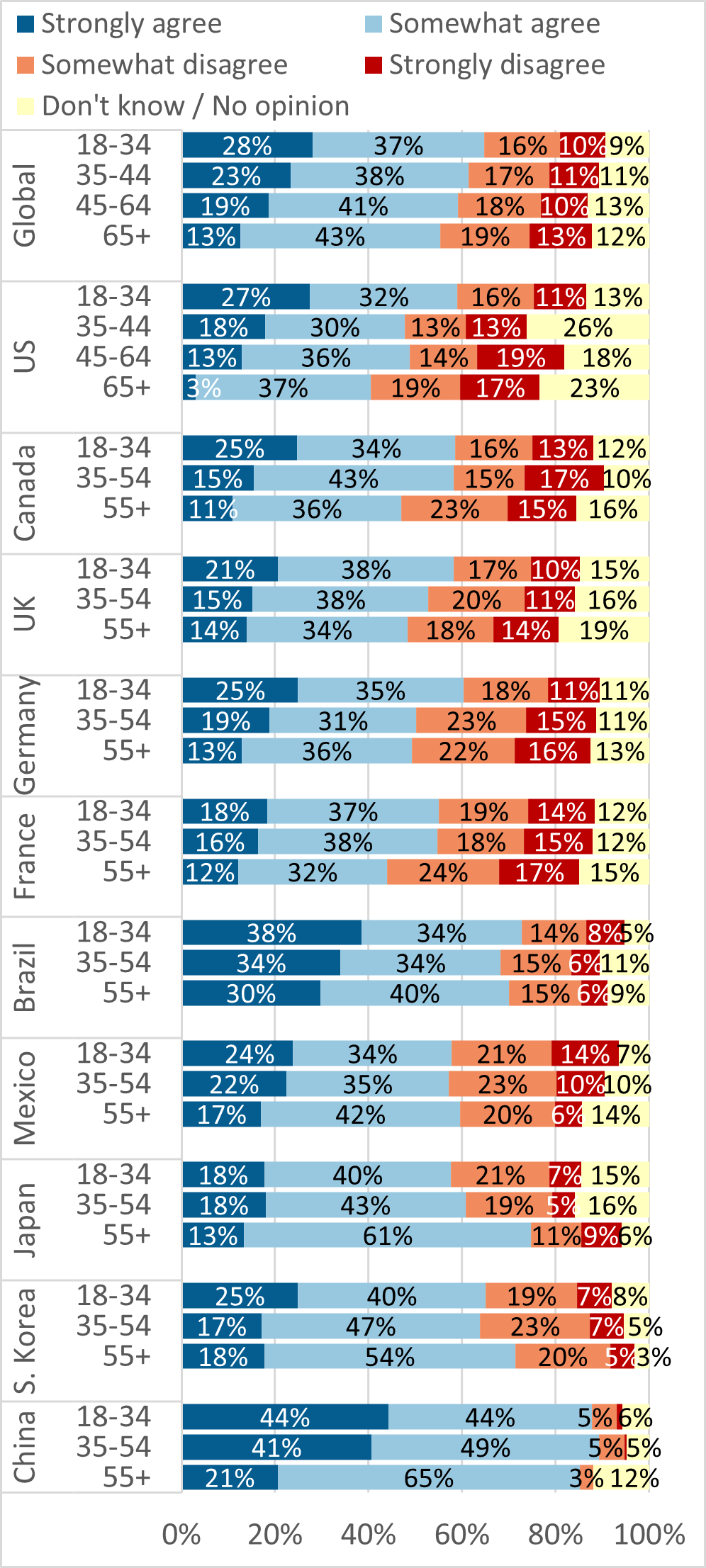}}
\fbox{\includegraphics[width=0.4689\linewidth]{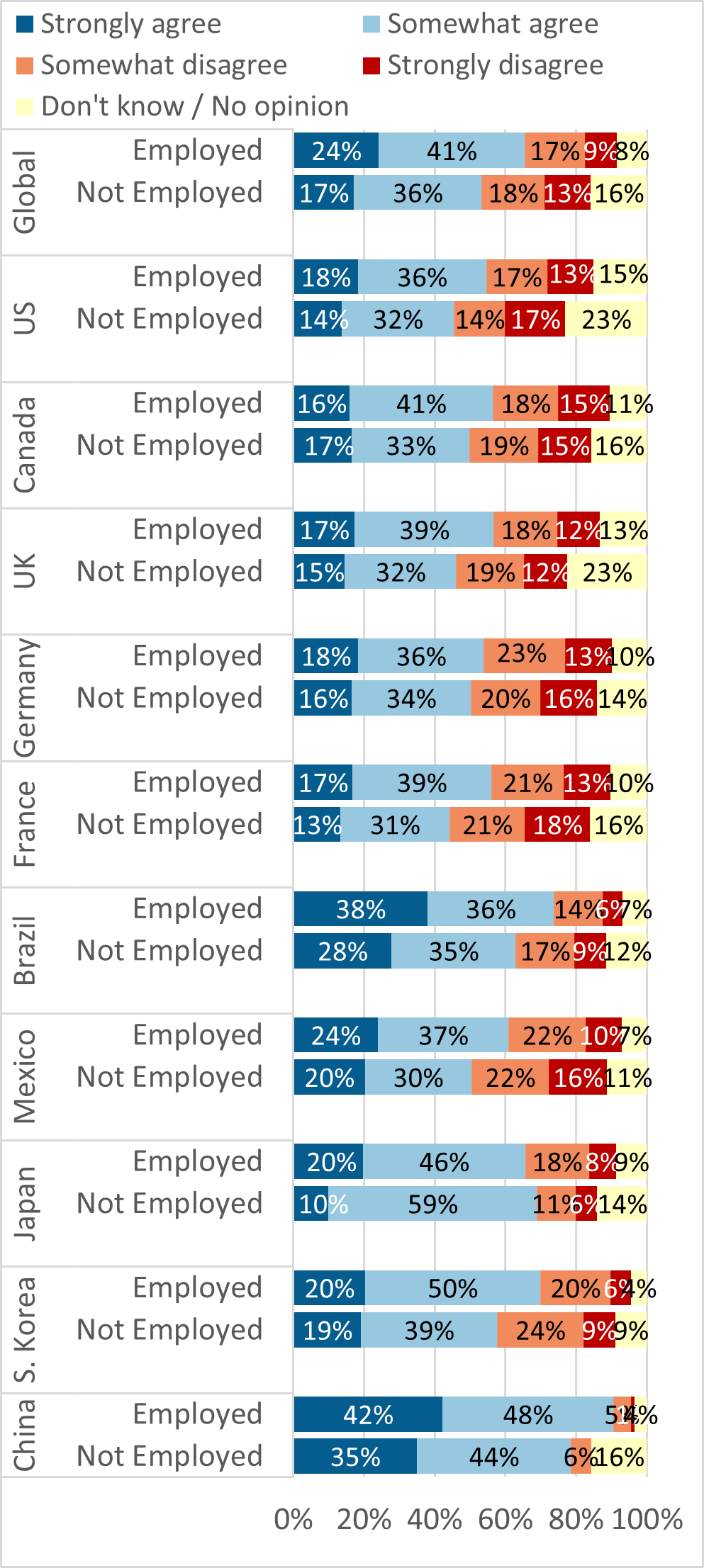}}
\caption*{\textbf{Figure A5C and A5D: Responses for “Thinking about AI, how much do you agree or disagree with the following?--— AI has an increased presence in my daily life.” } Figure A5C(left): Results grouped by age. Figure A5D(right): Results grouped by employment. (Question 21\_6 from the 2024 3M State of Science Insights)  }
\end{figure}

\begin{figure}[H]
\fbox{\includegraphics[width=0.47\linewidth]{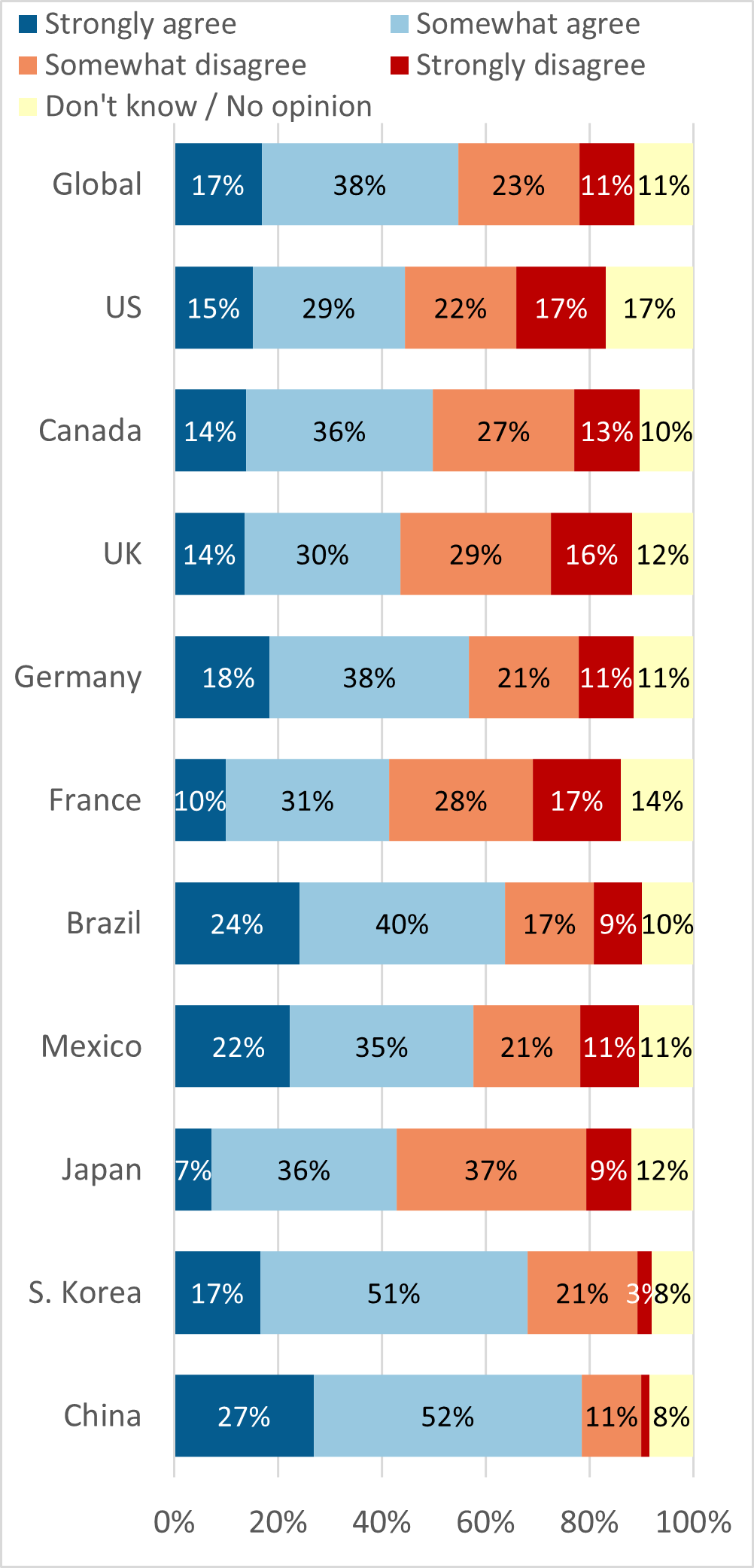}}
\fbox{\includegraphics[width=0.4659\linewidth]{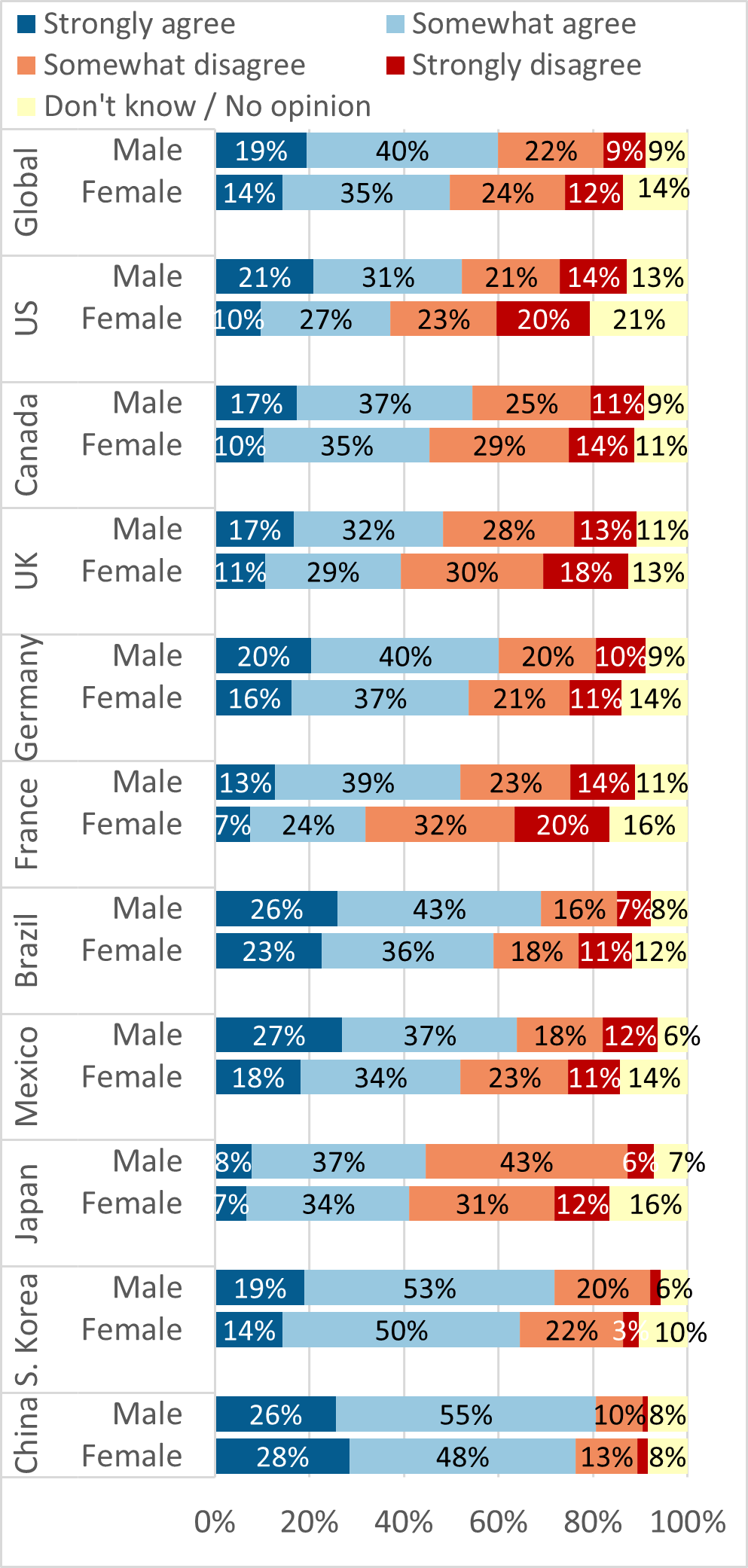}}
\caption*{\textbf{Figure A6A and A6B: Responses for “Thinking about AI, how much do you agree or disagree with the following?--— I have a strong understanding of where AI is being used in everyday life.” }Figure A6A(left): Results from general population. Figure A6B(right): Results grouped by gender. (Question 21\_7 from the 2024 3M State of Science Insights) }
\end{figure}
\begin{figure}[H]
\fbox{\includegraphics[width=0.47\linewidth]{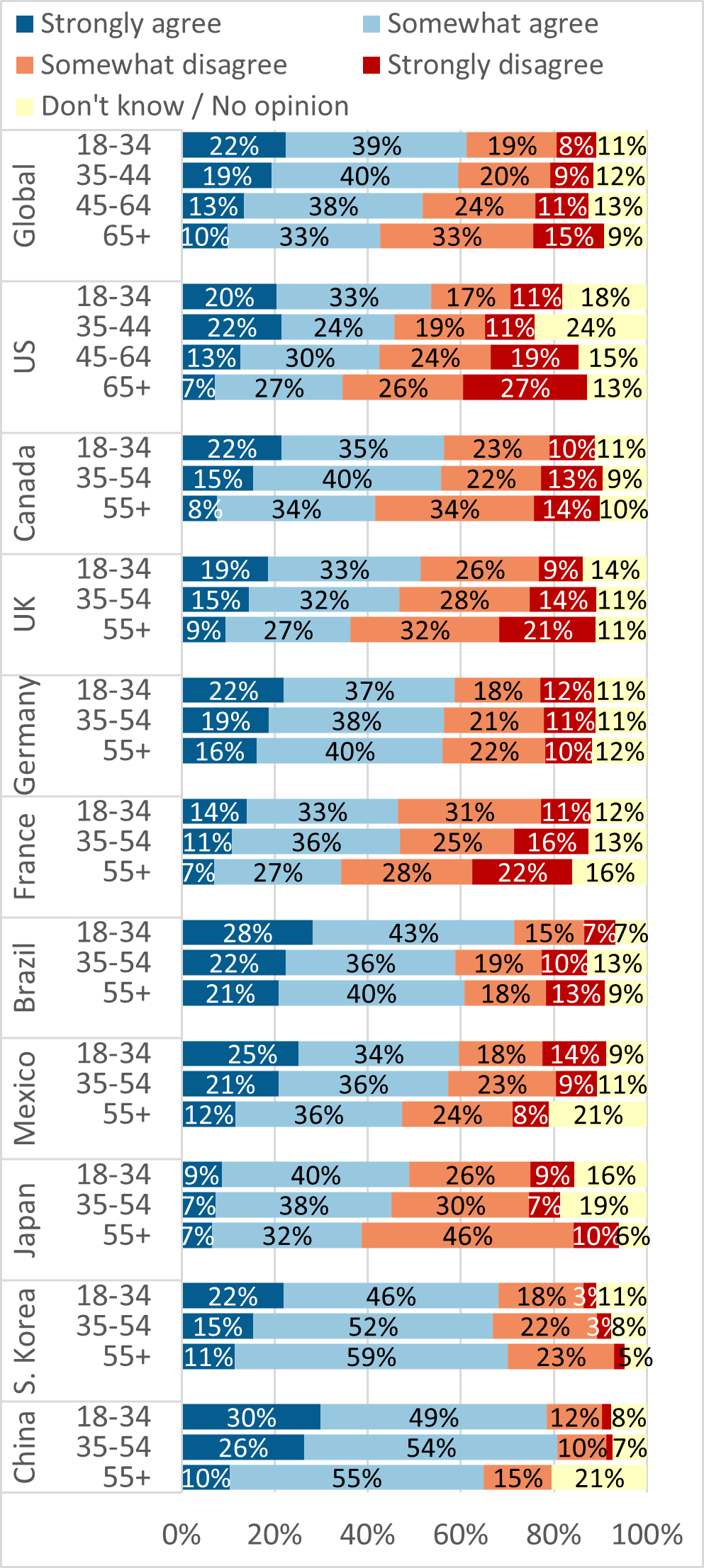}}
\fbox{\includegraphics[width=0.47\linewidth]{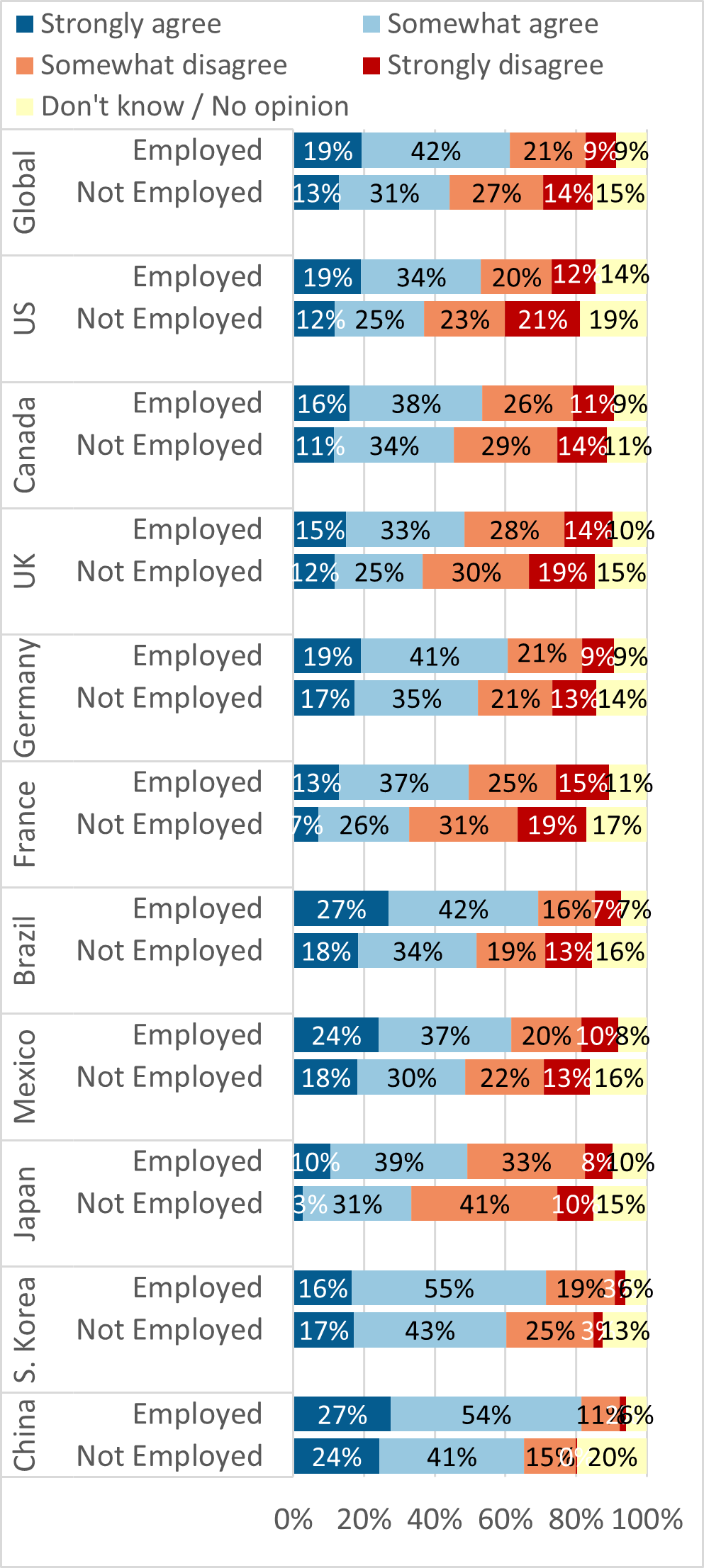}}
\caption*{\textbf{Figure A6C and A6D: Responses for “Thinking about AI, how much do you agree or disagree with the following?--— I have a strong understanding of where AI is being used in everyday life.” }Figure A6C(left): Results grouped by age. Figure A6D(right): Results grouped by employment. (Question 21\_7 from the 2024 3M State of Science Insights) }
\end{figure}

\begin{figure}[H]
\fbox{\includegraphics[width=0.47\linewidth]{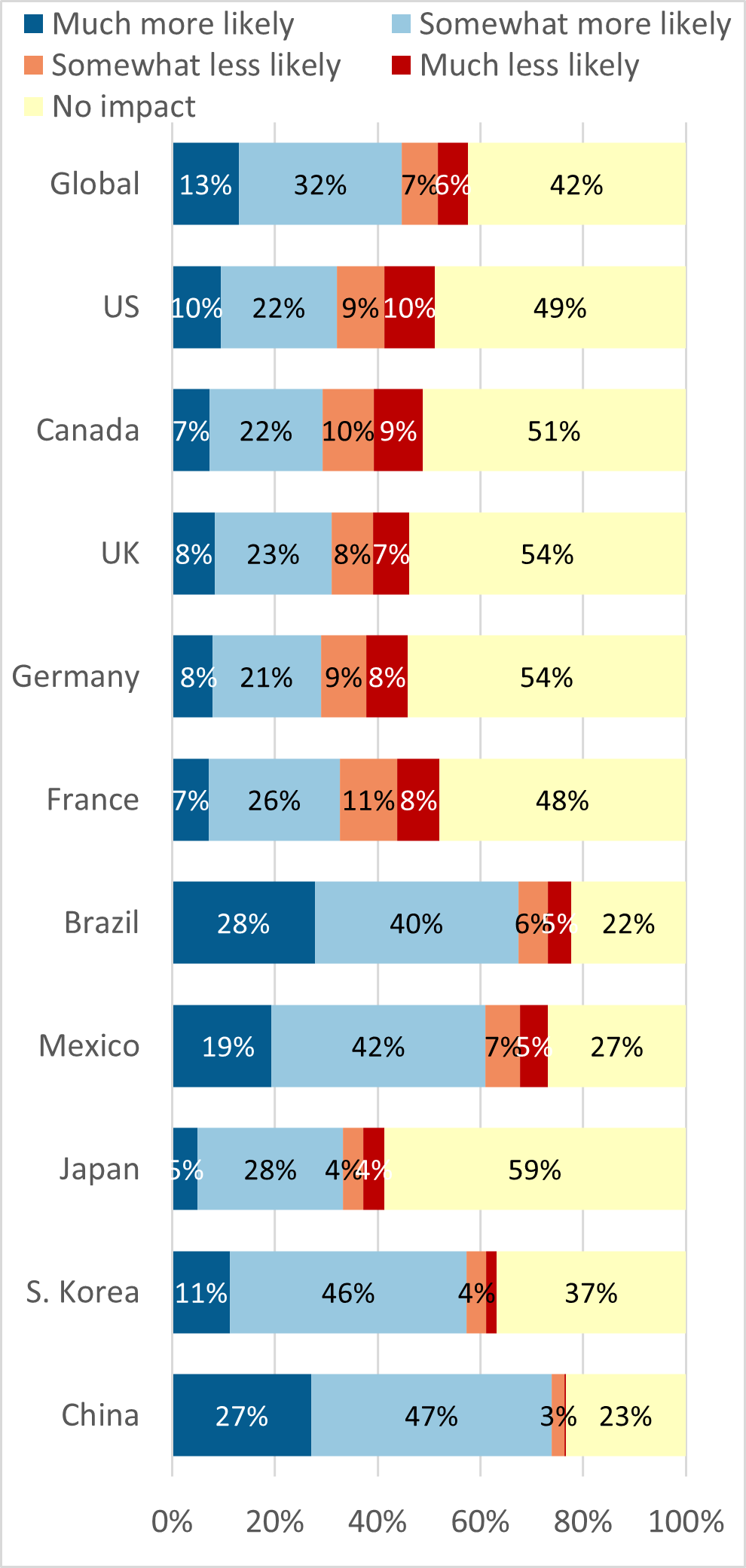}}
\fbox{\includegraphics[width=0.47\linewidth]{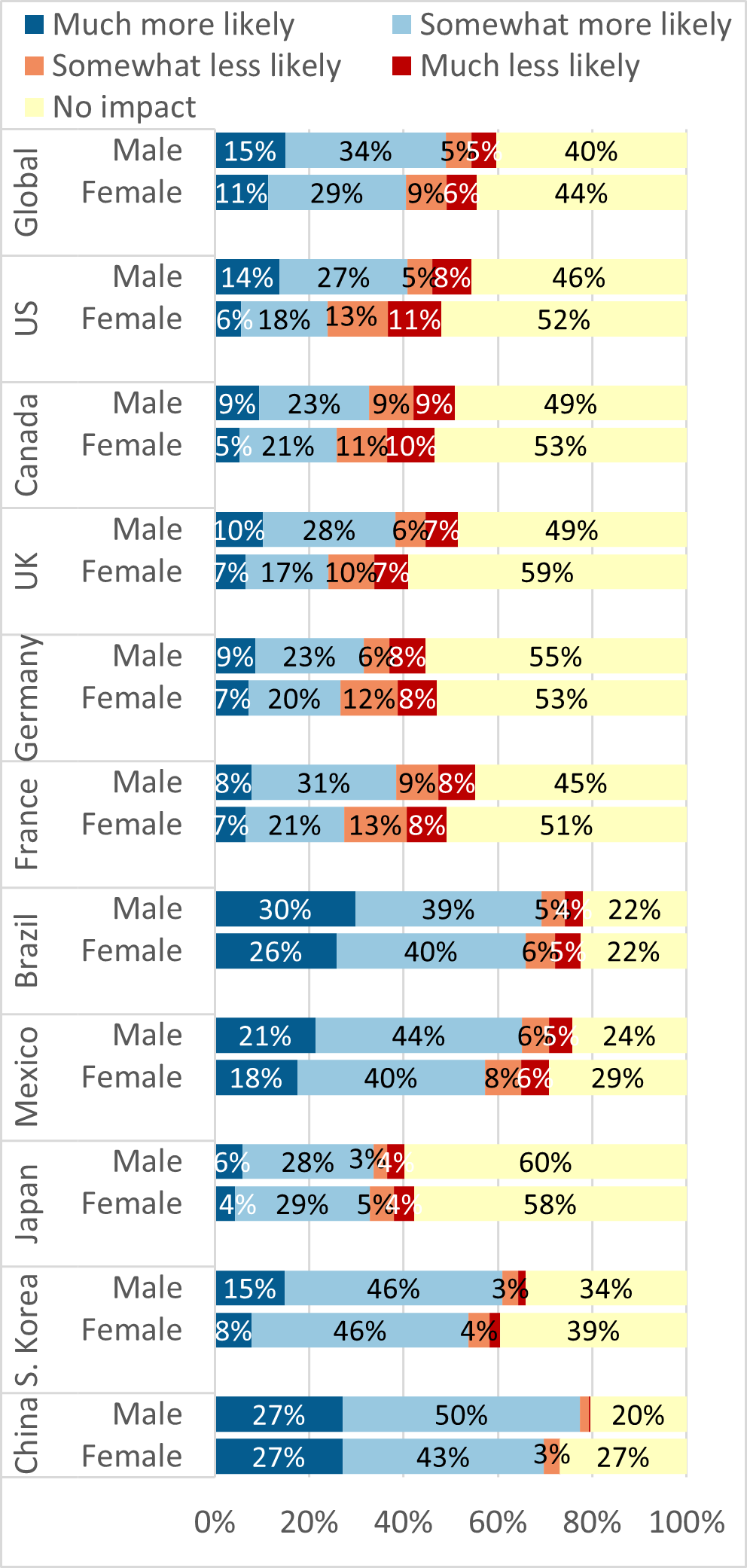}}
\caption*{\textbf{Figure A7A and A7B: Responses for “If a company were to use AI to innovate the products and services you use on a daily basis to improve your life, would you be more or less likely to --- Think highly of the company” }Figure A7A(left): Results from general population. Figure A7B(right): Results grouped by gender. (Question 22\_1 from the 2024 3M State of Science Insights) }
\end{figure}
\begin{figure}[H]
\fbox{\includegraphics[width=0.47\linewidth]{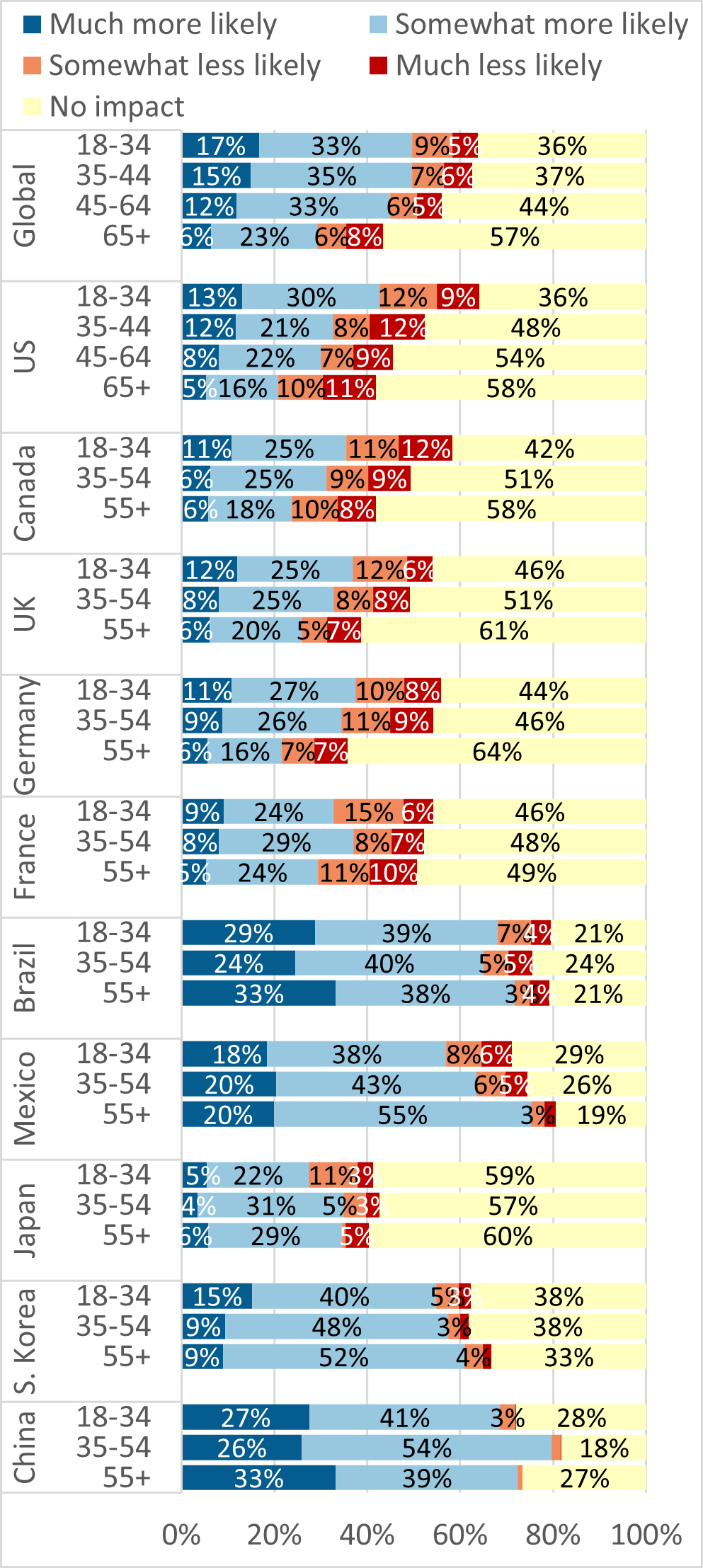}}
\fbox{\includegraphics[width=0.47\linewidth]{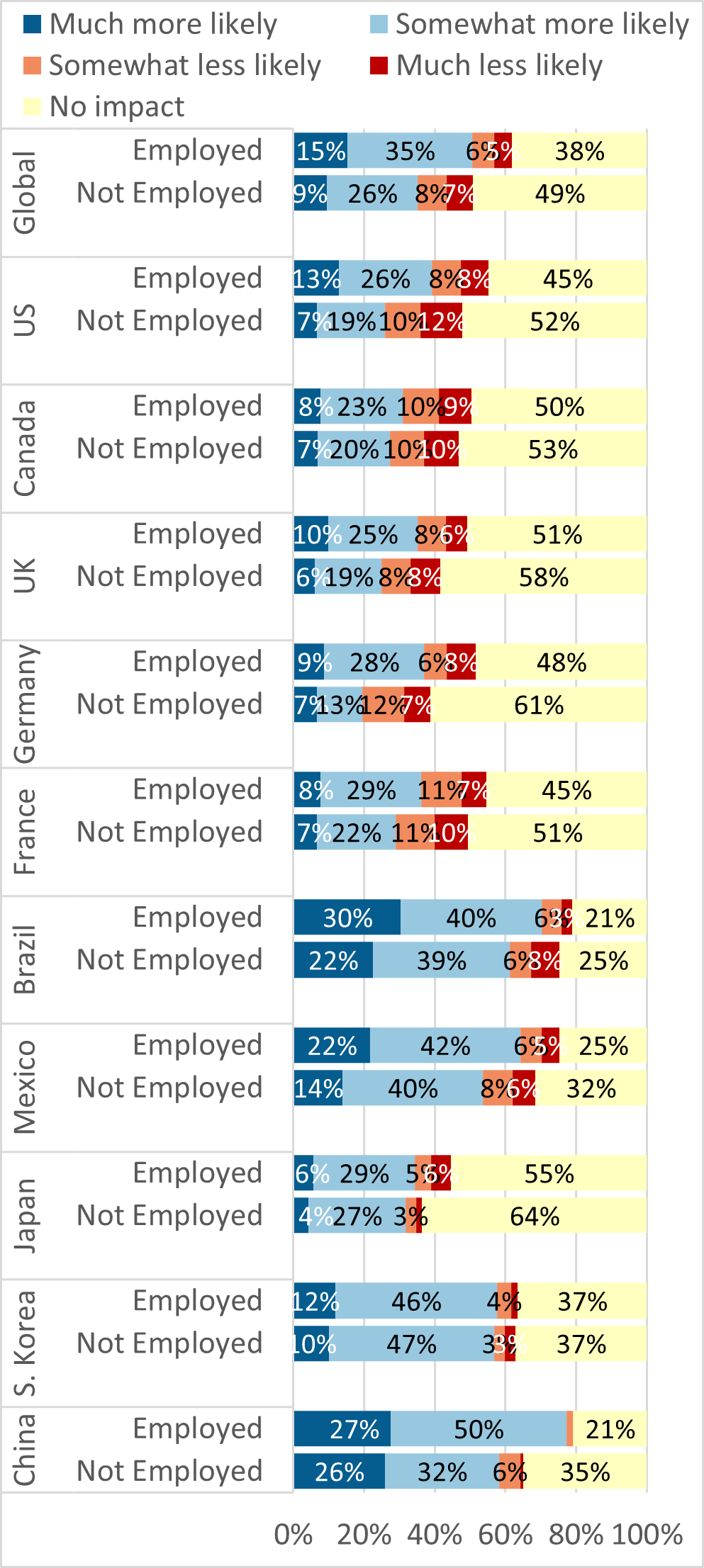}}
\caption*{\textbf{Figure A7C and A7D: Responses for “If a company were to use AI to innovate the products and services you use on a daily basis to improve your life, would you be more or less likely to --- Think highly of the company.” }Figure A7C(left): Results grouped by age. Figure A7D(right): Results grouped by employment. (Question 22\_1 from the 2024 3M State of Science Insights) }
\end{figure}

\begin{figure}[H]
\fbox{\includegraphics[width=0.47\linewidth]{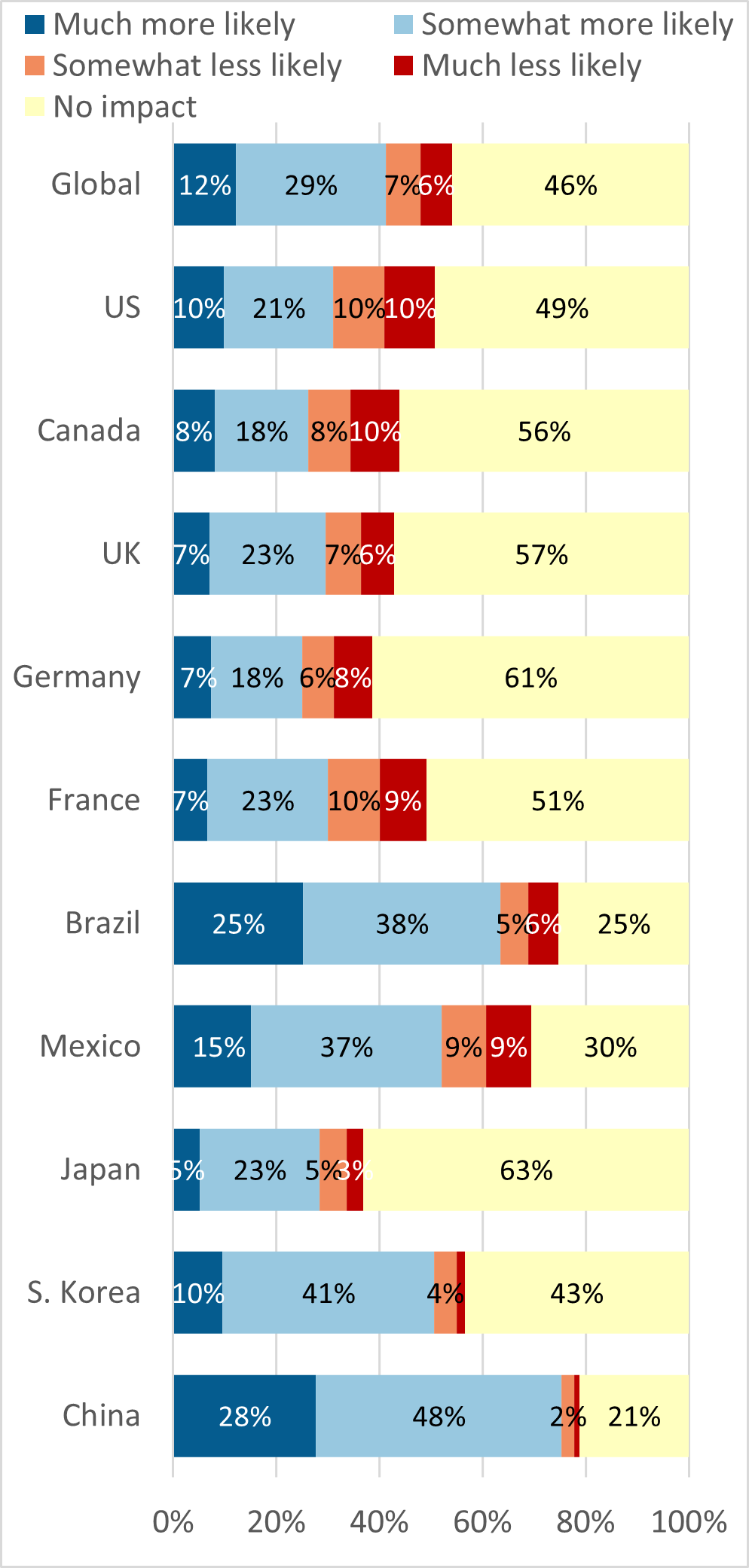}}
\fbox{\includegraphics[width=0.47\linewidth]{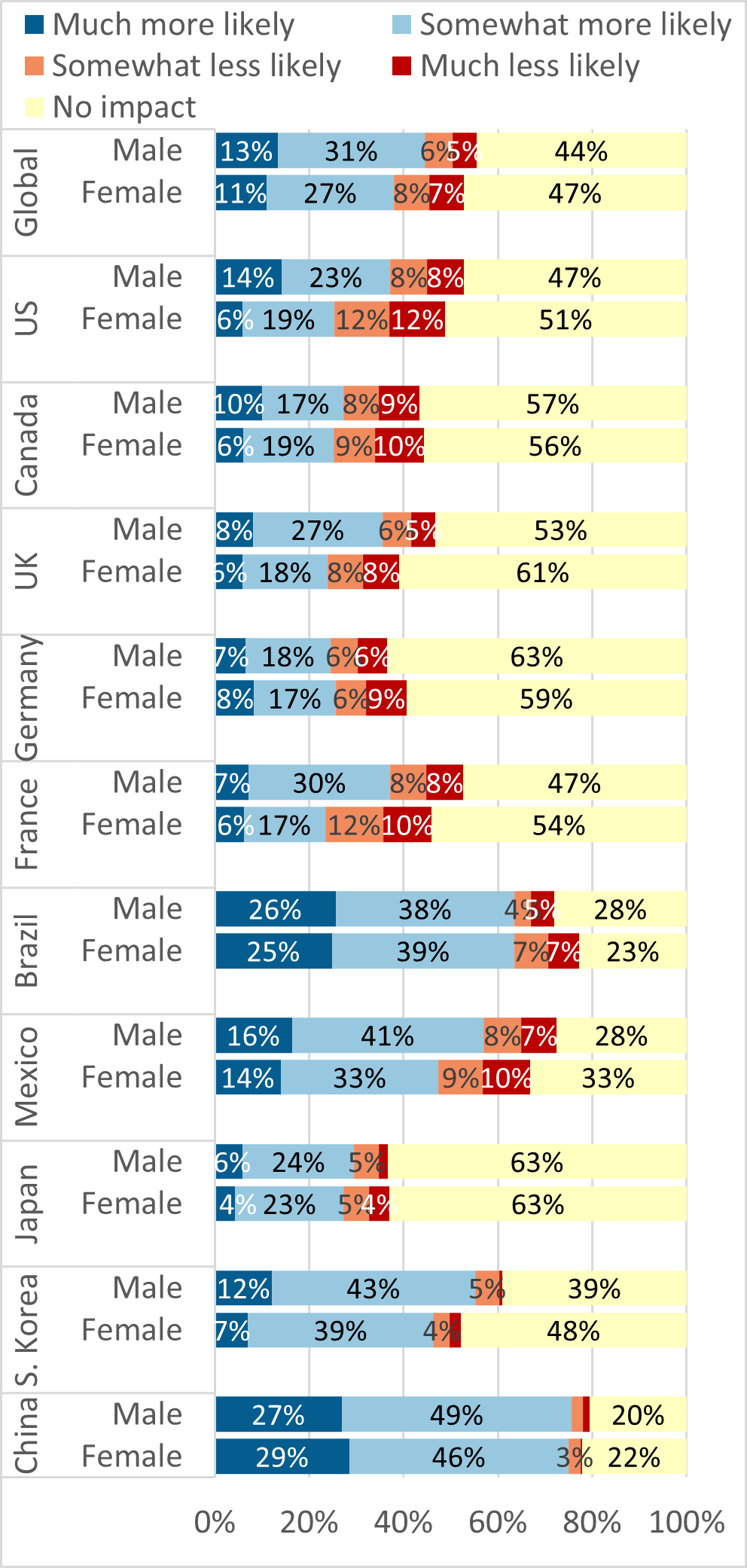}}
\caption*{\textbf{Figure A8A and A8B: Responses for “If a company were to use AI to innovate the products and services you use on a daily basis to improve your life, would you be more or less likely to --- Purchase from the company” }Figure A8A(left): Results from general population. Figure A8B(right): Results grouped by gender. (Question 22\_2 from the 2024 3M State of Science Insights) }
\end{figure}
\begin{figure}[H]
\fbox{\includegraphics[width=0.47\linewidth]{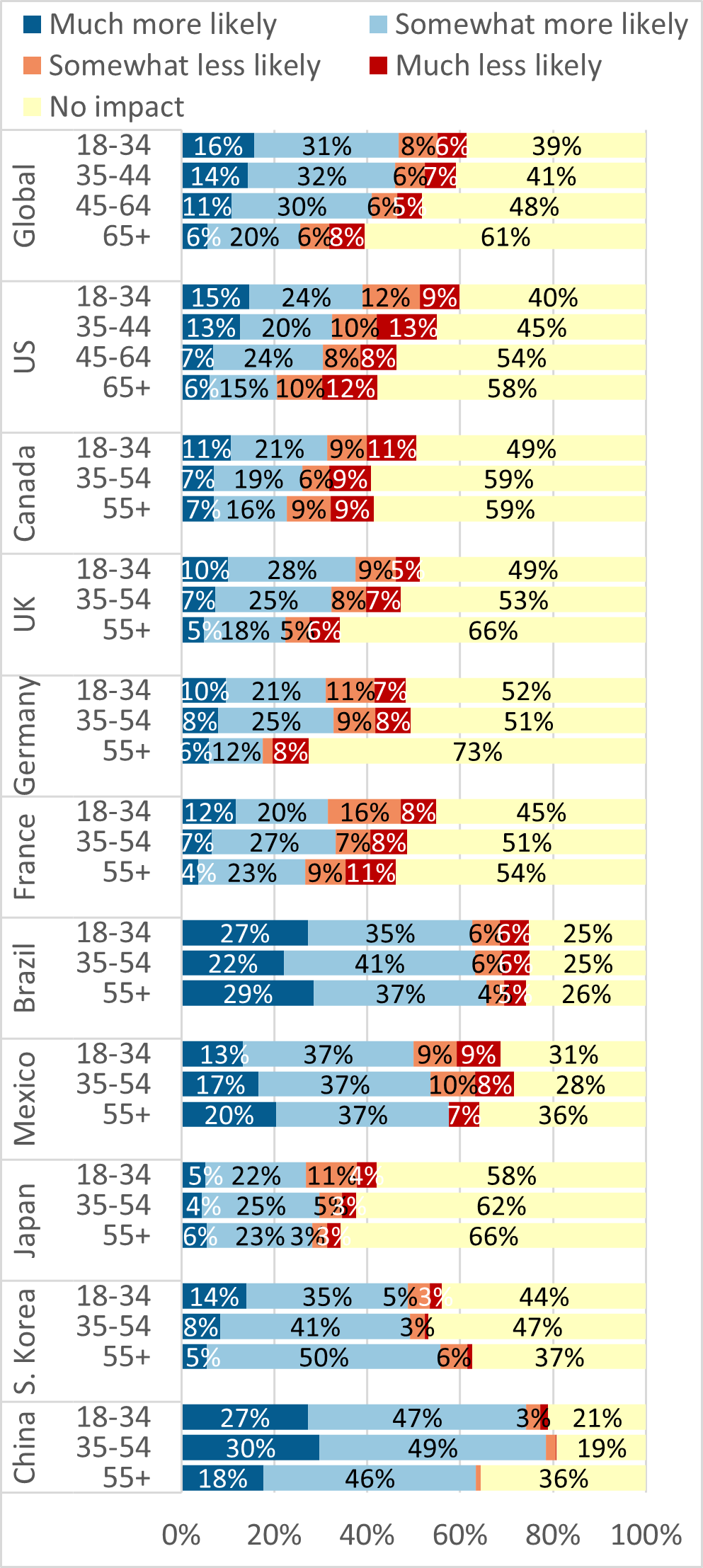}}
\fbox{\includegraphics[width=0.47\linewidth]{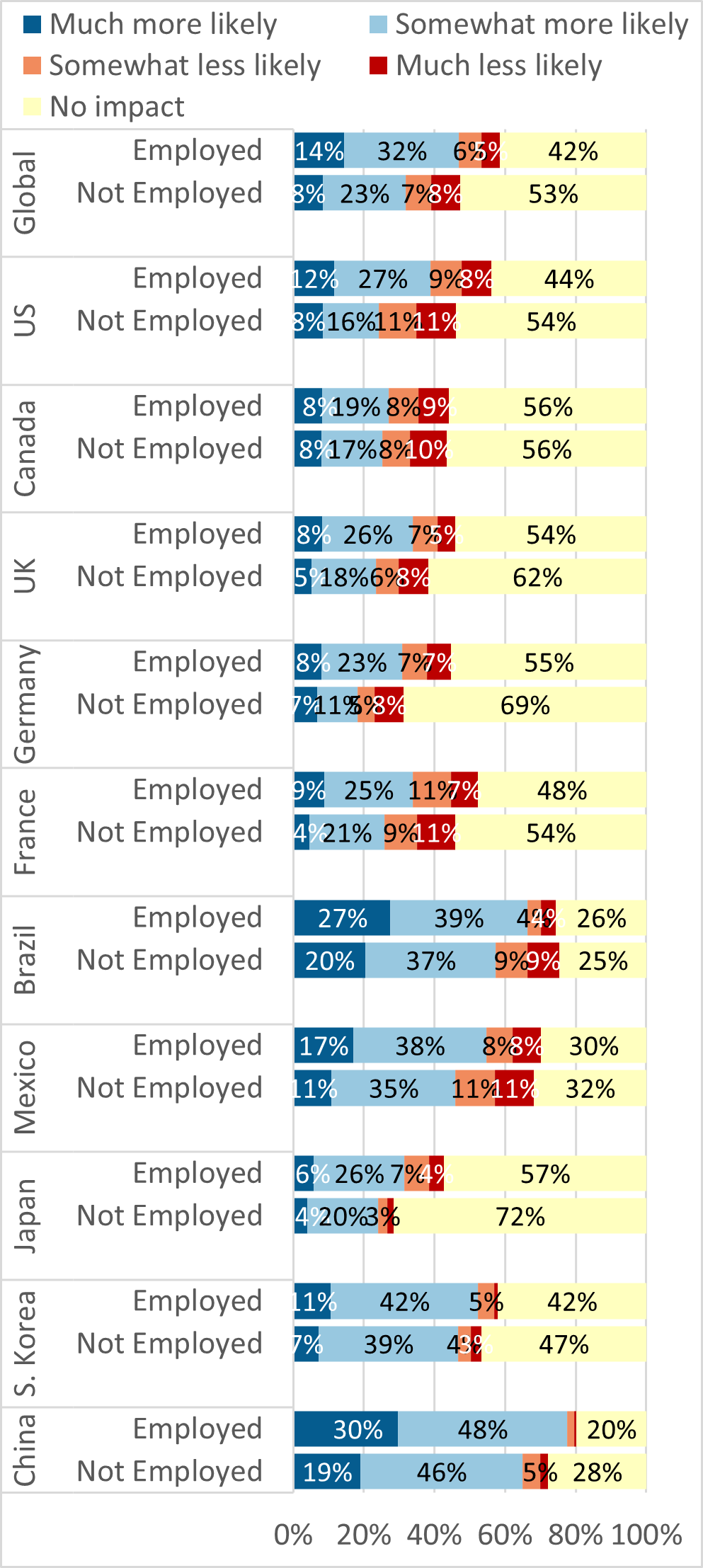}}
\caption*{\textbf{Figure A8C and A8D: Responses for “If a company were to use AI to innovate the products and services you use on a daily basis to improve your life, would you be more or less likely to --- Purchase from the company.” }Figure A8C(left): Results grouped by age. Figure A8D(right): Results grouped by employment. (Question 22\_2 from the 2024 3M State of Science Insights) }
\end{figure}

\begin{figure}[H]
\fbox{\includegraphics[width=0.47\linewidth]{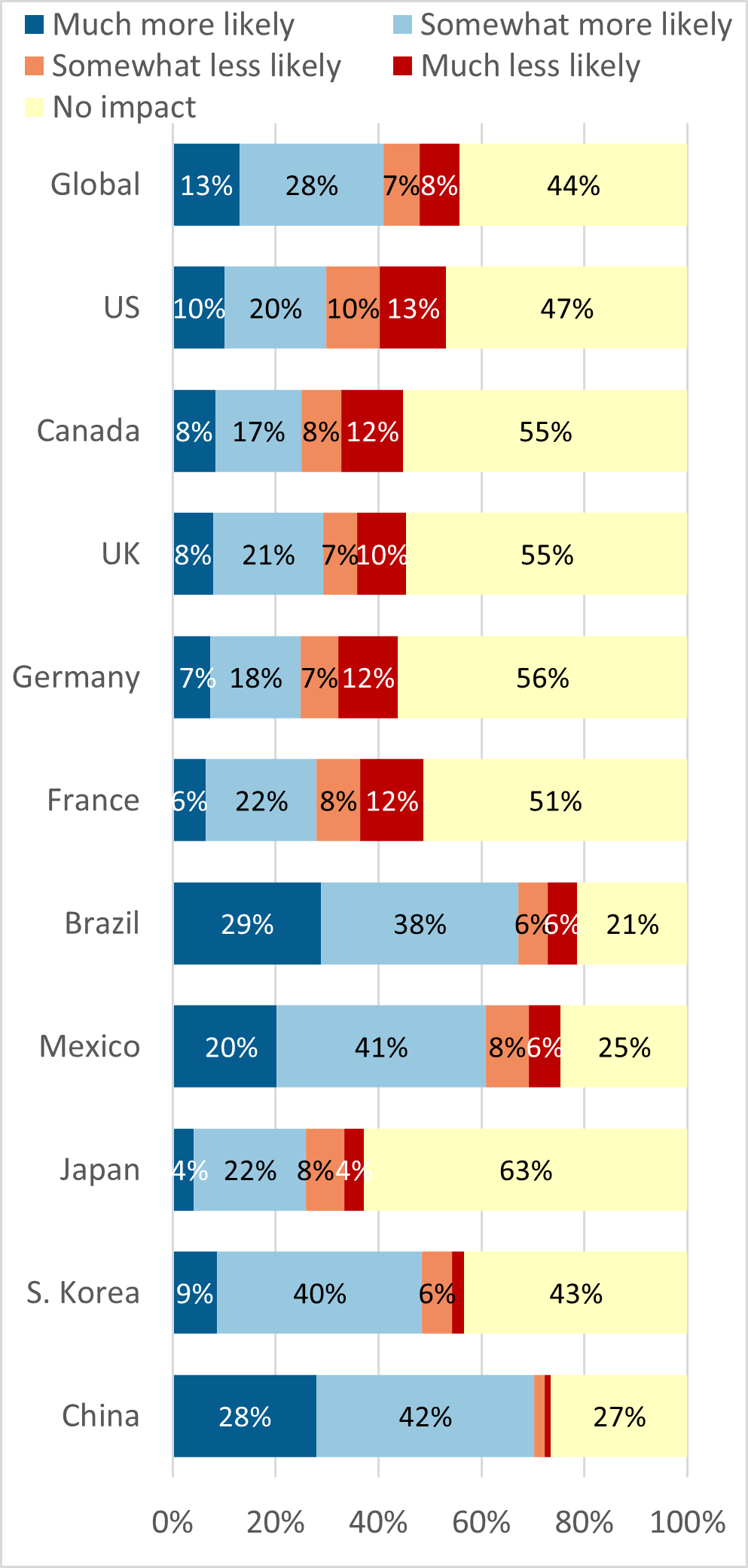}}
\fbox{\includegraphics[width=0.469\linewidth]{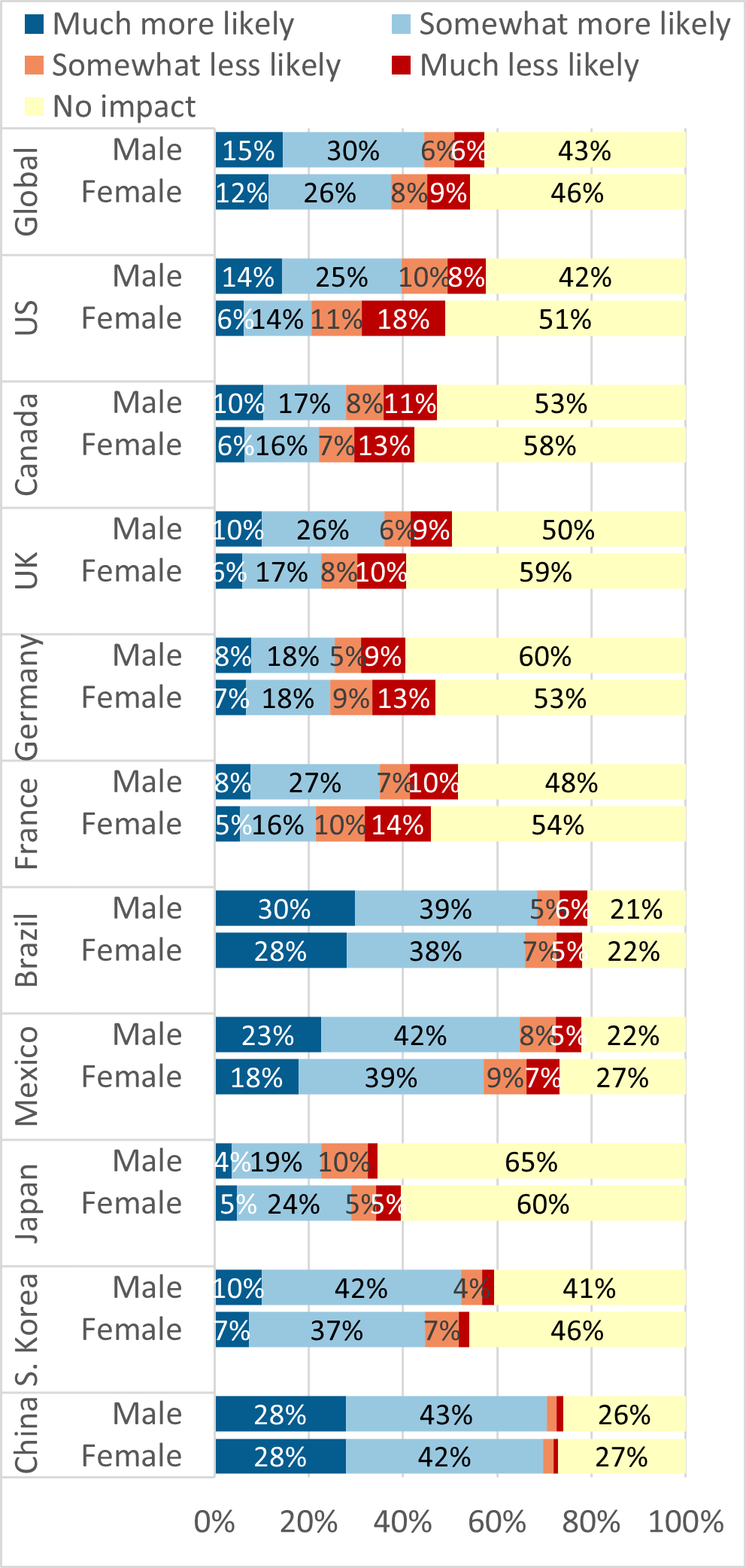}}
\caption*{\textbf{Figure A9A and A9B: Responses for “If a company were to use AI to innovate the products and services you use on a daily basis to improve your life, would you be more or less likely to --- Be interested in pursuing a job with the company” }Figure A9A(left): Results from general population. Figure A9B(right): Results grouped by gender. (Question 22\_4 from the 2024 3M State of Science Insights).  }
\end{figure}
\begin{figure}[H]
\fbox{\includegraphics[width=0.47\linewidth]{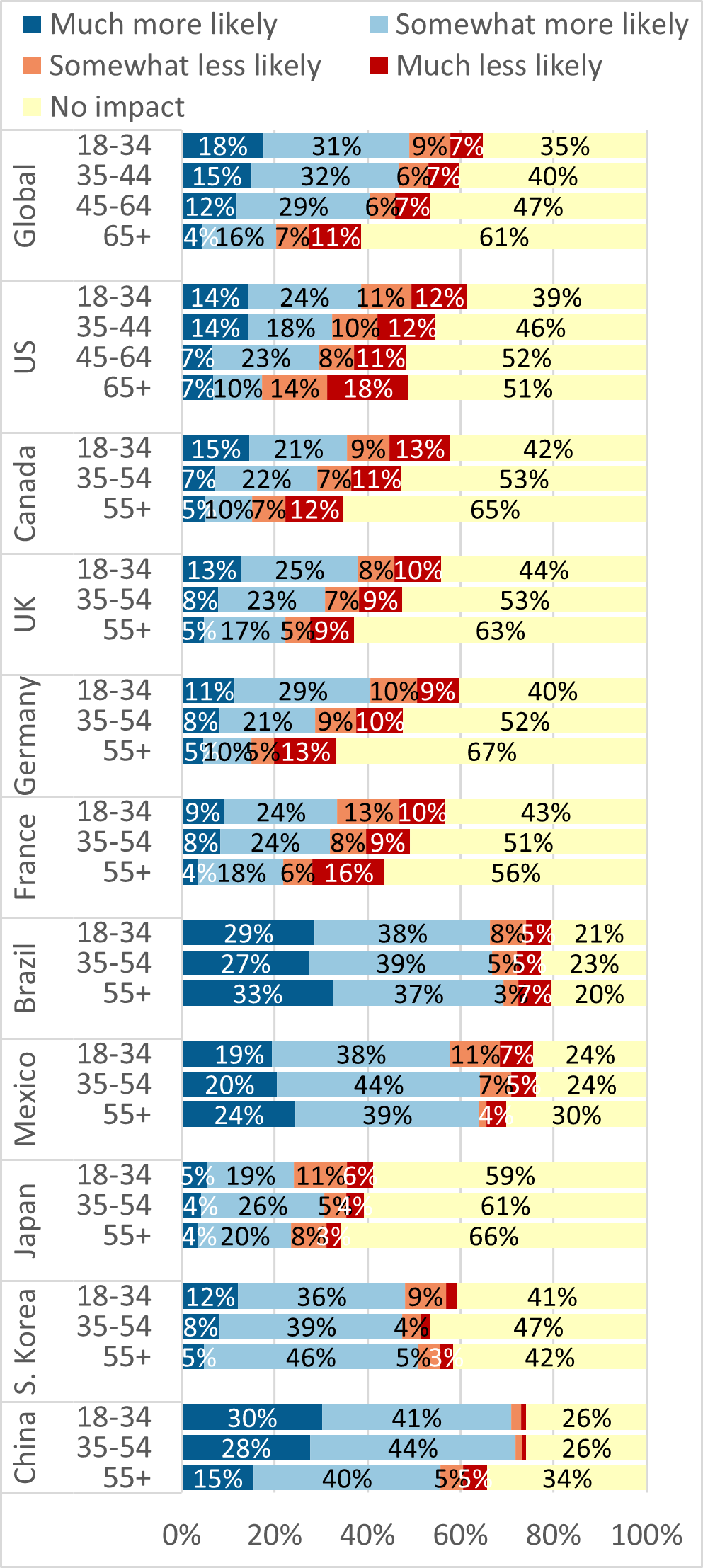}}
\fbox{\includegraphics[width=0.47\linewidth]{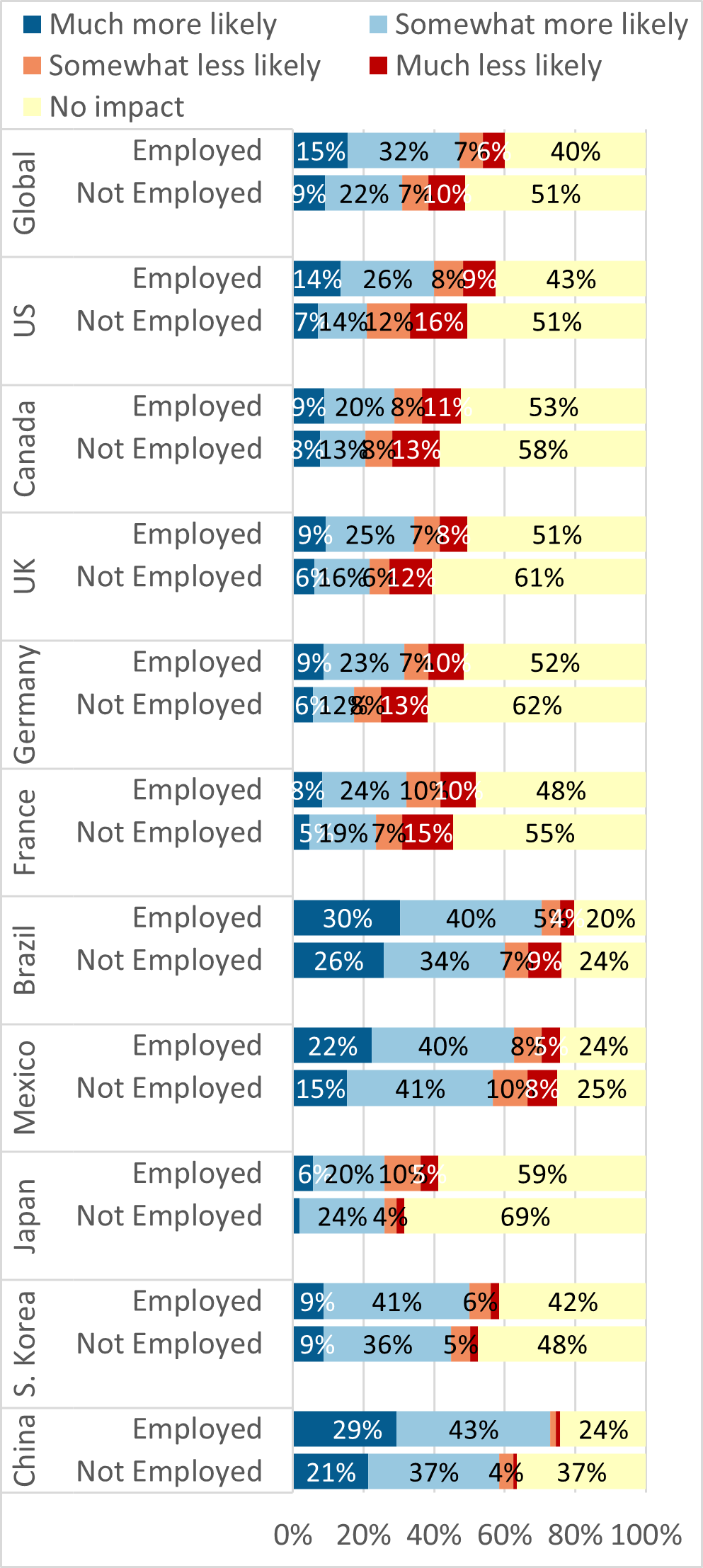}}
\caption*{\textbf{Figure A9C and A9D: Responses for “If a company were to use AI to innovate the products and services you use on a daily basis to improve your life, would you be more or less likely to --- Be interested in pursuing a job with the company” } Figure A9C(left): Results grouped by age. Figure A9D(right): Results grouped by employment. (Question 22\_4 from the 2024 3M State of Science Insights)}
\end{figure}

\clearpage
\begin{center}
\section*{Appendix A4}
AI/automation/robotics related questions from 3M State of Science Index (2018-2023)
\end{center}

\subsubsection*{2023 SOSI \cite{6}}

\begin{figure}[H]
\fbox{\includegraphics[width=0.95\linewidth]{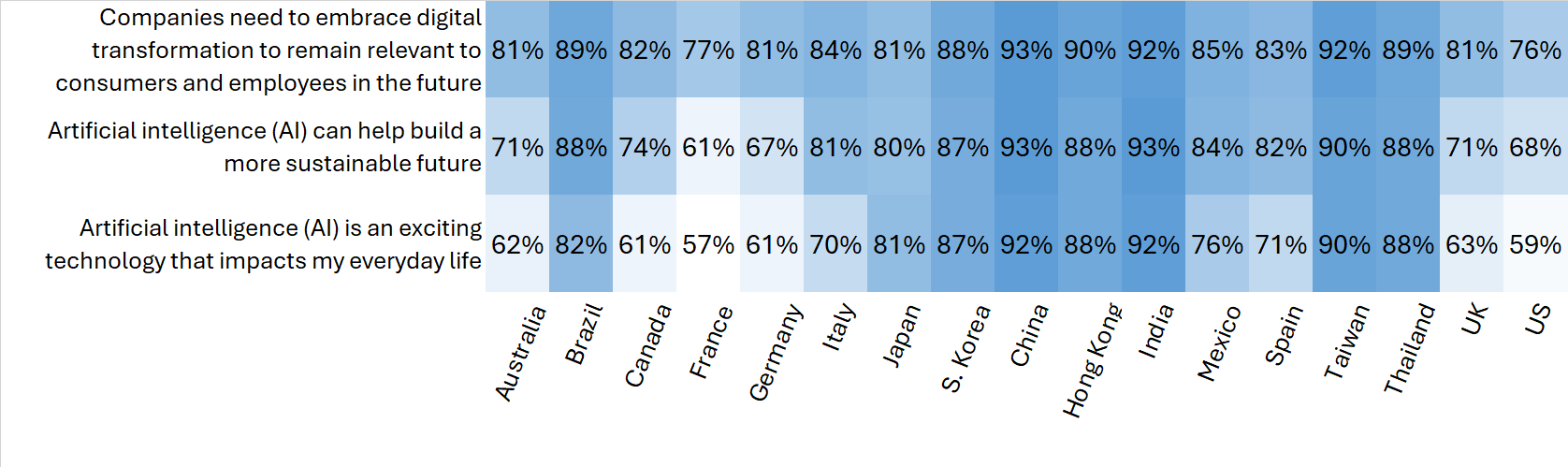}}
\caption*{\textbf{Figure A10: Responses for “How much do you agree or disagree with the following statements?”}Percent of respondents who “agree.” (Question 46 from the 2023 3M State of Science Index) }
\end{figure}

\subsubsection*{2022 SOSI \cite{5}}

\begin{figure}[H]
\fbox{\includegraphics[width=0.95\linewidth]{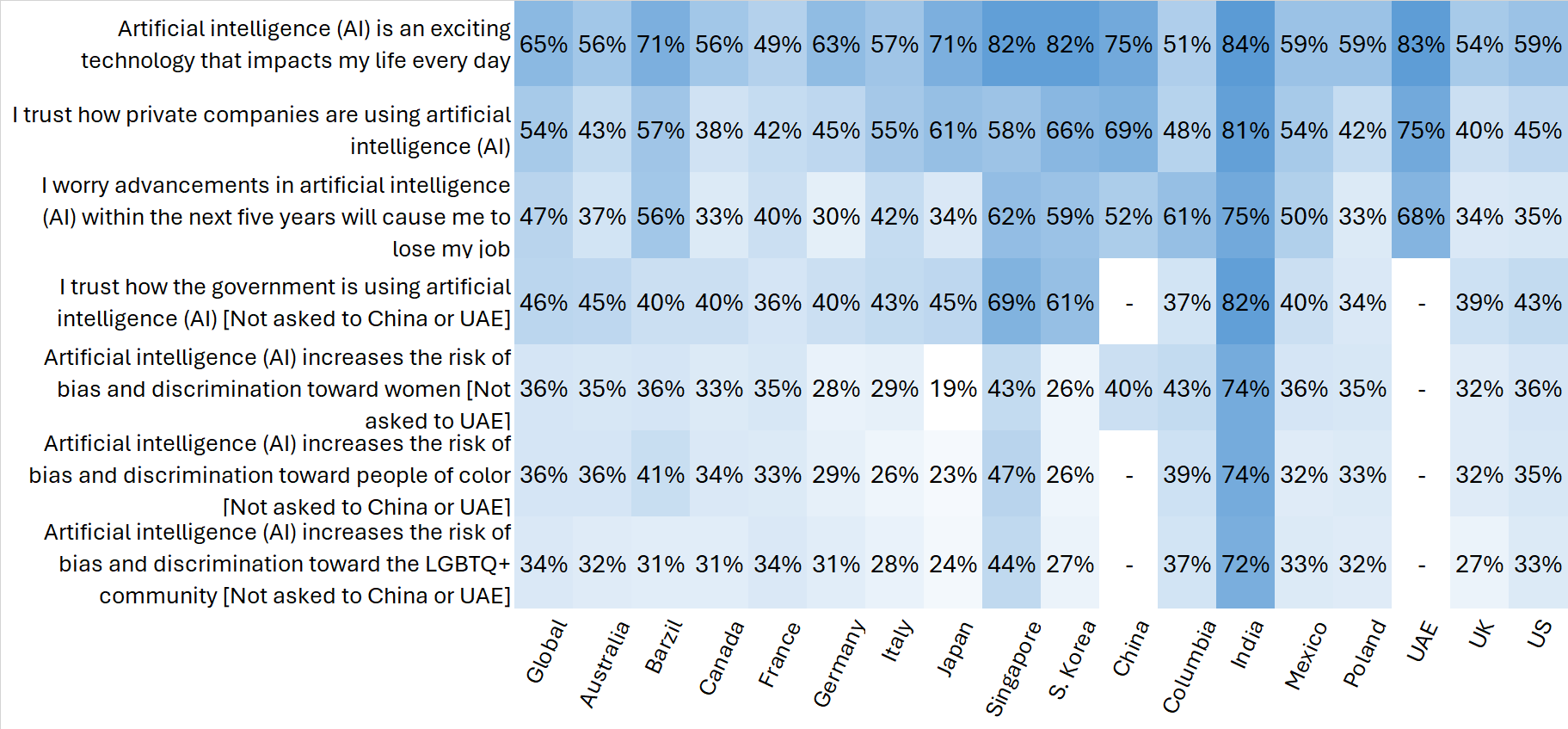}}
\caption*{\textbf{Figure A11: Responses for “How much do you agree or disagree with each of the following statements related to artificial intelligence (AI)? “ }Percent of respondents who “agree.” (Question 15 from the 2022 3M State of Science Index) }
\end{figure}

\subsubsection*{2021 SOSI \cite{4}}

\begin{figure}[H]
\fbox{\includegraphics[width=0.95\linewidth]{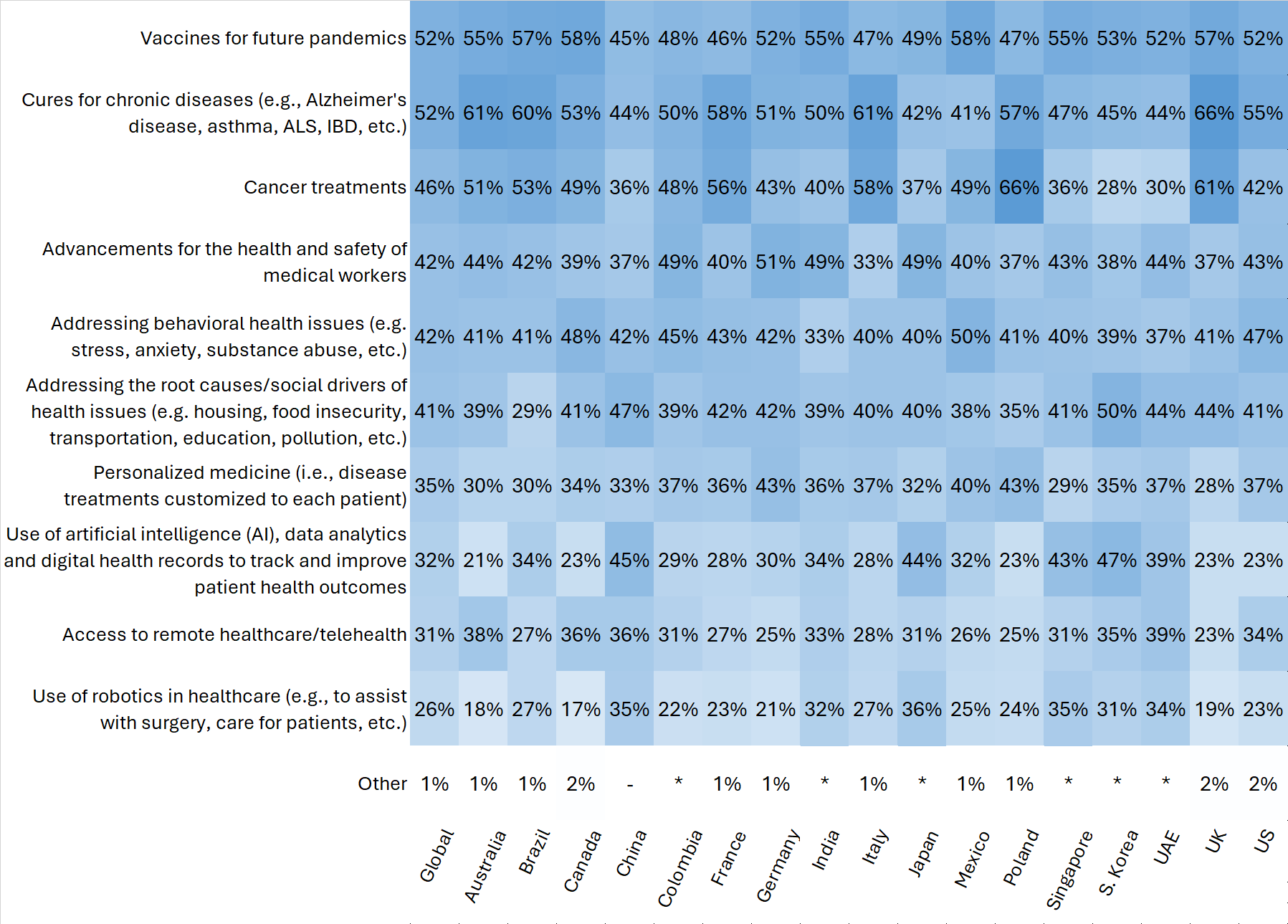}}
\caption*{\textbf{Figure A12: Responses for “Beyond the coronavirus/COVID-19, which FOUR of the following healthcare advancements should science prioritize? Select top four.” }(Question 40 from the 2021 3M State of Science Index) }
\end{figure}

\subsubsection*{2020 SOSI \cite{3}}
Two waves of data were released at once in 2020: one from our typical global survey, fielded pre-pandemic in August-October 2019, the other from a pulse survey conducted in the summer of 2020, after the pandemic hit.

\begin{figure}[H]
\fbox{\includegraphics[width=0.95\linewidth]{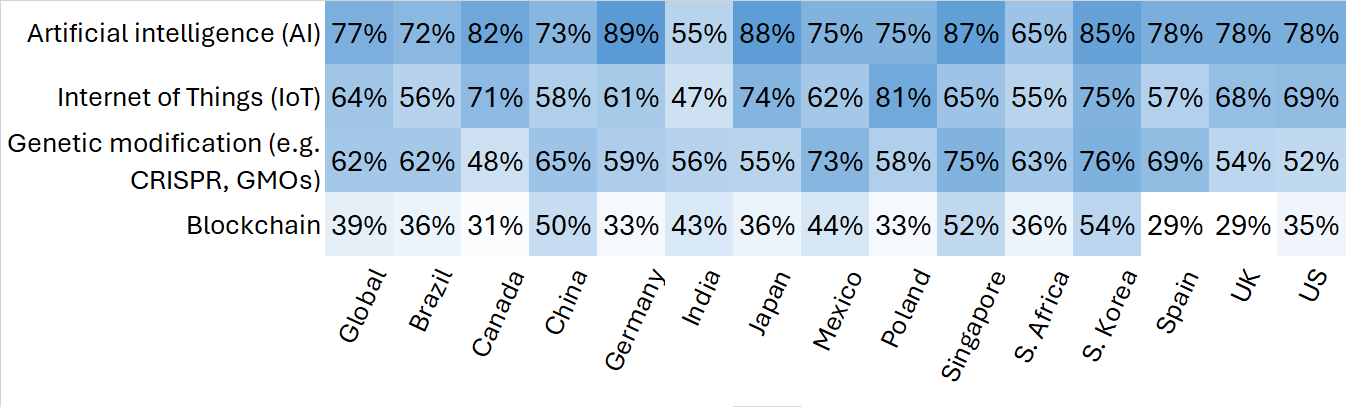}}
\caption*{\textbf{Figure A13: Responses for “How much do you know about each of the following terms? Artificial intelligence, Internet of Things, Genetic modification, and Block chain}Percent of respondents (pre-pandemic) who answered “know some” or “a lot” to each option.  (Question 7 from the 2020 pre-pandemic 3M State of Science Index [3]).  }
\end{figure}

\begin{figure}[H]
\fbox{\includegraphics[width=0.95\linewidth]{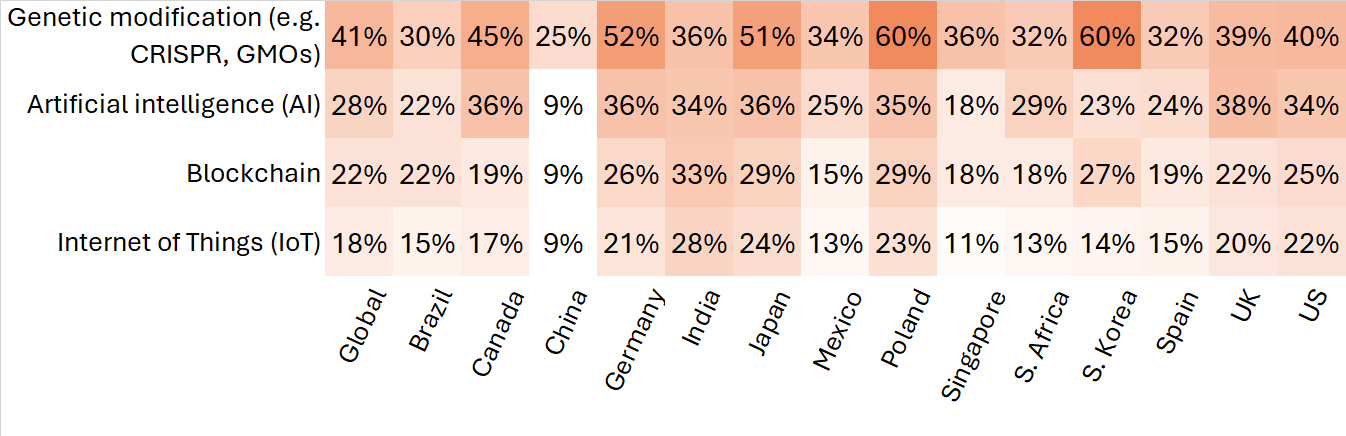}}
\fbox{\includegraphics[width=0.95\linewidth]{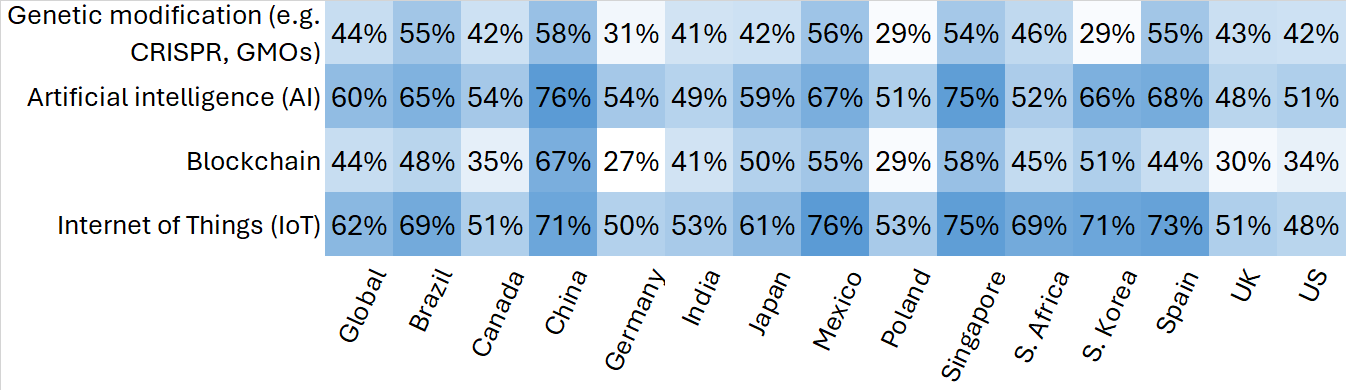}}
\caption*{\textbf{Figure A14A and A14B: Responses for “Do you consider each of the following a threat or improvement to society? CRISPR/GMOs, Artificial intelligence, Block chain, and Internet of Things.” }Figure A14A(top): Percent of respondents (pre-pandemic) who answered “threat.” Figure A14B(bottom): Percent of respondents (pre-pandemic) who answered “improvement” to each option. (Question 8 from the 2020 pre-pandemic 3M State of Science Index) 
  }
\end{figure}

\begin{figure}[H]
\centering
\fbox{\includegraphics[width=0.9\linewidth]{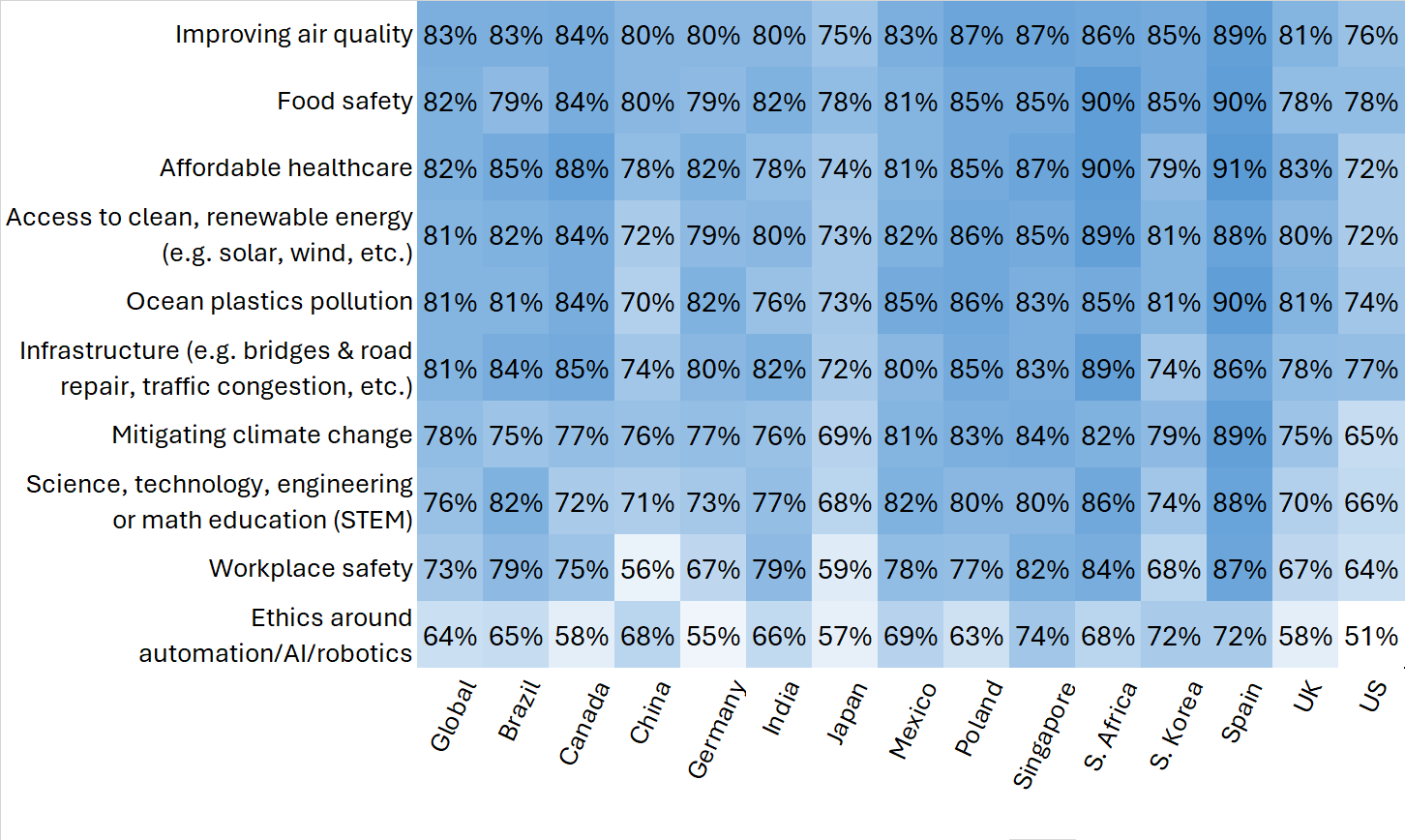}}
\fbox{\includegraphics[width=0.9\linewidth]{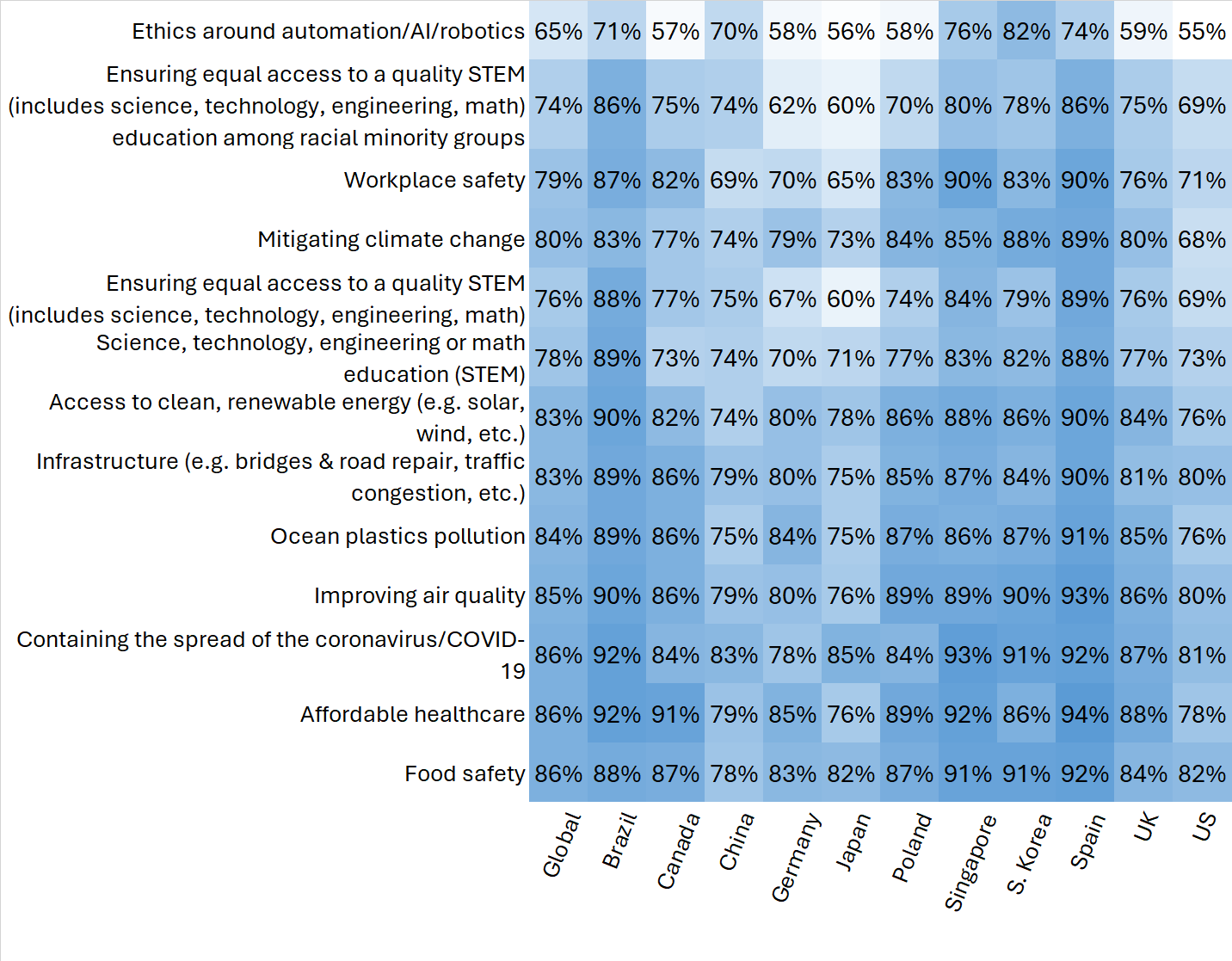}}
\caption*{\textbf{Figure A15A and A15B: Responses for “Do you think the government should be more or less involved in the following issues?” Percent of respondents who answered “more involved.” }Figure A15A(top): Pre-pandemic data. Figure A15B(bottom): Pandemic pulse data. (Question 22 from the 2020 pre-pandemic 3M State of Science Insights and question 8 from the 2020 pandemic pulse 3M State of Science Index) }
\end{figure}

\begin{figure}[H]
\centering
\fbox{\includegraphics[width=0.95\linewidth]{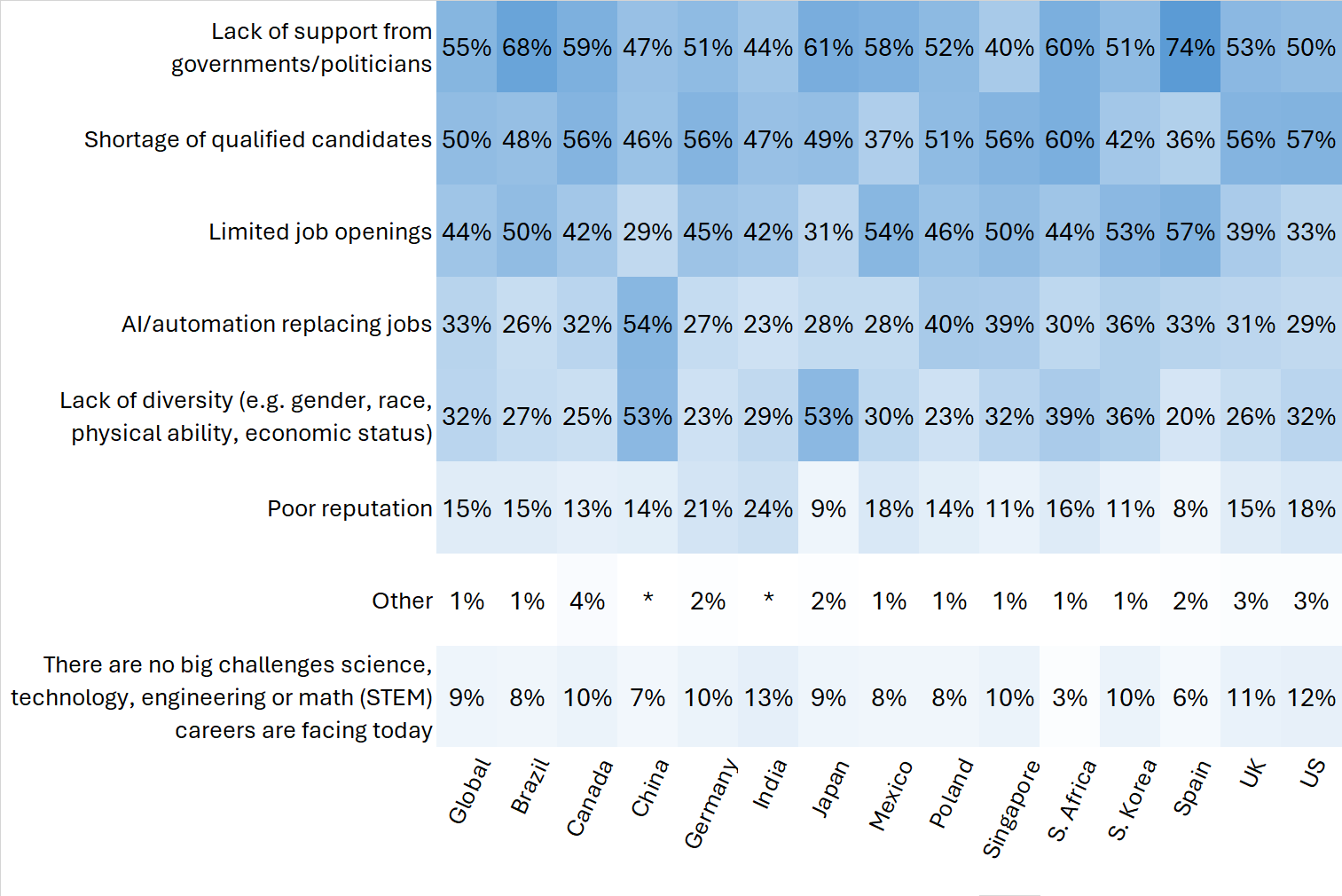}}
\caption*{\textbf{Figure A16: Responses for “What are the biggest challenges science, technology, engineering or math (STEM) careers are facing today? Select top three.” }Percent of respondents (pre-pandemic) who selected each option in their top three. (Question 35 from the 2020 pre-pandemic 3M State of Science Index) }
\end{figure}

\begin{figure}[H]
\centering
\fbox{\includegraphics[width=0.95\linewidth]{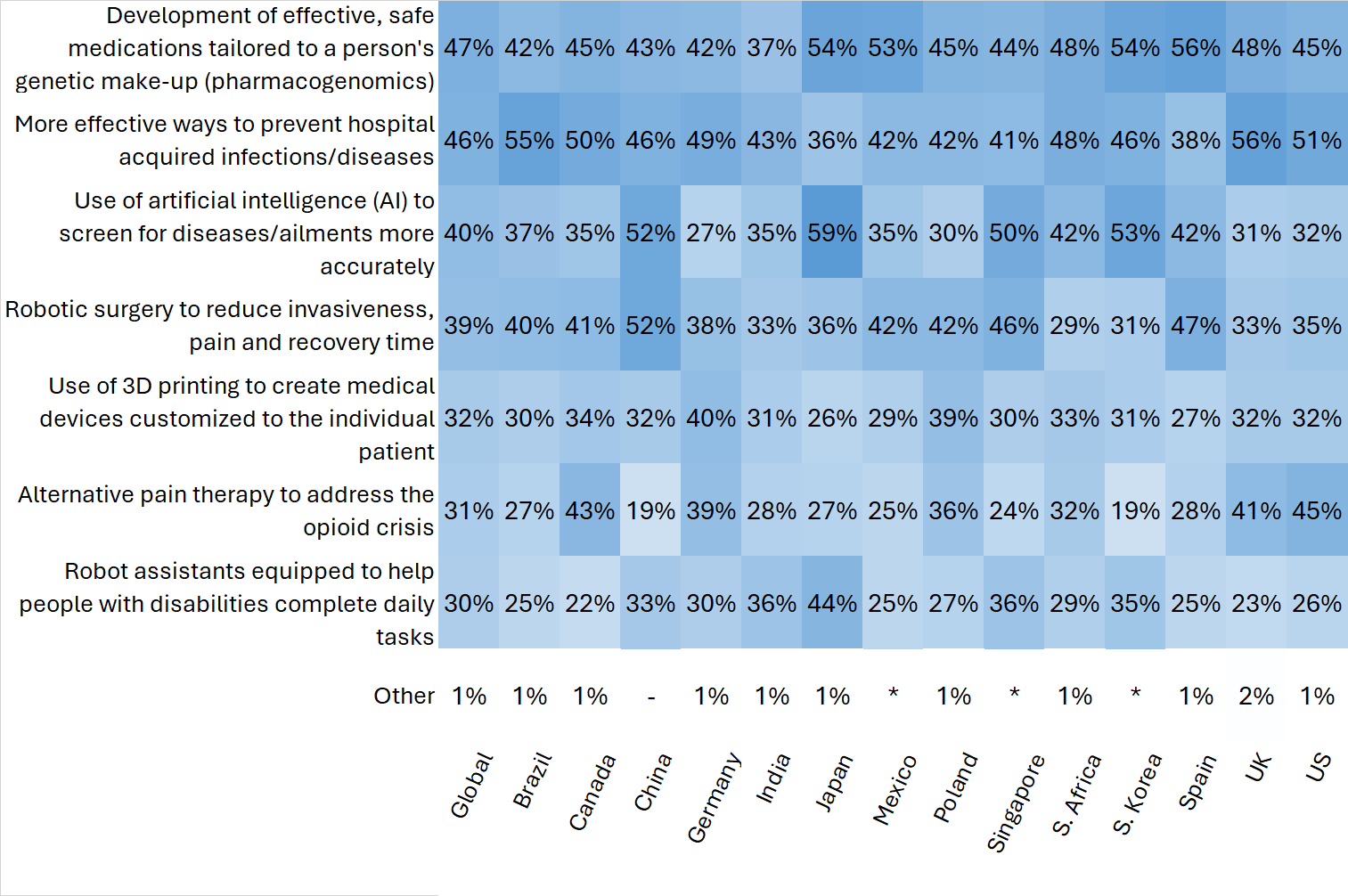}}
\caption*{\textbf{Figure A17: Responses for “Last year, we surveyed the general population and found that curing diseases is a top issue people want science to solve. Thinking about different healthcare advancements beyond disease cures, which, if any, of the following are you most excited about? Select top three.” }Percent of Respondents (pre-pandemic) who selected each option in their top three. (Question 45 from the 2020 pre-pandemic 3M State of Science Index) }
\end{figure}

\begin{figure}[H]
\centering
\fbox{\includegraphics[width=0.8\linewidth]{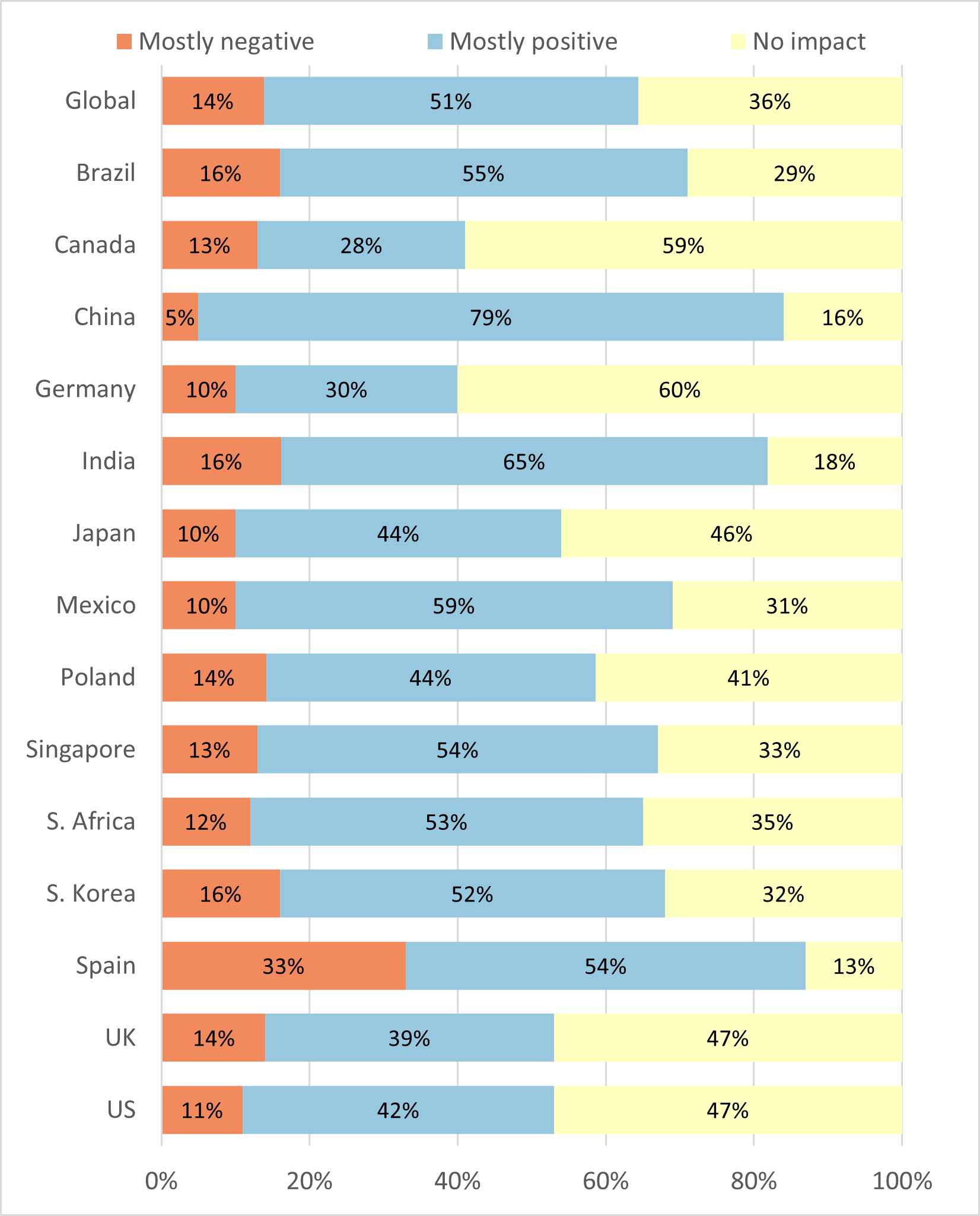}}
\caption*{\textbf{Figure A18: Responses for “What kind of impact, if any, has the integration of artificial intelligence(AI)/machine learning had on your job?” }(Question 52 from the 2020 pre-pandemic 3M State of Science Index)  }
\end{figure}

\begin{figure}[H]
\centering
\fbox{\includegraphics[width=0.95\linewidth]{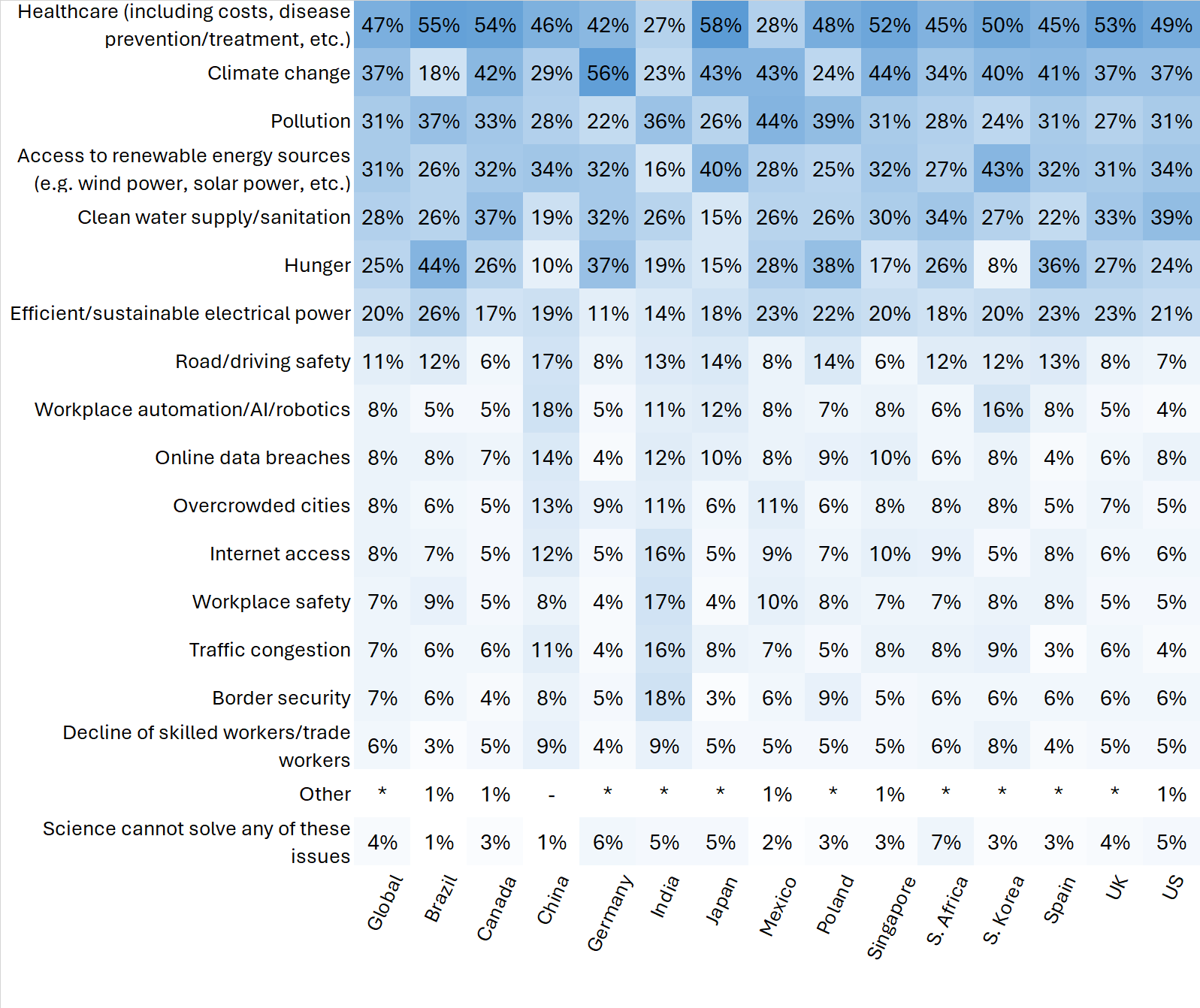}}
\caption*{\textbf{Figure A19: Responses for “Which, if any, of the following issues do you want science to most help solve? Please select top three.”}Percent of respondents who chose each option in their top three issues they most want science to help solve. (Question 36 from the 2019 3M State of Science Index) }
\end{figure}

\begin{figure}[H]
\centering
\fbox{\includegraphics[width=0.95\linewidth]{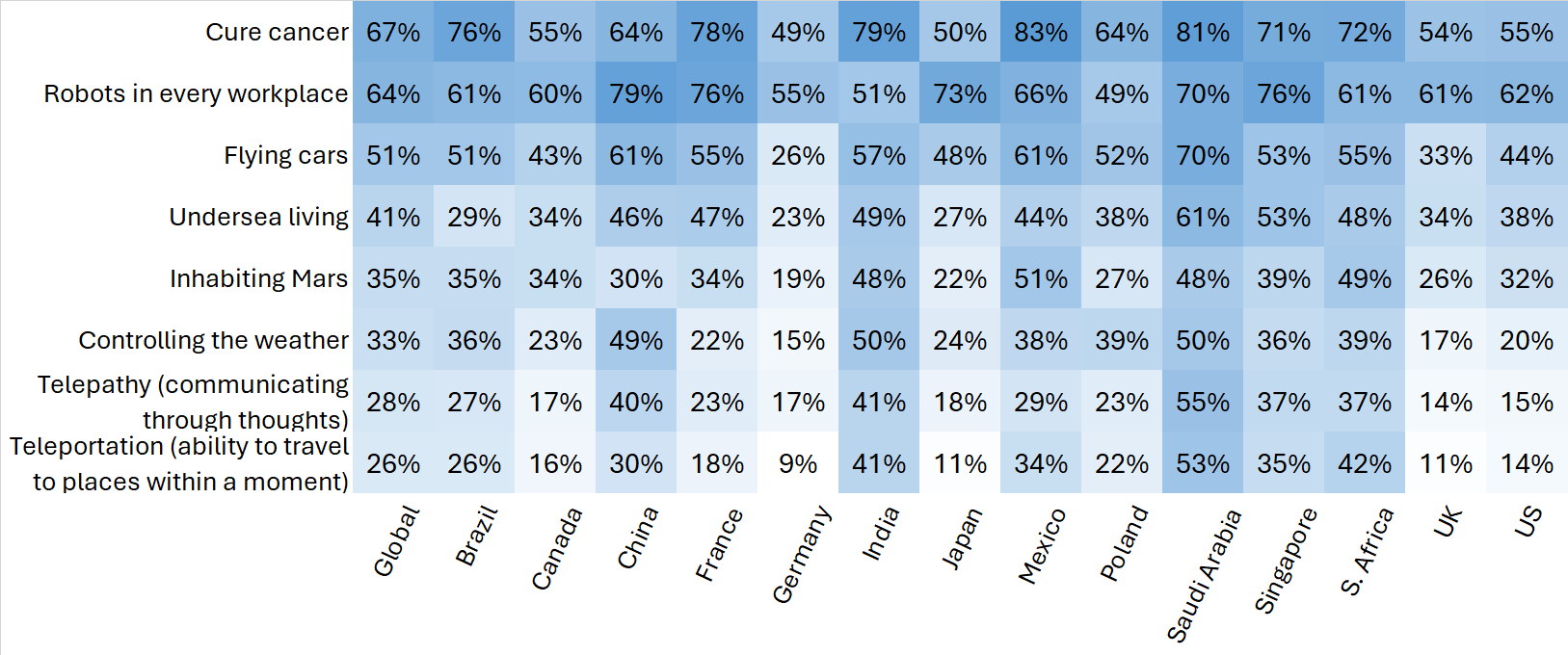}}
\caption*{\textbf{Figure A20: Responses for “Do you think science will achieve each of the following in your lifetime?”}Percent of respondents who answered yes to each option. (Question 22 from the 2018 3M State of Science Index)}
\end{figure}

\begin{center}
\section*{Appendix A5}
 Problems people most want science to help solve from 3M State of Science surveys (2018-2024)

 \begin{itemize}
     \item In 2018 we asked which issues people most wanted science to solve for society overall.
     \subitem Among the list of issues provided, the top two problems were disease treatment (53\%) and disease prevention (47\%). However, the next five top answers were all related to sustainability: access to affordable renewable energy sources (38\%), clean water supply and sanitation (38\%), energy supply (25\%), pollution (29\%), and climate change (36\%) appeared in the 7th spot from the top.
     \item In 2019 we asked which issues people most wanted science to help solve.
     \subitem Among the list of issues provided, after healthcare (47\%), the top problems were again all related to sustainability. In second place was climate change (37\%), followed by pollution (31\%), access to renewable energy sources (31\%), and clean water supply/sanitation (28\%).
     \item In 2020, pre-pandemic, we asked (beyond healthcare) which issues do you most want science to help solve.
     \subitem Among the list of issues provided, the number one problem respondents wanted to solve was climate change (49\%). Following this, the next top four issues were all related to sustainability. Air quality (45\%), ocean plastics pollution (42\%), clean water supply/sanitation (42\%), and access to renewable energy sources (39\%).
     \item In 2021, we asked which issues people most wanted science to help solve.
     \subitem Among the list of issues provided, the number one issue was the coronavirus/COVID-19 pandemic (51\%), and the next six were all related to sustainability challenges – climate change (47\%) in the number two spot, followed by ocean plastics pollution (41\%), access to renewable energy sources (38\%), air quality (38\%), clean water supply/sanitation (36\%), and reliance on fossil fuels (29\%).
     \item In 2022, we asked what issues (beyond the COVID-19/coronavirus pandemic) people most wanted science to help solve.
     \subitem Among the list of issues provided the top three problems were effects of climate change (58\%), followed by clean water supply/sanitation (56\%), and air quality (55\%).
     \item In 2023, we asked what the biggest priorities for science were to address in your country right now.
     \subitem Among the list of issues provided the top answers were minimizing and mitigating climate change (48\%), developing technology to predict and prevent natural disasters (45\%), and preparing for future global pandemics (45\%).
     \subitem In 2023, we also asked, “looking to the future, how concerned are you that you and/or a loved one may be displaced from where you live due to each of the following extreme weather conditions related to climate change?” Among the list of issues provided, people were somewhat/very concerned about droughts (86\%), fires (86\%), floods (85\%) and heat waves (84\%).
     \item In 2024, we asked “how well do each of the following align with what you believe science and technological innovation should focus on.”
     \subitem Science and technological innovation should advance people: End poverty and zero hunger,\\ health and well-being, gender and racial equality, quality education. (86\% answered “somewhat\\ well” or “very well”)
     \subitem Science and technological innovation should advance the planet: Clean water and sanitation, clean air, wildlife conservation, environmental conservation, sustainable consumption and production, natural resource management, and climate change solutions. (89\% answered “somewhat well” or “very well”)
     \subitem Science and technological innovation should advance prosperity: Economic growth, job creation, clean and affordable energy. Ensure that all human beings can enjoy prosperous and fulfilling lives and that economic, social, and technological progress occurs in harmony with nature. (88\% answered “somewhat well” or “very well”)
 \end{itemize}

 \end{center}
\end{document}